\documentclass[aps,pra,twocolumn,nofootinbib,superscriptaddress,preprintnumbers,floatfix,10pt]{revtex4-2}

\usepackage[english]{babel} 
\usepackage[utf8]{inputenc}
\usepackage[T1]{fontenc}
\usepackage{lmodern}
\usepackage{nameref}

\usepackage{amsmath,amsfonts,amssymb}
\usepackage{physics}
\usepackage{bbm}
\usepackage{graphicx}

\usepackage[colorlinks=true, linkcolor =blue, citecolor=green!70!black]{hyperref}

\usepackage{xcolor}
\usepackage{soul} 										%%changed ulem to soul cause ulem interferred with the bibliography

\newcommand{\I}{\mathrm{i}}
\newcommand{\E}{\mathrm{e}}

\DeclareMathOperator{\diag}{\mathrm{diag}}

\DeclareMathOperator{\arsinh}{\mathrm{arsinh}}
\newcommand*{\dg}{^{\dagger}}
\newcommand*{\tran}{^{\mkern-1.5mu\mathsf{T}}}
\newcommand{\ten}[1]{\boldsymbol{\mathbf{#1}}}

\newcommand{\rmF}{\mathrm{F}}
\newcommand{\rmmin}{\mathrm{min}}
\newcommand{\zt}{\tilde{\xi}}

\begin{document}

\title{Optimizing photon-number distributions of Gaussian states in the presence of loss: \\Towards minimizing the impact of loss in Gaussian boson sampling}

\author{Hendrik Ellenberg}
\email{hendrik.ellenberg@uni-jena.de}
\affiliation{Institute of Condensed Matter Theory and Optics, Friedrich-Schiller-University Jena, Max-Wien-Platz 1, 07743 Jena, Germany}

\author{René Sondenheimer}
\email{rene.sondenheimer@uni-jena.de}
\affiliation{Institute of Condensed Matter Theory and Optics, Friedrich-Schiller-University Jena, Max-Wien-Platz 1, 07743 Jena, Germany}
\affiliation{Fraunhofer Institute for Applied Optics and Precision Engineering IOF, Albert-Einstein-Str. 7, 07745 Jena, Germany}

\begin{abstract}
We analyze the impact of photon loss on the photon-number statistics of Gaussian states. 
Specifically, we propose and carefully evaluate several methods to mitigate deviations in the photon-number distributions of lossy (displaced) squeezed vacuum states from those of their lossless counterparts.  
These methods rely on appropriately redefining the parameters of Gaussian states when the loss budget is known in order to recover, as closely as possible, the desired photon-number distribution associated with each target state.
While it is intrinsically hard to directly optimize the photon-number distribution of high-dimensional, correlated multimode Gaussian states, the proposed methods are instead based on optimizing specific key properties such as fidelity, phase-space functions, low-order moments of the underlying photon-number statistics, or overlap with the vacuum state.
In particular, our results show that optimizing the fidelity between a pure Gaussian target state and a modified Gaussian state that has passed through a loss channel does typically not result in closeness of the corresponding photon-number distributions.
Furthermore, we show that correcting for the vacuum overlap minimizes the deviation in the photon-number distribution for large parameter ranges which we explicitly prove for single-mode squeezed vacuum and provide numerical evidence for general (displaced) squeezed vacuum states. 
As photon loss is a key limitation for Gaussian boson sampling, our results provide insights into the feasibility and limitations of such photonic quantum simulations in lossy environments and offer guidelines for mitigating these imperfections.
\end{abstract}

\maketitle

\section{Introduction}
Quantum simulators are tailored devices that offer a versatile approach to analyzing specific complex quantum systems.
The quantum hardware proposed for these simulations is as varied as the problems they aim to address.
These include trapped ions~\cite{blattQuantumSimulationsTrapped2012, lanyonUniversalDigitalQuantum2011, monroeProgrammableQuantumSimulations2021, porrasEffectiveQuantumSpin2004}, cold atoms~\cite{aidelsburgerColdAtomsMeet2021, schweizerFloquetApproachZ22019}, superconducting circuits~\cite{guoObservationDynamicalQuantum2019, mezzacapoNonAbelianSU2Lattice2015}, quantum dots~\cite{barthelemyQuantumDotSystems2013, vandiepenQuantumSimulationAntiferromagnetic2021}, and photons~\cite{aspuru-guzikPhotonicQuantumSimulators2012, harrisQuantumTransportSimulations2017, maQuantumSimulationWavefunction2011, xueExperimentalLinearopticsSimulation2017}.
The latter platform has addressed a diverse range of potential problems, spanning quantum random walks~\cite{broomeDiscreteSinglePhotonQuantum2010, peruzzoQuantumWalksCorrelated2010}, particle physics~\cite{matthewsObservingFermionicStatistics2013, matthewsSimulatingArbitraryQuantum2013} or quantum chemistry~\cite{babbushChemicalBasisTrotterSuzuki2015, kassalSimulatingChemistryUsing2011, lanyonQuantumChemistryQuantum2010}, for instance.

A notable development in this field is boson sampling~\cite{aaronsonComputationalComplexityLinear2010}, which utilizes photons to perform computations for which we have accumulated evidence that they are infeasible for classical computers~\cite{aaronsonBosonsamplingFarUniform2014}. 
Boson sampling is a nonuniversal model of quantum computation in which $N$ indistinguishable single-photon Fock states are injected into an $M$-mode passive linear-optical network. 
Sampling from the photon-number distribution of the output state is classically hard as the computation of the photon-number distribution is tied to computing the permanent of a matrix, a problem known to be \linebreak$\mathbf{\#P}$-complete~\cite{scheelPermanentsLinearOptical2004}.
Consequently, boson sampling is a reasonable candidate to falsify the extended Church-Turing thesis~\cite{aaronsonComputationalComplexityLinear2010}.
Although there have been explorative demonstrations of boson sampling~\cite{springBosonSamplingPhotonic2013, brodPhotonicImplementationBoson2019, wangScalableBosonSampling2018, wangBosonSampling202019, maxtillmannExperimentalBosonSampling2013, matthewa.broomePhotonicBosonSampling2013}, implementing it remains difficult.
This challenge arises because the original protocol relies on deterministic single-photon sources, which are currently the subject of intense investigations~\cite{daehyunahnBroadbandHighbrightnessQuantumdot2023, huiwangOndemandSemiconductorSource2019, niccolosomaschiNearoptimalSinglephotonSources2016, xingdingOnDemandSinglePhotons2016}.
The challenge to create and interfere multiple indistinguishable single-photon states led to variants of the original boson sampling protocol, in which Gaussian states, specifically squeezed states of light, were used as a nonclassical resource.
Initially, this was proposed in scattershot boson sampling~\cite{lundBosonSamplingGaussian2014, chenglongyouMultiparameterEstimationSingle2017, sonjabarkhofenDrivenBosonSampling2017} and then further developed to Gaussian boson sampling (GBS)~\cite{hamiltonGaussianBosonSampling2017, kruseDetailedStudyGaussian2019, madsenQuantumComputationalAdvantage2022}. 
It has been shown that the resulting output photon-number probabilities are proportional to hafnians of matrices determined by the system parameters, yielding a sampling problem that is hard to simulate classically. 
First experimental investigations can be found in Refs.~\cite{paesaniGenerationSamplingQuantum2019, arrazolaQuantumCircuitsMany2021, zhongExperimentalGaussianBoson2019, zhongQuantumComputationalAdvantage2020, zhongPhaseProgrammableGaussianBoson2021, liBenchmarking50PhotonGaussian2022, liuRobustQuantumComputational2025}.

Besides conceptual aspects of computational complexity theory, GBS can also be used to simulate specific problems in an efficient manner.
Huh et al.~showed that GBS can be used to compute Franck-Condon factors (FCFs)~\cite{huhBosonSamplingMolecular2015}, a quantum chemical problem that is intractable for classical computers. 
Moreover, GBS has applications in graph optimization~\cite{bradlerGaussianBosonSampling2018, arrazolaUsingGaussianBoson2018, cazalisGaussianBosonSampling2024a, ohQuantumInspiredClassicalAlgorithm2024, zhangEfficientClassicalSampling2025, heTimedomainmultiplexedGaussianBoson2025}, graph similarity~\cite{bradlerDualityHeartGaussian2019, schuldQuantumHardwareinducedGraph2020, schuldMeasuringSimilarityGraphs2020a}, point processes~\cite{jahangiriPointProcessesGaussian2020}, and molecular docking~\cite{banchiMolecularDockingGaussian2020, yuUniversalProgrammableGaussian2023}. 
In addition, hybrid photonic sampling platforms have been proposed to systematically compare Gaussian and non-Gaussian input states~\cite{stefszkyBenchmarkingGaussianNonGaussian2025}.

Both the complexity theory problems and the intricate physical or mathematical problems simulated by (Gaussian) boson sampling are subject to imperfections in any experimental realization of such a system. 
A central question is whether the resulting noisy sampling distribution can be simulated classically in an efficient manner, which is particularly important in the complexity-theoretic setting~\cite{aaronsonBosonSamplingLostPhotons2016, rahimikeshariSufficientConditionsEfficient2016, oszmaniecClassicalSimulationPhotonic2018, garcia-patronSimulatingBosonSampling2019, brodClassicalSimulationLinear2020, qiRegimesClassicalSimulability2020, ohClassicalSimulationLossy2021, bulmerBoundaryQuantumAdvantage2022,umanskiiClassicalModellingLossy2024}. 
Beyond these complexity-theoretic considerations, Clements~et~al.~investigated the relationship between system noise and the resulting photon-number distributions of a particular two-mode GBS system used to calculate FCFs of tropolone~\cite{clementsApproximatingVibronicSpectroscopy2018}. 
To account for imperfections, the fidelity was maximized between a lossy Gaussian state and a target Gaussian state where the latter implements the desired GBS protocol for an imperfection-free realization.
Furthermore, a classicality benchmark was introduced in this context that any experimental realization has to outperform.
In general, various sources of photon loss, e.g., scattering and absorption, may contribute to a possible reduction of the precision of the simulation results in optical platforms.
A detailed understanding of the impact of these imperfections on the photon-number distribution is indispensable to avoid misinterpreting results from a GBS experiment.

In this work, we further explore the impact that photon losses have on the photon-number statistics of pure Gaussian states. This detrimental influence might be mitigated by finding another Gaussian state whose photon-number distribution after loss approximate that of the desired target state in the loss-free setting, i.e., by appropriately redefining the experimental input parameters of a GBS experiment. 
While directly optimizing for the photon-number distribution is a computationally hard problem, we are able to exploit the fact that Gaussian states remain Gaussian under loss.
Therefore, we can use other measures that can be computed efficiently to optimize the similarity between a lossy Gaussian state and a desired Gaussian target state.
In addition to the fidelity of two Gaussian states, we investigate various other schemes that are related to the phase-space distributions of these states, low-order moments of the photon-number statistics, the overlap with the vacuum state, or perform classical corrections of the displacements.
We carefully compare these schemes regarding the total variation distance of the photon-number distributions of the resulting states with the desired target distribution and show that vacuum overlap correction provides an optimal choice for large parameter ranges.

In Sec.~\ref{sec:Mitigation}, we give a short recap of the Wigner function formalism for Gaussian states, loss channels, and photon-number distributions for the sake of briefly introducing our nomenclature and conventions.
Then, we introduce several mitigation schemes to reduce the impact of photon loss on the photon-number distribution. 
In Sec.~\ref{sec:smSqueezedState}, we apply these methods to the simplest nontrivial case, single-mode squeezed vacuum.
Subsequently, we discuss particularities for single-mode displaced squeezed vacuum states in Sec.~\ref{sec:smDisplacedSqueezedState}.
A broadening of the discussion to the multimode case is given in Sec.~\ref{sec:MultiMode}.
In particular, we demonstrate the advantage of our methods by applying it to the calculation of FCFs of simple molecules for lossy GBS devices.
Section~\ref{sec:Conclusions} concludes this work.

\section{Strategies for mitigating the impact of loss on the photon-number distributions of Gaussian states}
\label{sec:Mitigation}
Throughout this paper, we examine bosonic systems comprising of $M$ harmonic oscillators (modes) with annihilation and creation operators $\hat{a}_j$ and $\hat{a}\dg_j$ for the \mbox{$j$-th} mode. By defining the quadrature operators ${\hat{x}_j=\frac{1}{\sqrt{2}}\big(\hat{a}_j+\hat{a}\dg_j\big)}$ and $\hat{p}_j=\frac{1}{\I\sqrt{2}}\big(\hat{a}_j-\hat{a}\dg_j\big)$ and collecting these operators in
\begin{equation}
	\hat{\vb{r}}=\qty(\hat{x}_1,\,\hat{p}_1,\,\cdots,\,\hat{x}_M,\,\hat{p}_M)\tran
\end{equation}
results in a $2M$-dimensional vector. 
Any Gaussian state $\hat{\rho}$ can be characterized using only the first and second moment of this combined quadrature operator $\hat{\vb{r}}$.
These are given by the mean $\bar{\vb{r}}=\expval{\hat{\vb{r}}}$ and the covariance matrix $\sigma_{jk}=\frac{1}{2}\expval{\{\hat{r}_j,\hat{r}_k\}}-\expval{\hat{r}_j}\expval{\hat{r}_k}$, where $\{\cdot,\cdot\}$ denotes the anticommutator.

As Gaussian transformations map Gaussian states to Gaussian states, they can be fully characterized by how they affect the first and second moments. At this level, these operations act linearly, yielding an efficient description of the state's evolution~\cite{braunsteinQuantumInformationContinuous2005}.
However, when non-Gaussian transformations or measurements are applied, modeling the state or obtaining measurement outcomes becomes intricate for large $M$ due to the unavoidable exponential growth of the Hilbert space with the number of modes and the resulting computational complexity of tracking the full photon-number distribution.
In particular, applications of GBS rely on the photon-number distributions of the output states, which are obtained by photon-number detection, a non-Gaussian measurement. 
Obtaining the photon-number distribution of an $M$-mode Gaussian state classically can be done by computing a $2M$-variate Hermite polynomial as discussed in~\cite{dodonovMultidimensionalHermitePolynomials1994}. 
Similarly, the probabilities, moments, and factorial moments of the photon-number statistics can be obtained by differentiation of generating functions depending on the covariance matrix and mean of multimode Gaussian states~\cite{fitzkeSimulatingPhotonStatistics2023}.
While both formulations face computational complexity for large mode and photon numbers, the latter enables practical implementations for sufficiently small photon numbers.

Restricting ourselves to the case of pure states, any multimode Gaussian state can be described by a multimode displacement and squeezing operator acting on the vacuum, $\hat{D}\qty(\vb*{\alpha})\hat{S}\qty(\ten{\xi})\ket{0}$ where $\ten{\xi}$ is an $M\times M$-dimensional complex symmetric matrix encoding the multimode squeezing, $\hat{S}(\ten{\xi}) = \E^{\frac{1}{2}\xi_{ij}^{*}\hat{a}_i \hat{a}_j - \frac{1}{2}\xi_{ij}\hat{a}_i^\dagger\hat{a}_j^\dagger }$, and $\vb*{\alpha}$ being an $M$-dimensional complex vector defining the amount of displacement, $\hat{D}(\vb*{\alpha}) = \E^{\alpha_i\hat{a}_i^\dagger -  \alpha_{i}^{*}\hat{a}_i}$. Such a state can be experimentally generated by preparing $M$ uncorrelated single-mode displaced squeezed vacuum states and mixing them via a linear interferometer implementing a rotation $\hat{R}\qty(\ten{U}) = \E^{\I \hat{a}_i^{\dagger}H_{ij} \hat{a}_j}$ with $\hat{R}\qty(\ten{U})^{\dagger} \hat{a}_i \hat{R}\qty(\ten{U}) = (\E^{-\I \ten{H}})_{ij} \hat{a}_{j} \equiv (\ten{U})_{ij} \hat{a}_{j}$ among the modes where $\ten{U}$ is a unitary matrix. Thus before any measurement process, the output state of an ideal GBS system can be described with three Gaussian operations~\cite{maMultimodeSqueezeOperators1990},
\begin{equation}\label{eqn:GBS}
	\ket{\mathrm{GBS} (\ten{\xi},\vb*{\alpha},\ten{U})} = \hat{R}\qty(\ten{U}) \hat{D}\qty(\vb*{\alpha})\hat{S}\qty(\ten{\xi})\ket{0} \equiv \hat{U}_\mathrm{GBS}\ket{0},
\end{equation}
where $\ten{U}$ can be a general element of the unitary group $\mathrm{U}(M)$, $\ten{\xi}$ is now a complex diagonal $M\times M$ matrix, and $\vb*{\alpha}$ still denotes an arbitray $M$-dimensional complex displacement vector. 
However, for specific GBS applications, the arguments of the transformations can be further restricted. For instance, for the case of FCFs and perfect matchings in graph theory, $\ten{U}$ is an orthogonal matrix, $\ten{\xi}$ can be restricted to a real diagonal matrix, and $\vb*{\alpha}$ to a real vector~\cite{huhBosonSamplingMolecular2015,bradlerGaussianBosonSampling2018}.

\begin{figure}[t]
 \centering
 \includegraphics[width=0.95\columnwidth]{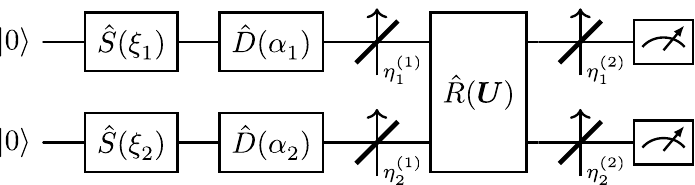}
 \caption{Two-mode GBS system with additional beam splitters modeling loss at relevant locations. Each $\eta_i^{(l)}$ gives the transmissivity for an incoming photon, while $1-\eta_i^{(l)}$ is the corresponding loss probability. The index $i$ denotes the mode and the superscript $l$ the layer in the system.}
 \label{fig:DoktorovSO2MitLoss}
\end{figure}

\begin{figure}[t]   
    \centering
    \includegraphics[width=1\columnwidth]{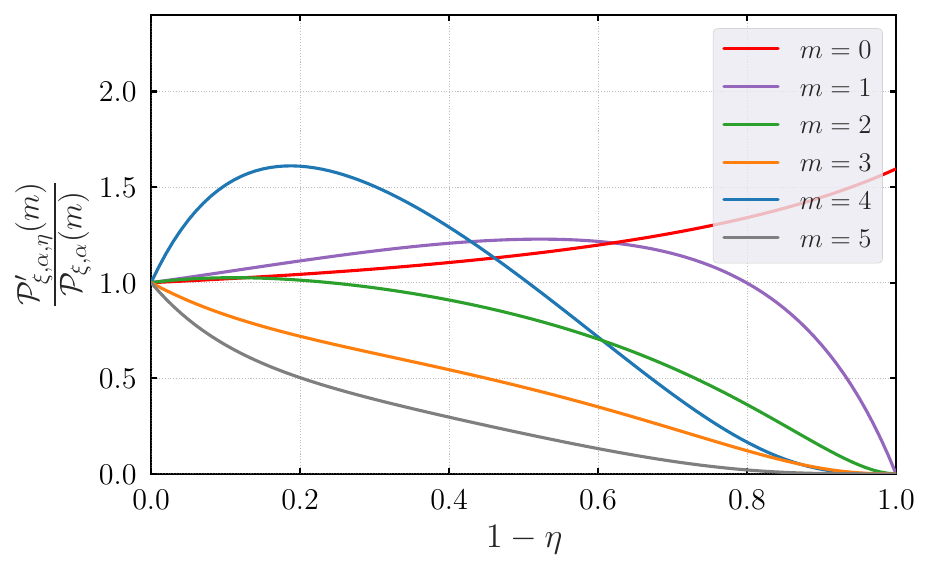}
    \caption{Effect of loss on photon-number distribution. For a displaced squeezed vacuum state with $\alpha = 0.25$ and $\xi = 0.9$, we plot, the ratio $\mathcal{P}'_{\xi,\alpha;\eta}(m)/\mathcal{P}_{\xi,\alpha}(m)$ as a function of loss ($1-\eta$) for selected photon numbers $m=0,\dots,5$. $\mathcal{P}_{\xi,\alpha}(m)$ denotes the probabilities of the target distribution given by the displaced squeezed vacuum state while $\mathcal{P}'_{\xi,\alpha;\eta}(m)$ is the corresponding probability after a pure-loss channel with transmissivity $\eta$. Ratios larger than unity arise because loss redistributes probability weight from higher to lower photon numbers, thereby potentially increasing the latter.} 
    \label{fig:FdF}
\end{figure}

In this paper, we investigate the impact of photon loss as an imperfection in the system which can also be modeled as a Gaussian process.
If the system exhibits a certain probability $1-\eta^{(l)}_i$ that a photon is lost in mode $i$ at a specific position (layer) $l$ in the system, we model this effect as an additional beam splitter with transmissivity $\eta^{(l)}_i$, see~Fig.~\ref{fig:DoktorovSO2MitLoss}.
At this beam splitter, the incoming photons have a probability of $1-\eta^{(l)}_i$ to scatter into the environment.
Therefore, we formally introduce ancillary loss modes that are initially in the vacuum state and subsequently trace over all such modes in the output state.
As all these operations and involved states are Gaussian, the final state will remain Gaussian (but will be no longer a pure state) and can efficiently be calculated.
However, the resulting photon-number distribution will now be affected by the parameters $\eta^{(l)}_i$ parametrizing the loss, cf.~Fig.~\ref{fig:FdF}.

In order to quantify the dissimilarity of two photon-number distributions, e.g., of an ideal situation without imperfections in the system and its lossy version, we use the total variation distance as an appropriate measure,
\begin{align}
	\delta(\hat{\rho}, \hat{\rho}')
	&\equiv\frac{1}{2}\sum_{\vb{m}}\abs{\mathcal{P}(\vb{m})-\mathcal{P}'(\vb{m})} \notag	\\
	&=\frac{1}{2}\sum_{\vb{m}}\abs{\Tr\big(\hat{\rho}\op{\vb{m}}\big)-\Tr\big(\hat{\rho}'\op{\vb{m}}\big)},
\end{align}
where $\mathcal{P}(\vb{m})$ and $\mathcal{P}'(\vb{m})$ denote the probabilities to measure the multimode Fock state $\ket{\vb{m}}$ of the two density matrices $\hat{\rho}$ and $\hat{\rho}'$, respectively.
The vector $\vb*{m}\in\mathbb{N}_0^M$ characterizes a multimode number state where each component of that vector corresponds to one of the $M$ modes of the system. In our case, $\hat{\rho}$ is a pure multimode Gaussian state with the desired target photon-number distribution given by 
$\hat{\rho}(\tilde{\vb*{\xi}},\tilde{\vb*{\alpha}},\tilde{\ten{U}}) = \op{\mathrm{GBS} (\tilde{\vb*{\xi}},\tilde{\vb*{\alpha}},\tilde{\ten{U}})}$, where we introduced the notation that quantities with a tilde indicate the parameters of the lossless target state. The state $\hat{\rho}'$ will be some multimode Gaussian state that results from a Gaussian state that has been affected by all potential system losses. Thus it will be a mixed state depending on the parameters of the GBS setup as well as the loss parameters $\vb*{\eta}$, ${\hat{\rho}'(\vb*{\xi},\vb*{\alpha},\ten{U};\vb*{\eta}) = \mathcal{E}_{\vb*{\eta}}(\op{\mathrm{GBS(\vb*{\xi},\vb*{\alpha},\ten{U})}})}$, with $\mathcal{E}_{\vb*{\eta}}$ the quantum channel describing the entire system loss being a composition of individual loss channels $\mathcal{E}_{\eta_i^{(l)}}$ acting on mode $i$ at layer $l$.

In the lossless case, we have $\hat{\rho}'(\tilde{\vb*{\alpha}},\tilde{\vb*{\xi}},\tilde{\ten{U}};\eta_i^{(l)}=1) = \hat{\rho} $ and thus obtain $\delta=0$. 
However, the measured photon-number distribution of $\hat{\rho}'$ deviates from the target distribution of $\hat{\rho}$ for nonzero loss, $\delta(\hat{\rho}, \hat{\rho}')>0$.
In order to mitigate the effects of loss on the photon-number distribution, the goal is to decrease $\delta$ to get as close as possible to the photon-number distribution of the target state $\hat{\rho}$. 
This might be achieved by changing the input parameters (squeezing $\ten{\xi}$, displacement $\vb*{\alpha}$, interferometer $\ten{U}$) to adjust $\hat{\rho}'$ in such a way that $\delta$ is minimized. 
However, since $\delta$ is intricately tied to the photon-number distribution of the output state, no standard optimization algorithm (such as gradient descend) for minimizing $\delta$ is efficient as the number of modes in the GBS system increases. 
By contrast, other measures can efficiently be optimized due to the Gaussian nature of the involved states which we group into five main classes. These are the fidelity, a method that we will denote as displacement correction, phase-space distributions, low-order moments of the photon-number distribution, as well as vacuum overlap.
As these methods increase the similarity of the involved states with respect to some specific notion, it is worthwhile to examine if such optimizations can also significantly decrease $\delta$. 
In the following, we distinguish between correction methods and optimization methods.
Correction refers to choosing parameters that compensate loss so the quantity of interest matches the target value (e.g., displacements, mean photon number, vacuum overlap).
Optimization refers to choosing parameters that minimize the deviation from the target-state quantity of interest (fidelity, phase-space distributions).

\subsection{Fidelity optimization}
\label{sec:Fidelity}
It is possible to compute the fidelity between two Gaussian states using only the first two moments of the quadrature operators. If at least one state is pure, it reduces to
\begin{equation}\label{eqn:Fidelity}
	F(\hat{\rho},\hat{\rho}') = \frac{1}{\sqrt[4]{\det(\ten{\sigma}+\ten{\sigma}')}}\E^{-\frac{1}{4}(\bar{\vb{r}}'-\bar{\vb{r}})\tran(\ten{\sigma}+\ten{\sigma}')^{-1}(\bar{\vb{r}}'-\bar{\vb{r}})}.
\end{equation}
While unprimed moments are fixed by the desired Gaussian state $\hat{\rho}$ corresponding to the target photon-number distribution of interest, we can tune the primed moments $\ten{\sigma}'$ and $\bar{\vb{r}}'$ via redefining the input parameters of a state that is impacted by loss. 
From Eq.~\eqref{eqn:Fidelity}, we can already infer that the fidelity maximization decouples into independent optimizations over the displacement and squeezing parameters, since the first moments depend only on the displacements (and $\vb*{\eta}$) due to the chosen operator-ordering convention in Eq.~\eqref{eqn:GBS}, while the covariance matrix is only affected by the squeezing parameters and loss.
Any mismatch in the means reduces the fidelity. As loss merely rescales displacement in a classical fashion, $\alpha \to \sqrt{\eta} \alpha$, we can straightforwardly determine the optimal displacement maximizing the fidelity by matching first moments, i.e., enforcing $\bar{\vb r}'-\bar{\vb r} = 0$. For the single-mode case, we have to adjust the displacement as $\alpha_{\rmF}=\tilde\alpha/\sqrt{\eta}$ and in general one has to solve an $M$-dimensional system of linear equations to obtain $\vb*{\alpha}_{\rmF}$. Determining the optimal squeezing parameters now reduces to the task to minimize $\det(\ten{\sigma}+\ten{\sigma}')$.

This error mitigation strategy has already been applied in Ref.~\cite{clementsApproximatingVibronicSpectroscopy2018} for the simulation of FCFs.
At first sight, one would expect that finding a new set of initial parameters $(\vb*{\alpha}_\rmF,\vb*{\xi}_\rmF,\ten{U}_\rmF)$, that maximizes the fidelity $F\big( \hat{\rho}(\tilde{\vb*{\alpha}},\tilde{\vb*{\xi}},\tilde{\ten{U}}), \hat{\rho}'(\vb*{\alpha}_\rmF,\vb*{\xi}_\rmF,\ten{U}_\rmF;\vb*{\eta}) \big)$ for a given loss budget, decreases the total variation distance in contrast to the case where the initial parameters of the Gaussian state are unchanged.
However, we will see in the following sections that this is not the case over a wide parameter range. 
Even worse, it is usually the case that $\delta\big( \hat{\rho}(\tilde{\vb*{\alpha}},\tilde{\vb*{\xi}},\tilde{\ten{U}}), \hat{\rho}'(\vb*{\alpha}_\rmF,\vb*{\xi}_\rmF,\ten{U}_\rmF;\vb*{\eta}) \big) > \delta\big( \hat{\rho}(\tilde{\vb*{\alpha}},\tilde{\vb*{\xi}},\tilde{\ten{U}}), \hat{\rho}'(\tilde{\vb*{\alpha}},\tilde{\vb*{\xi}},\tilde{\ten{U}};\vb*{\eta}) \big)$ meaning that fidelity-optimized parameters yield a larger dissimilarity in the photon-number distribution than leaving the initial parameters unchanged. 
Thus, alternative mitigation strategies are required.

\subsection{Displacement correction}
As discussed in the previous subsection, the effect of a loss channel on the displacement of a Gaussian state is purely classical and can be corrected straightforwardly. 
Thus, we propose a mitigation method, termed displacement correction (DC), which chooses the input displacement such that the post-loss displacement matches the displacement of the target state, $\alpha_{\mathrm{DC}} = \tilde{\alpha}/\sqrt{\eta}$ in the single-mode case and analogously in the multimode case via an $M$-dimensional linear system, while keeping the squeezing parameters unchanged. Thus, we only correct the loss-induced diminished displacement but accept the full loss impact on the remaining properties of the state. While this seems to be a rather naive attempt, it turns out that this simple-to-implement strategy can provide a successful method to reduce $\delta$. Finally note that fidelity optimization automatically accounts for displacement correction as $\vb*{\alpha}_{\rmF} = \vb*{\alpha}_{\mathrm{DC}}$.

\subsection{Phase-space distribution optimization}
\label{sec:PhaseSpaceOptimization}
As a further alternative method, we investigate a family of strategies that we denote as phase-space distribution optimizations.
These strategies leverage the description of Gaussian states in phase space as the Wigner function $W(\vb{r})$ of these states takes the form of a multivariate Gaussian distribution, which can be interpreted as an usual probability distribution. 
In particular, statistical measures exist that are able to quantify the (dis)similarity between two Gaussian probability distributions in an efficient manner by performing standard matrix operations for which efficient algorithms are known. Thus, they can be evaluated even if the number of modes grows large. 
The most common quantities with this respect are the Wasserstein metric (WAS)
\begin{align}
 D_\mathrm{WAS}(\hat{\rho},\hat{\rho}')&=(\bar{\vb{r}}-\bar{\vb{r}}')\tran (\bar{\vb{r}}-\bar{\vb{r}}') + \Tr[\ten{\sigma}] + \Tr[\ten{\sigma}'] \notag\\
 &\quad -2\Tr[(\ten{\sigma}\ten{\sigma}')^{\frac{1}{2}}],
\label{eqn:WAS}
\end{align}
the Kullback-Leibler divergence (KLD)
\begin{align}
 D_\mathrm{KLD}(\hat{\rho},\hat{\rho}') &= \frac{1}{2}(\bar{\vb{r}}-\bar{\vb{r}}')\tran\ten{\sigma}'^{-1}(\bar{\vb{r}}-\bar{\vb{r}}') + \frac{1}{2}\Tr[\ten{\sigma}'^{-1}\ten{\sigma}] \notag \\
 &\quad + \frac{1}{2}\ln(\frac{\det(\ten{\sigma}')}{\det(\ten{\sigma})}) - M,
\label{eqn:KLD}
\end{align}
the Bhattacharyya distance (BHA)
\begin{align}
 D_\mathrm{BHA}(\hat{\rho},\hat{\rho}') &= \frac{1}{8}(\bar{\vb{r}}-\bar{\vb{r}}')\tran(\ten{\sigma}+\ten{\sigma}')^{-1}(\bar{\vb{r}}-\bar{\vb{r}}') \notag \\
 &\quad+\frac{1}{2}\ln(\frac{\det(\ten{\sigma}+\ten{\sigma}')}{\sqrt{\det(\ten{\sigma})\det(\ten{\sigma}')}}),
\label{eqn:BHA}
\end{align}
and the Hellinger distance. 
As the latter is closely related to the Bhattacharyya distance via the Bhattacharyya coefficient, optimizing both quantities leads to the same initial parameters $(\vb*{\xi}_{\mathrm{BHA}},\vb*{\alpha}_{\mathrm{BHA}},\ten{U}_{\mathrm{BHA}})$ for $\hat{\rho}'$.

All measures can take values between 0 and $\infty$ and are exactly 0 if and only if the phase-space distributions are the same.
Therefore, minimizing the respective measures effectively reduces the dissimilarity between two Gaussian states in phase space.
As these measures compare different (but closely related) properties of the phase-space distributions, minimizing them will lead to slightly different initial parameters for $\hat{\rho}'$. 
Heuristically, the Wasserstein metric measures the minimal cost of shifting one phase-space distribution into another by capturing geometric differences in means and covariances. The Bhattacharyya distance measures how much both Gaussian distributions overlap in volume, i.e., their core overlap.
In contrast to these two symmetric measures, the Kullback-Leibler divergence is an asymmetric information loss. It penalizes how surprised one would be sampling from $\hat{\rho}'$ instead of having the state $\hat{\rho}$, substantially weighting mismatches in support and tails.

A priori it is not known, which of these different phase-space optimization methods leads to the smallest total variation distance of the photon-number distributions.
In this respect, we would like to also emphasize that we have additional choices due to the asymmetry of the Kullback-Leibler divergence.
Thus, we will also include the alternative nonsymmetric choice $D_\mathrm{KLD}(\hat{\rho}',\hat{\rho})$ and a symmetrized version of both, $D_\mathrm{KLD}(\hat{\rho},\hat{\rho}') + D_\mathrm{KLD}(\hat{\rho}',\hat{\rho})$, in our study. 
At least for the single-mode case, we can systematically prove a fixed relation between the different initial parameters minimizing the different phase-space quantities, see Sec.~\ref{sec:smSqueezedState}. 
Nonetheless, as for the fidelity, an important simplification of Eqs.~\eqref{eqn:WAS}-\eqref{eqn:BHA} can already be made for an arbitrarily large system. 
All phase-space measures depend on the first moments only through the mean mismatch, which increases each of them.
However, we can always enforce $\bar{\vb r}'=\bar{\vb r}$ by applying displacement correction.
In this sense, phase-space optimization automatically accounts for displacement correction.
The remaining nontrivial task is the optimization of the covariances.

In addition to these phase-space strategies, we will also include an optimization analysis with respect to the total variation distance of the phase-space distributions of a pure Gaussian target state of interest and an initially pure Gaussian state that passed through some loss channel.
This leads to an integration over the absolute value of a difference of two Gaussian distributions,
\begin{align}
\delta_{\mathrm{PS}}(\hat{\rho},\hat{\rho}') = \int_{\mathbb{R}^{2M}}\abs{W(\vb{r})-W'(\vb{r})}\dd \vb{x}^M\dd \vb{p}^M. 
\label{eq:TVDPS}
\end{align}
This integral can still be solved analytically in terms of Bessel functions in the case of single-mode states, i.e., two dimensional probability distributions~\cite{waughEvaluationIntegralElliptic1961}.
However for large mode numbers, this will not be a practical approach as intricate high-dimensional integrals have to be solved~\cite{devroyeTotalVariationDistance2023, bhattacharyyaApproximatingTotalVariation2025}.
Nonetheless, the low dimensional analysis might provide further insights into the difference of the phase-space and photon-number distribution in terms of the total variation distance such that we include this analysis for single-mode systems for completeness.

\subsection{Low-order photon-number moments correction}
For this class of methods, we exploit that low-order moments of the photon-number statistics can be calculated efficiently for Gaussian states. Any Gaussian state is uniquely defined via its mean $\vb{\bar{r}}$ and covariance matrix $\ten{\sigma}$ with respect to the quadrature operators $\hat{x}_j$ and $\hat{p}_j$. 
In particular, any higher-order quadrature moment in phase space can be expressed as some function of $\vb{\bar{r}}$ and $\ten{\sigma}$. 
Similarly, any expectation value of ladder operators can be decomposed into sums of products of first- and second-order expectation values for Gaussian states~\cite{valloneMeansCovariancesPhoton2019,heinzelExploitingHigherorderCorrelation2026}, since ladder-operator correlations are obtained from quadrature moments by a unitary change of basis. This provides efficient access to the mean photon numbers $\bar{n}_j=\langle \hat{a}_j^\dagger\hat{a}_j\rangle$ and the photon-number covariance matrix $\mathcal{N}_{jk}=\langle \hat{n}_j\hat{n}_k\rangle-\langle \hat{n}_j\rangle\langle \hat{n}_k\rangle$, even when the full multimode photon-number distribution is intractable to compute. We have~\cite{valloneMeansCovariancesPhoton2019},
\begin{align}
 \bar{n}_j &= \frac{1}{2} \big( \langle \hat{x}_j^2 \rangle + \langle \hat{p}_j^2 \rangle - 1 \big) \notag\\
 \ten{\mathcal{N}} &= \ten{A} \circ \ten{A}^* + \ten{B} \circ \ten{B}^* - \frac{1}{4} \ten{\mathbb{I}} + \notag\\
 &\quad 2 \mathrm{Re} \left[ (\vb*{\alpha}^* \vb*{\alpha}^T) \circ \ten{A} + (\vb*{\alpha}^* \vb*{\alpha}^T) \circ \ten{B} \right],  
\end{align}
where $\circ$ denotes the Hadamard product and 
\begin{align}
	\vb*{\alpha}&=\frac{1}{\sqrt 2}(\langle\hat{\vb*{x}}\rangle+\I\langle\hat{\vb*{p}}\rangle),  \notag\\
	A_{jk}&=\frac{1}{4}\Big[\langle\{\hat{x}_j,\hat{x}_k\}\rangle - 2\langle\hat{x}_j\rangle\langle\hat{x}_k\rangle
				+\langle\{\hat{p}_j,\hat{p}_k\}\rangle  \notag \\
				&\qquad - 2\langle\hat{p}_j\rangle\langle\hat{p}_k\rangle +\I\langle\{\hat{x}_k,\hat{p}_j\}\rangle - \I\langle\{\hat{x}_j,\hat{p}_k\}\rangle\Big], \notag \\
	B_{jk} &= \frac12 \Big[\big(\langle \hat x_j \hat x_k\rangle - \langle \hat x_j\rangle \langle \hat x_k\rangle\big)- \big(\langle \hat p_j \hat p_k\rangle - \langle \hat p_j\rangle \langle \hat p_k\rangle\big)  \notag \\
				&\qquad + \I\big(\langle \hat x_j \hat p_k\rangle - \langle \hat x_j\rangle \langle \hat p_k\rangle\big)+ \I\big(\langle \hat p_j \hat x_k\rangle - \langle \hat p_j\rangle \langle \hat x_k\rangle\big)\Big].
\end{align}
Now, it is possible to construct measures utilizing the photon-number means
\begin{equation}\label{eqn:Delta1}
	\Delta_{\bar{n}} = \frac{\sqrt{(\vb{\bar{n}}-\vb{\bar{n}}')\tran(\vb{\bar{n}}-\vb{\bar{n}}')}}{\sqrt{\vb{\bar{n}}\tran\vb{\bar{n}}}},
\end{equation}
or the photon-number variances 
\begin{equation}\label{eqn:Delta2}
	\Delta_{\mathcal{N}} = \frac{\sqrt{\Tr[(\ten{\mathcal{N}}-\ten{\mathcal{N}}')^2]}}{\sqrt{\Tr[\ten{\mathcal{N}}^2]}}.
\end{equation}
Although one could construct further measures that include higher-order moments up to a chosen order, the computation becomes intractable beyond some order for large mode numbers, since using sufficiently many moments is effectively equivalent to reconstructing the full photon-number distribution. We therefore restrict our analysis to moments up to second order of the photon-number operators.

For single-mode displaced squeezed vacuum as well as for general multimode scenarios, Eq.~\eqref{eqn:Delta1} or \eqref{eqn:Delta2} provide only a scalar constraint that couples the various input parameters. As a result, the correction problem is generally underdetermined and admits ambiguities. These ambiguities can be resolved in several ways. For instance, one may enforce mean or variance matching on individual modes, combine mean and variance constraints, or combine these conditions with additional correction or optimization criteria discussed in this section.

\subsection{Vacuum overlap correction}
To establish a further method to potentially reduce the impact of loss, we use the fact that the vacuum state itself can be viewed as a Gaussian state. Thus, the overlap of any Gaussian state with the vacuum state can be efficiently calculated,
\begin{align}
 \bra{\vb{0}} \hat\rho \ket{\vb{0}} = \frac{1}{\sqrt{\det (\ten{\sigma} + \frac{1}{2}\mathbbm{1} ) }} \E^{-\frac{1}{2} \bar{\vb{r}}\tran (\ten{\sigma} + \frac{1}{2}\mathbbm{1})^{-1}  \bar{\vb{r}} } \equiv \mathcal{P}(\vb{0}).
\label{eq:VacuumOverlap}
\end{align}
Choosing the initial parameters of a lossy Gaussian state such that the vacuum overlap of this state is equal to the vacuum overlap of the target state, this method corrects directly for a specific local property of the photon-number distribution which might be particularly useful for zero-mean Gaussian states. This can be motivated heuristically by the fact that the photon-number distribution of any single- or multimode squeezed vacuum state has the largest weight on the vacuum. In such a scenario, the correction reduces to determining the initial squeezing parameters from 
\begin{align}
\det (\ten{\sigma'} + \frac{1}{2}\mathbbm{1} ) = \det (\ten{\sigma} + \frac{1}{2}\mathbbm{1} ). 
\label{eq:VacuumOverlapNonDisplaced}
\end{align}
In particular, we have $\det (\ten{\sigma} + \frac{1}{2}\mathbbm{1} ) = \prod_{i=1}^{M} \cosh^{2}\abs{\zt_i}$ for the target state.

Similar to the photon-number moments correction, we have ambiguities for
a general multimode scenario (or for single-mode displaced squeezed vacuum), if we only focus on equal global vacuum overlap of the entire multimode state, $ \bra{\vb{0}} \hat\rho' \ket{\vb{0}} =  \bra{\vb{0}} \hat\rho \ket{\vb{0}}$. However, we also obtain additional possibilities to resolve these ambiguities. For example, similarly to the global vacuum overlap, we straightforwardly obtain the vacuum overlap of individual modes by tracing out the other modes and applying Eq.~\eqref{eq:VacuumOverlap} to the corresponding subspace of the mode of interest. Equivalently, we might also consider the joint vacuum overlap of a subset of modes. This leads in total to $2^M-1$ distinct vacuum-overlap objective functions, one for each nonempty subset $S\subseteq\{1,\ldots,M\}$, $\bra{\vb{0}_S}\hat\rho_S\ket{\vb{0}_S} = \bra{\vb{0}_S}\hat\rho'_S\ket{\vb{0}_S}$. Alternatively, we could implement vacuum-overlap correction by scaling all squeezing parameters by a common factor until the vacuum probability matches, or introduce weighted modifications based on the transmissivities of the individual modes. 
Furthermore, we may combine vacuum overlap correction with one of the other proposed schemes such as mean photon-number correction in each mode. For displaced squeezed vacuum, including displacement correction provides a further natural way to reduce the residual ambiguities.

\section{Single-mode squeezed vacuum state}
\label{sec:smSqueezedState}
As a first test, we examine the effectiveness of each mitigation method in reducing loss-induced deviations in the photon-number distribution for the simplest nontrivial example, a single-mode squeezed vacuum state. Thus, the desired target state is characterized by a single parameter $\tilde \xi$. In this section, we choose it to be real and positive without loss of generality, since the photon-number distribution does not depend on the squeezing phase for a single-mode squeezed vacuum state. Introducing a photon loss channel with loss probability $1-\eta$, we first search for a squeezed vacuum state $\hat{\rho}(\xi)$, with $\xi\geq0$, that is able to mitigate the influence of loss on the photon-number distribution such that we ideally minimize $\delta\big(\hat{\rho}(\tilde{\xi}),\hat{\rho}'(\xi;\eta)\big)$ for a specific 
\begin{align}
\xi_\mathrm{min} = \underset{\xi \geq 0}{\arg\min}\, \delta\big(\hat{\rho}(\tilde \xi), \hat{\rho}'(\xi;\eta)\big),
\end{align}
where $\xi_\mathrm{min} = \xi_\mathrm{min}(\tilde{\xi};\eta)$. Throughout this section, we analyze how close the proposed mitigation schemes introduced in Sec.~\ref{sec:Mitigation} approximate this optimal initial parameter.

Using the fact that any initial squeezed vacuum state $\hat{\rho}(\xi)$ will be transformed into a squeezed thermal state via a loss channel, $\hat{\rho}'(\xi;\eta) = \mathcal{E}_\eta(\hat{\rho}(\xi)) = \hat{S}(\zeta)\hat{\rho}(\mu)\hat{S}^\dagger(\zeta)$, with thermal mean occupation number $\mu = \mu(\xi;\eta)$ and squeezing parameter $\zeta = \zeta(\xi;\eta)$ given by,
\begin{align}
 \mu &= \frac{1}{2} \sqrt{ \eta^2 + (1-\eta)^2 + 2\eta(1-\eta)\cosh(2\xi)} - \frac{1}{2}, 
 \label{eq:ThermalMean}\\
 \label{eq:zeta}\zeta &= \frac{1}{2} \arsinh\bigg(\frac{\eta \sinh(2\xi)}{2\mu + 1} \bigg),
\end{align}
we can express the photon-number distribution of a lossy squeezed vacuum state $\hat{\rho}'(\xi;\eta)$ in terms of a hypergeometric function
\begin{align}
 \mathcal{P}'_{\xi;\eta}(m) &= \frac{\left[ \frac{\mu (1+\mu)}{\mu^2 + (1+2\mu)\cosh^2\zeta } \right]^m}{\sqrt{(1+\mu)^2 + (1+2\mu)\sinh^2\zeta }}  
 \label{eq:PNSsqth} \\
&\quad \times \!{}_2F_1\Bigg(\frac{1-m}{2},-\frac{m}{2},1,\frac{(1+2\mu)^2\sinh^2(2\zeta)}{4\mu^2(1+\mu)^2}\Bigg). \notag
\end{align}
For the target photon-number distribution of a single-mode squeezed vacuum state $\hat{\rho}(\tilde{\xi})$, we have
\begin{align}
 \mathcal{P}_{\tilde{\xi}}(m) = \frac{1+(-1)^m}{2} \frac{m!}{2^m (\frac{m}{2}!)^2} \frac{\tanh^m(\tilde \xi)}{\cosh(\tilde \xi)}
 \label{eq:PNSsqvac}
\end{align}
and the total variation distance reads
\begin{align}
\delta\big(\hat{\rho}(\tilde \xi), \hat{\rho}'(\xi;\eta)\big) = 
\frac{1}{2} \sum_{m=0}^\infty \abs{\mathcal{P}_{\tilde{\xi}}(m) - \mathcal{P}'_{\xi;\eta}(m)}.
 \label{eq:deltaSMSV}
\end{align}
Note that we also obtain Eq.~\eqref{eq:PNSsqvac} from Eq.~\eqref{eq:PNSsqth} in the limit $\eta \to 1$ and $\xi = \tilde{\xi}$.

\begin{figure*}[t]   
    \centering
    \includegraphics[width=0.9\textwidth]{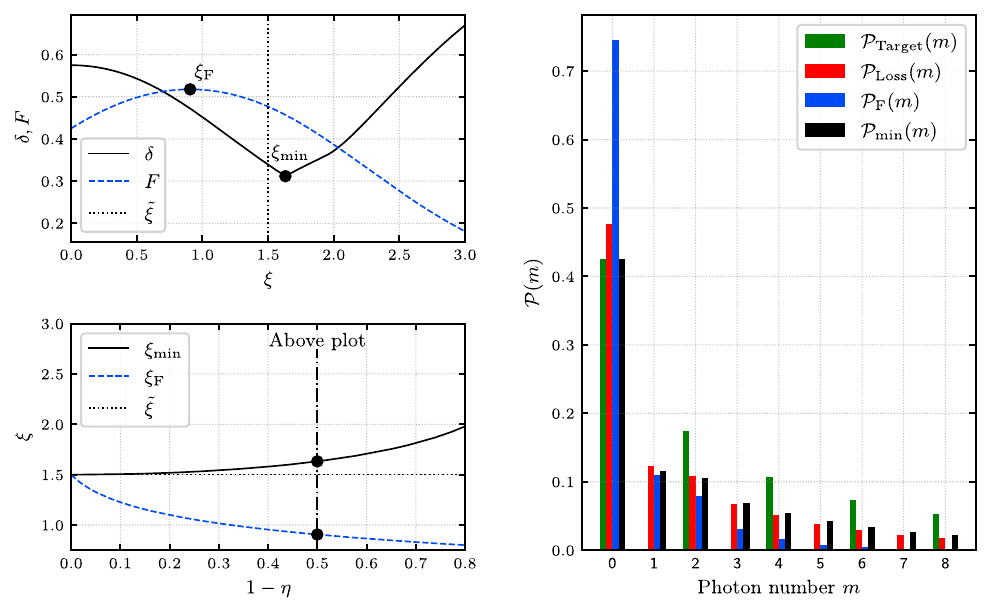}
    \caption{The upper left plot shows $\delta$ (black solid line) and $F$ (blue dashed line) for an target squeezing pa\-ra\-me\-ter $\tilde{\xi}=1.5$ under variation of $\xi$ at a loss parameter of $1-\eta=0.5$. Both $\xi_\mathrm{min}$ and $\xi_\mathrm{F}$ are depicted with a black dot and labeled. The vertical dotted black line indicates $\xi=\zt$. The lower left plot shows $\xi_\mathrm{min}$ (solid black line) and $\xi_\mathrm{F}$ (dashed blue line) for different loss parameters at a fixed $\zt=1.5$. The values for $\xi_\mathrm{min}$ and $\xi_\mathrm{F}$ from the upper plot are also depicted in the lower plot by black dots. The right plot shows the photon-number distribution of the target state (green), its lossy version (red), the fidelity-optimized state (blue), and the $\delta$-minimized state (black) from left to right for a target squeezing parameter of $\zt=1.5$ and $\eta = 0.5$. The typical even-odd signature of a perfect squeezed vacuum state can be seen.} 
    \label{fig:zminzmax}
\end{figure*}

Using Eq.~\eqref{eq:deltaSMSV}, we can numerically determine the squeezing parameter $\xi_\mathrm{min}$ of the lossy state that minimizes $\delta$. 
Typically, we make the observation that $\xi_\mathrm{min} > \tilde{\xi}$ which is intuitively clear as loss removes photons from the system and, thus, redistributes higher occupation numbers to lower occupation numbers. 
Only in a particular region in parameter space for large squeezing and low loss, we find examples with $\xi_\mathrm{min} \lesssim \tilde{\xi}$, see Sec.~\ref{sec:VacOver} for further details.
We plot $\delta$ as a function of the initial parameter $\xi$ for the example $\tilde{\xi} = 1.5$ and $1-\eta = 0.5$ in the left upper part of Fig.~\ref{fig:zminzmax} as a solid black line. 
We further analyze how $\xi_\mathrm{min}$ alters for different loss by varying $\eta$. Also as expected, we find that $\xi_\mathrm{min}$ is monotonically increasing for increasing system loss and fixed target distribution parametrized by $\tilde{\xi}$. Exemplarily, this behavior is depicted in the left lower part of Fig.~\ref{fig:zminzmax} as a solid black line. 
We also exemplarily depict the photon-number distribution $\mathcal{P}_{\tilde\xi}(m)$ (green) for $\tilde{\xi} = 1.5$ and compare it with its lossy counterpart $\mathcal{P}'_{\tilde\xi;\eta}(m)$  (red), the fidelity optimized version $\mathcal{P}'_{\xi_{\rmF};\eta}(m)$ (blue, see next subsection for details), as well as the $\delta$-minimized version $\mathcal{P}'_{\xi_{\mathrm{min}};\eta}(m)$ (black) for $\eta=0.5$ in the right part of Fig.~\ref{fig:zminzmax}.

Finally, we would like to highlight, that we are able to lower bound the total variation distance for a given target distribution and transmissivity $\eta$ for any $\xi$. First, we split the total variation distance in its even and odd parts,
\begin{align}
 \delta &= 
\frac{1}{2} \sum_{m=0}^\infty \abs{\mathcal{P}_{\tilde{\xi}}(2m) - \mathcal{P}'_{\xi;\eta}(2m)} 
+ \frac{1}{2} \sum_{m=0}^\infty  \mathcal{P}'_{\xi;\eta}(2m+1) \notag \\
&\equiv \delta_\mathrm{even} + \delta_\mathrm{odd},
\end{align}
where we used that $\mathcal{P}_{\tilde{\xi}}(2m+1) = 0$. Second, we can lower bound the even part by the vacuum contribution, $\delta_\mathrm{even} \geq \delta_\mathrm{vac} = \frac{1}{2} \abs{\mathcal{P}_{\tilde{\xi}}(0) - \mathcal{P}'_{\xi;\eta}(0)}$. The bound is tight only for the trivial cases. Equality holds for $\eta=1$ (no loss) or $\tilde\xi=0$. Thus, we obtain 
\begin{align}
 \delta \geq \frac{1}{2} \abs{\mathcal{P}_{\tilde{\xi}}(0) - \mathcal{P}'_{\xi;\eta}(0)}  + \frac{1}{2} \sum_{m=0}^\infty \mathcal{P}'_{\xi;\eta}(2m+1).
\end{align}
The vacuum probabilities can be calculated in an efficient manner for any Gaussian state and read for our current example, $\mathcal{P}_{\zt}(0) = \frac{1}{\cosh \zt}$ and $\mathcal{P}'_{\xi;\eta}(0) = \frac{1}{\sqrt{1+\eta(2-\eta)\sinh^2 \xi}}$. Also the summation over the odd contributions to the photon-number distribution of a lossy squeezed state can be brought into a closed-form expression such that we finally obtain,
\begin{align}
 \delta &\geq \delta_\mathrm{vac} + \delta_\mathrm{odd} \notag\\ 
 &= \frac{1}{2} \abs{ \frac{1}{\cosh \zt} - \frac{1}{\sqrt{1+\eta(2-\eta)\sinh^2 \xi}} }  \notag \\ 
 &\quad + \frac{1}{4} \Bigg(1 - \frac{1}{\sqrt{1 + 4\eta(1-\eta)\sinh^2 \xi}} \Bigg).
\label{eq:lowerbounddeltaSMSV}
\end{align}
The contribution from the odd photon-number probabilities is monotonically increasing in $\xi$ and bounded by $\frac{1}{4}$ for $\xi\to\infty$.
This reflects that for fixed loss strength any increase in squeezing (and hence in the mean photon number) raises the probability that the channel removes an odd number of photons, thereby monotonically building up population in the odd-photon sector that is absent in the squeezed-vacuum target state.

\subsection{Fidelity optimization}
While optimizing the fidelity is a common approach in quantum information science to reduce dissimilarities of quantum states, it proves of limited utility considering the photon-number distribution of squeezed vacuum under loss. In fact, the squeezing parameter $\xi=\xi_\mathrm{F}$ that maximizes the fidelity $F\big(\hat{\rho}(\tilde{\xi}),\hat{\rho}'(\xi;\eta)\big)$ for a given transmissivity $\eta$ can be determined analytically,
\begin{align}\label{eq:zFidelity}
  \xi_\mathrm{F} = \frac{1}{4} \ln(\frac{1-\eta + \E^{2\tilde \xi}}{1-\eta + \E^{-2\tilde \xi}}), \quad \text{thus}\quad \frac{1}{2}\tilde{\xi} \leq \xi_\mathrm{F} \leq \tilde{\xi}.
\end{align}
Note, that one could in principle allow for complex squeezing parameters $\xi$. Nevertheless, the photon-number distribution of lossy squeezed vacuum does not depend on the phase of $\xi$ as already discussed. Further, the maximization of the fidelity given in Eq.~\eqref{eqn:Fidelity} for complex $\xi$ directly leads to the fact that the phases of $\xi_\mathrm{F}$ and $\tilde{\xi}$ have to be aligned such that the Wigner functions of the states $\hat\rho$ and $\hat\rho'$ have the same orientation in phase space in order to generate the maximal overlap implying $\xi_\mathrm{F} \geq 0$ for our convention.

The result that $\xi_\mathrm{F} \leq \tilde{\xi}$ seems to be surprising at first sight as we expect that methods that mitigate the impact of loss yielding larger squeezing. Calculating the resulting total variation distances for various target squeezing parameters and losses even reveals ${\delta\big(\hat\rho(\tilde{\xi}), \hat\rho'(\xi_\mathrm{F};\eta) \big) \geq \delta\big(\hat\rho(\tilde{\xi}), \hat\rho'(\tilde{\xi};\eta) \big)}$, i.e., the dissimilarity between the photon-number distribution of the target state $\hat\rho(\tilde{\xi})$ and the photon-number distribution of the lossy state $\hat\rho'(\xi_\mathrm{F};\eta)$ that maximizes the fidelity with the target state is larger than the dissimilarity of the target photon-number distribution with its own lossy counterpart. Therefore, the fidelity is not a suitable measure for mitigating the influence of loss on the underlying photon-number distributions of pure single-mode squeezed vacuum states.
This result is also depicted in the left upper part of Fig.~\ref{fig:zminzmax} for the example $\tilde{\xi} = 1.5$ and $\eta=0.5$. To further illustrate the discrepancy, we plotted the functional dependence of the fidelity $F\big(\hat{\rho}(\tilde{\xi}),\hat{\rho}'(\xi;\eta)\big)$ as a function of the input squeezing for the state $\rho'$ as a dashed blue line as well.

The fact that the fidelity does not provide a useful guidance for our purpose can be traced back to a crucial difference in the fidelity and the total variation distance of the photon-number distribution. 
From a Fock basis perspective, there are two main differences that might be responsible for this effect, odd photon-number contributions and off-diagonal terms. 
While $\delta$ contains only information of the diagonal elements of the density matrices in Fock basis (photon-number probabilities), $F$ receives contributions from diagonal and off-diagonal terms.
In particular, loss introduces nonzero weight on odd photon numbers in the squeezed-vacuum distribution whereas the pure squeezed vacuum target state has support only on even $m$, cf. Eq.~\eqref{eq:PNSsqvac} and Eq.~\eqref{eq:PNSsqth}. These odd-$m$ contributions enter $\delta$ directly since $\delta$ measures absolute differences of probabilities.
However, these terms play no role for $F$ being the overlap of $\rho$ with $\rho'$ because the Fock-basis matrix elements between states of different parity vanish. 
Interestingly, although the odd-photon leakage in a lossy channel does contribute to $\delta$, its quantitative effect on the fidelity-optimized squeezing parameter $\xi_\rmF$ that maximizes $F$ is surprisingly small.
One finds that the dominant corrections come from how the even-number probabilities deform under loss.

By contrast, the fidelity is sensitive to off-diagonal terms of the density matrix encoding how well coherence and purity are preserved. These structures are completely absent from $\delta$, which does not encode any information about the interference properties of the state and retains only photon-count statistics. Optimizing the fidelity will also minimize the degradation of off-diagonal elements. As the initial squeezing $\xi$ grows, the average photon number of the state increases, so that the purity of the state and coherence is more fragile to loss. This is also apparent in Eq.~\eqref{eq:ThermalMean} as the thermal mean occupation number is a monotonically increasing function of the initial squeezing $\xi$. Thus, the smaller $\xi$ is, the smaller $\mu$ becomes, implying higher purity of $\hat\rho'$. The fidelity-based optimum therefore prefers a smaller initial squeezing to penalizing loss-induced decoherence in the off-diagonal sector which has more weight than to match diagonal photon-number profiles. To preserve purity and interference visibility, $\xi_\mathrm{F}$ is driven below $\tilde{\xi}$. This effect is visualized in the left lower part of Fig.~\ref{fig:zminzmax} where we plot $\xi_\mathrm{F}$ as a function of the system loss $1-\eta$ for fixed $\tilde{\xi}$ as a dashed blue curve. The fidelity optimization effectively trades some statistical overlap in the diagonal probabilities in order to maintain the quantum interference properties.

\subsection{Displacement correction}
For a nondisplaced squeezed state, this correction method is trivial as it keeps the input squeezing unchanged and we obtain $\xi_\mathrm{DC} = \tilde{\xi}$. Thus, we usually find $\xi_\mathrm{DC} < \xi_\mathrm{min}$, apart from the particular parameter-space region for large squeezing and low loss, see Sec.~\ref{sec:VacOver}.
Nonetheless, this method leads to smaller deviations in the photon-number distribution than the fidelity optimization, as $\xi_{\rmF} < \xi_\mathrm{DC}$.
A detailed discussion of this ordering, together with a comparison to the other mitigation methods, is provided in Sec.~\ref{sec:ComparisonSMSV}.

\subsection{Phase-space distribution optimization}
\label{sec:SMSV-PSO}
In contrast to the fidelity optimization and the displacement correction, there is no compact analytical solution for the various phase-space distribution optimizations proposed in Sec.~\ref{sec:PhaseSpaceOptimization}. Considering the Wasserstein metric, the Bhattacharyya distance, and the three different versions of the Kullback-Leibler divergence as functions of the initial squeezing parameter $\xi$ characterizing the state $\rho'$ and determining the first derivatives with respect to $\xi$, one can reformulate the optimization task into a root finding problem of an at least fourth-order polynomial in $\E^{2\xi}$ (Bhattacharyya distance, both asymmetric Kullback-Leibler divergences) or an even higher-order polynomial (symmetric Kullback-Leibler divergence, Wasserstein metric). Thus, for some phase-space optimizations an analytical solution exists in principle but its actual solution is nonilluminating for any further practical evaluation. Nonetheless, finding respective solutions for all these phase-space distribution optimization strategies can be done by any standard numerical solver. Similar to the fidelity case, it is straightforward to prove that any nontrivial phase of the squeezer has to vanish as the derivative of the phase-space measures will be proportional to the sine function of that phase. Thus, aligned phases correspond to the minimum.

Although we eventually solve the phase-space optimization problems only in a numerical fashion, one can already draw important conclusions on the properties of the solutions on structural grounds of the underlying root finding problems. 
In Appendix~\ref{sec:RelationsPSO}, we proof that each root finding problem has a unique positive solution and that these roots satisfy
\begin{align}\label{eq:IneqPhaseSpace}
 \xi_\mathrm{WAS} < \xi_\mathrm{KLDup} < \xi_\mathrm{BHA} < \xi_\mathrm{KLDsym} < \xi_\mathrm{KLDpu},
\end{align}
where $\xi_\mathrm{KLDup}$ and $\xi_\mathrm{KLDpu}$ minimize $D_\mathrm{KLD}(\hat{\rho},\hat{\rho}')$ and $D_\mathrm{KLD}(\hat{\rho}',\hat{\rho})$, respectively. Furthermore, we proof ${\zt < \xi_\mathrm{WAS}}$, i.e., the phase-space optimizations increase the initial squeezing parameters as naively expected.

\begin{figure}[t]   
    \centering
    \includegraphics[width=1\columnwidth]{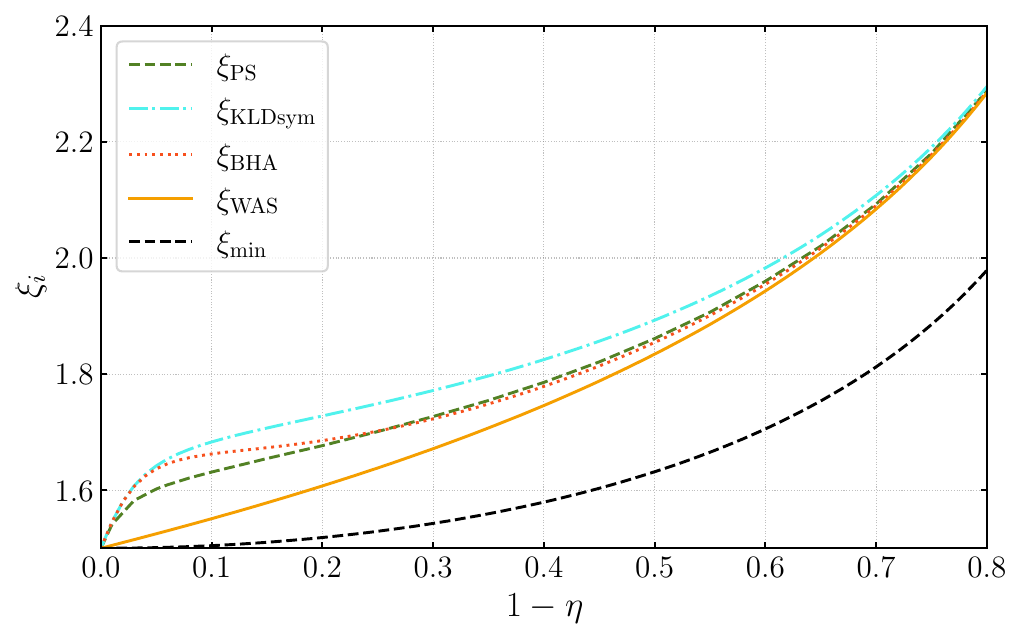}
	\caption{Squeezing parameters obtained from different phase-space distribution optimization methods as a function of loss for a target state with $\zt = 1.5$. Among the phase-space approaches, the Wasserstein metric (solid yellow) increases more gradually at small losses, whereas the other methods [$\mathrm{PS}$ (dashed dark green), $\mathrm{BHA}$ (dotted orange), and $\mathrm{KLDsym}$ (dash-dotted turquoise)] show a steeper initial increase. For larger losses, all phase-space methods converge to similar squeezing values.}
    \label{fig:PhaseSpaceMethods}
\end{figure}

Besides these analytical results, we also made several numerical observations by analyzing the parameter range $0<\zt<2$ and arbitrary system loss. First of all, we find that as the system loss increases, the differences between the various squeezing parameters $\xi_i$ that optimize the different phase-space measures shrink. This effect increases with $\zt$, leading to differences among the $\xi_i$ of only a few per mille for $0.5<\zt<2$. Second, we also analyzed the phase-space optimization based on the total variation distance of the Wigner functions, see Eq.~\eqref{eq:TVDPS}. Albeit, we have no strict bound on $\xi_\mathrm{PS}$ that is minimizing $\delta_\mathrm{PS}(\hat\rho,\hat\rho')$ as for the other squeezing parameters $\xi_i$, we have acquired numerical evidence that $\xi_\mathrm{WAS} < \xi_\mathrm{PS} < \xi_\mathrm{KULsym}$. For high transmissivity and sufficiently small target squeezing, e.g., $\eta>0.9$ and $\zt < 1.5$, $\xi_\mathrm{PS} < \xi_\mathrm{KULup}$ while $\xi_\mathrm{PS}$ grows larger than $\xi_\mathrm{KULup}$ for $\zt=2$. For small transmissivity, we observe at least $\xi_\mathrm{PS} < \xi_\mathrm{KULsym}$. Third, and most importantly, $\xi_\mathrm{min} < \xi_\mathrm{WAS}$. Thus, all squeezing parameters optimizing some of the introduced phase-space measures which can be calculated in an efficient manner are overshooting the actual initial squeezing of the lossy state $\hat\rho'$ that is minimizing the total variation distance with respect to the photon-number distribution. Therefore, the best available choice would be the optimization of the Wasserstein metric for the phase-space distributions of the target state $\hat\rho$ and the lossy state $\hat\rho'$. This is exemplarily shown in Fig.~\ref{fig:PhaseSpaceMethods} for $\zt = 1.5$.

Finally, we would like to note that one could equally formulate the phase-space optimization using the Husimi function, $Q(\alpha) = \frac{1}{\pi} \langle\alpha|\hat{\rho}|\alpha\rangle$, as an alternative approach in general.
For Gaussian states, $Q(\alpha)$ is a Gaussian distribution in phase space that is closely related to the Wigner function via a convolution with a vacuum Gaussian kernel.
As a consequence, it is expected that the squeezing parameters that optimize the corresponding phase-space measures for $Q$ are very close to those obtained from the Wigner-function-based optimization.
We tested this hypothesis numerically for the Wasserstein metric, obtaining only small deviations of typically less than one percent for all loss parameters and $\zt\leq 5$.
In particular, the difference becomes negligible for large $\zt$. We therefore restrict the discussion to the Wigner representation.

\subsection{Low-order photon-number moments correction}
In contrast to the phase-space distribution optimization methods, the parameters $\xi_{\Delta_{\bar{n}}}$ and $\xi_{\Delta_\mathcal{N}}$, minimizing the low-order photon-number moments metrics, can be calculated analytically, resulting in $\Delta_{\bar{n}} = 0$ and $\Delta_\mathcal{N} = 0$, respectively.
Setting the photon-number means of the target and lossy state equal, $\Delta_{\bar{n}} = 0$, leads to 
\begin{equation}\label{eq:SMSV-MeanMatching}
	\xi_{\Delta_{\bar{n}}}=\arsinh \left(\frac{\sinh (\tilde \xi)}{\sqrt{\eta }}\right)\geq \tilde \xi.
\end{equation}
Alternatively, we may optimize for the photon-number variance.
For the target squeezed vacuum state, we have $\tilde{\mathcal{N}} = 2 \sinh^2(\tilde \xi) \cosh ^2(\tilde \xi)$.
Whereas the photon-number variance of a lossy squeezed vacuum state is given by $\mathcal{N}'(\xi) = 2 \eta^2 \sinh^4(\xi) + (\eta^2+\eta) \sinh^2(\xi)$.
Demanding that $\tilde{\mathcal{N}}=\mathcal{N}'$ for $\xi = \xi_{\Delta_\mathcal{N}}$, we obtain a quadratic equation in $\sinh^2(\xi_{\Delta_\mathcal{N}})$ which is solved by
\begin{align}
 \sinh^2(\xi_{\Delta_\mathcal{N}}) = \frac{1+\eta}{4\eta} \Bigg[ \sqrt{1+ \frac{8\tilde{\mathcal{N}}}{(1+\eta)^2} } -1 \Bigg].
 \label{eq:xiDeltaVar}
\end{align}
Moreover, we are able to proof that $\xi_{\Delta_\mathcal{N}} > \xi_{\Delta_{\bar{n}}}$ by expressing $\tilde{\mathcal{N}}$ in terms of the variable $\sinh^2(\xi_{\Delta_{\bar{n}}}) = \frac{1}{\eta} \sinh^2(\zt)$, 
\begin{align}
 \mathcal{N}'(\xi_{\Delta_\mathcal{N}}) &= \tilde{\mathcal{N}} = 2\eta^2 \sinh^4(\xi_{\Delta_{\bar{n}}}) + 2\eta \sinh^2(\xi_{\Delta_{\bar{n}}}) \notag \\
 &> \mathcal{N}'(\xi_{\Delta_{\bar{n}}}).
\end{align}
As $\mathcal{N}'$ is a monotonically increasing function in $\xi$, we obtain $\xi_{\Delta_{\bar{n}}} < \xi_{\Delta_\mathcal{N}}$.

Furthermore, we proof in Appendix~\ref{sec:RelationWAS-Var} that $\xi_{\Delta_{\mathcal{N}}} < \xi_\mathrm{WAS}$.
Therefore, we obtain the following chain of relations 
\begin{align}
 \xi_{\Delta_{\bar{n}}} < \xi_{\Delta_{\mathcal{N}}} < \xi_\mathrm{WAS}, %< \xi_\mathrm{KLDup} < \xi_\mathrm{BHA} < \xi_\mathrm{KLDsym} < \xi_\mathrm{KLDpu}.
\end{align}
where $\xi_\mathrm{WAS}$ is the smallest squeezing parameter that we obtain from the phase-space distribution minimizations. 
Moreover, numerical evidence shows again that also the low-order photon-number moments optimizers provide an upper bound on  $\xi_\mathrm{min} <  \xi_{\Delta_{\bar{n}}}$.

\subsection{Vacuum-overlap correction}\label{sec:VacOver}
Also the parameter $\xi_\mathrm{vac}$, defined by vacuum matching of the target distribution and the photon-number distribution of a lossy squeezed vacuum state $\mathcal{P}_{\tilde{\xi}}(0) = \mathcal{P}'_{\xi_\mathrm{vac};\eta}(0)$, can be determined analytically,
\begin{align}
 \xi_\mathrm{vac} = \arsinh \bigg( \frac{\sinh \tilde{\xi}}{\sqrt{\eta(2-\eta)}} \bigg).
\end{align}
As $0\leq \eta \leq 1$, we have $\tilde{\xi} \leq \xi_\mathrm{vac} \leq \xi_{\Delta_{\bar{n}}}$. 
Remarkably, we find two numerical observations that hold over a wide parameter range. 
First, at $\xi=\xi_\mathrm{vac}$, the even and odd subsectors of the total variation distance satisfy $\delta_\mathrm{even} = \delta_\mathrm{odd}$ (note that $\delta_\mathrm{even}$ only contains deviations in $\mathcal{P}(2m)$ for $m>0$ as $\delta_\mathrm{vac} = 0$ by construction).
Second, and crucial to our strategy for mitigating the impact of loss, comparing $\xi_\mathrm{vac}$ with the numerically obtained $\xi_\mathrm{min}$ shows that, up to small numerical inaccuracies, $\xi_\mathrm{vac} = \xi_\mathrm{min}$. 
To elucidate these observations, we first discuss the relation $\delta_\mathrm{even} = \delta_\mathrm{odd}$ at $\xi = \xi_\mathrm{vac}$, and then argue how it leads to $\xi_\mathrm{vac} = \xi_\mathrm{min}$.
A systematic proof is provided in Appendix~\ref{sec:VacEqMin}.

\begin{figure}[t]   
    \centering
    \includegraphics[width=1\columnwidth]{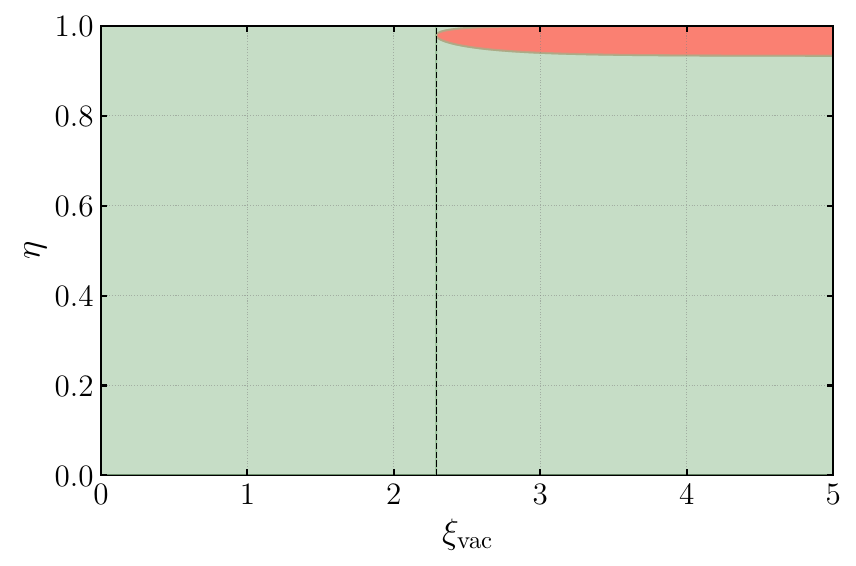}
	\caption{Parameter regions for the location of the minimizer $\xi_{\rmmin}$. In the green region, $\xi_{\min} = \xi_{\mathrm{vac}}$ as proven analytically. In particular, for all $\xi_\mathrm{vac}<2.290047\ldots$ (dashed black line) we have $\xi_{\min} = \xi_{\mathrm{vac}}$ for all loss parameters $\eta$. For $\xi_\mathrm{vac}>2.290047\ldots$, the minimizer coincides with vacuum-overlap correction for $\eta < 14/15$. For $\eta > 14/15$, $\xi_\mathrm{vac}$ is no longer optimal in the red region where we find $\xi_{\min} < \xi_{\mathrm{vac}}$.}
    \label{fig:regions}
\end{figure}

The numerical observations indicate that matching the vacuum overlap turns out to have a more sophisticated impact than being a simple local constraint.
Writing $\delta=\delta_{\mathrm{even}}+\delta_{\mathrm{odd}}$ and noting that the target distribution has strictly even support as $\mathcal{P}_{\tilde{\xi}}(2m+1)=0$, the odd contribution arises entirely from loss-induced leakage of probability weight into odd photon numbers.
From a heuristic perspective, loss affects the even-parity sector in two ways. 
First, it transfers weight from the even to the odd sector, which is the sole source of $\delta_{\mathrm{odd}}$. 
Second, it reshapes the distribution within the even sector by redistributing weight among even photon numbers. 
One would therefore naively expect $\delta_{\mathrm{even}}>\delta_{\mathrm{odd}}$ in general. 
However, at the point $\xi=\xi_{\mathrm{vac}}$ defined by vacuum matching, numerical evidence shows that the minimum satisfies $\delta_{\mathrm{even}}=\delta_{\mathrm{odd}}$, and hence
\begin{align}
 \delta =2 \delta_{\mathrm{odd}}.
\end{align}
This relation indicates that once the vacuum probabilities are matched, the loss-induced redistribution within the even sector is effectively balanced by inflow from higher-lying even Fock states, such that the net deviation of the lossy state in the even sector mirrors the deviation built up in the odd sector.

To make this balance transparent, we separate $\delta$ in contributions for which $\mathcal{P}_{\zt}(m)\geq \mathcal{P}_{\xi;\eta}'(m)$ and in contributions for which $ \mathcal{P}_{\zt}(m)<  \mathcal{P}_{\xi;\eta}'(m)$ and define the index sets $I_{+}:=\{m:\mathcal{P}_{\tilde{\xi}}(m)\geq \mathcal{P}'_{\xi;\eta}(m)\}$ and ${I_{-}:=\{m:\mathcal{P}_{\tilde{\xi}}(m)< \mathcal{P}'_{\xi;\eta}(m)\}}$. 
Further, we define $S_{+} = \sum_{m\in I_{+}}[\mathcal{P}_{\zt}(m)-\mathcal{P}_{\xi;\eta}'(m)]$ and $S_{-} = \sum_{m\in I_{-}}[\mathcal{P}_{\xi;\eta}'(m)-\mathcal{P}_{\zt}(m)]$, with $0 = \sum_{m}[\mathcal{P}_{\zt}(m)-\mathcal{P}_{\xi;\eta}'(m)] = S_{+} - S_{-}$ implying $S_{+} = S_{-}$, i.e., the total positive and negative deviations balance because both distributions are normalized. 
Furthermore, we have 
\begin{align}
 \delta = \frac{1}{2}(S_{+} + S_{-}) = S_{-}.
\end{align}
At $\xi=\xi_{\mathrm{vac}}$, vacuum matching implies that $m=0 \in I_{+}$, and in fact one finds that all even terms satisfy $\mathcal{P}_{\tilde{\xi}}(2m)>\mathcal{P}'_{\xi;\eta}(2m)$ for $m>0$, so that $I_{-}$ contains only odd indices, see Appendix~\ref{sec:proofIminus}.
Hence the entire contribution from $I_{-}$ comes from the odd sector, yielding $\delta = S_{-} = 2\delta_{\mathrm{odd}}$ at $\xi=\xi_{\mathrm{vac}}$.

Moving away from $\xi_{\mathrm{vac}}$ reintroduces an additional vacuum contribution for $\xi<\xi_{\mathrm{vac}}$, giving $\delta \geq 2\delta_{\mathrm{vac}}+2\delta_{\mathrm{odd}}$, whereas for $\xi\ge\xi_{\mathrm{vac}}$ one has $\delta\ge 2\delta_{\mathrm{odd}}$, see Appendix~\ref{sec:VacEqMin} for details.
There, we also establish the required monotonicity on both sides, proving that $\xi_{\mathrm{vac}} = \xi_{\mathrm{min}}$ for single-mode squeezed vacuum in a broad parameter regime (in particular, for all $\xi_{\mathrm{vac}} < 2.290047\dots$ independently of $\eta$, and also for $\xi_{\mathrm{vac}} \geq 2.290047\dots$ provided $\eta<14/15$, see Fig.~\ref{fig:regions}). Thus, at $\xi_\mathrm{min}$ the dominant effect of loss is the unavoidable creation of odd-parity population while the residual reshaping within the even sector, constrained by the matched vacuum overlap, truly vanishes at the minimum.

A more detailed analysis of the special cases arising for $\xi_\mathrm{vac}>2.290047\dots$ is presented in Appendix~\ref{sec:EdgeCase}.
In particular, we discuss that for large squeezings additional even photon-number indices enter the set $I_{-}$ and we find $\xi_{\min} < \xi_{\mathrm{vac}}$.
However, the resulting shift of $\delta(\xi_{\mathrm{min}})$ relative to $\delta(\xi_{\mathrm{vac}})$ remains small.
For example, fixing $\eta=0.97$ and $\zt=2.38$, we obtain a relative deviation of $\delta(\xi_{\mathrm{vac}})$ to the real minimum of $\delta$ of below $1\%$ such that $\xi_{\mathrm{vac}}$ remains a good estimator for $\xi_{\mathrm{min}}$.

Based on these observations, we are able to efficiently obtain a simple estimator for the squeezing parameter that minimizes the impact of loss on the photon-number distribution and can even quantitatively lower bound the total variation distance for fixed system loss.
We emphasize that these quantities purely rely on Gaussian information which can efficiently be determined without calculating the actual distribution.

\subsection{Comparison of mitigation methods for single-mode squeezed vacuum}
\label{sec:ComparisonSMSV}
\begin{figure}[t]   
    \centering
    \includegraphics[width=1\columnwidth]{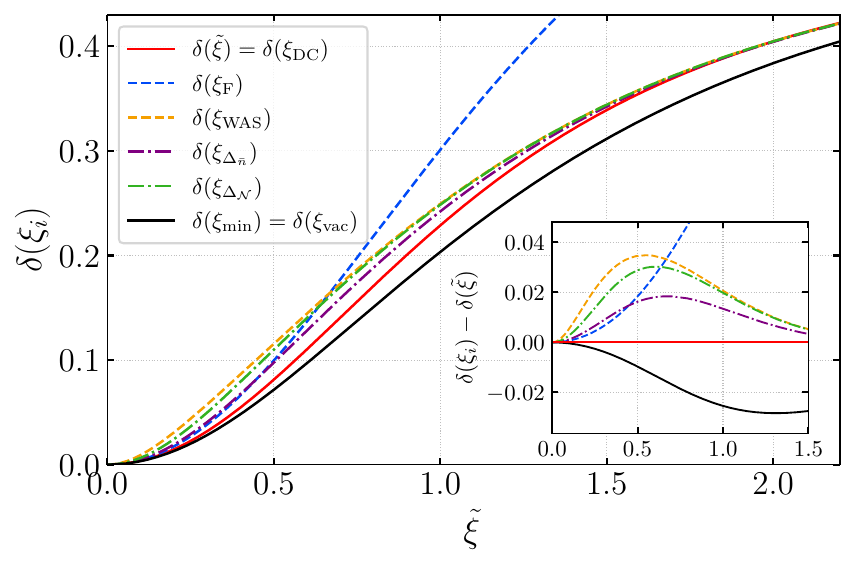}
	\caption{The total variation distance $\delta(\xi_i)$ between the photon-number distribution of a squeezed vacuum target state and a lossy squeezed vacuum state ($\eta=0.5$) is plotted as a function of the target squeezing $\tilde{\xi}$ for different optimization and correction methods. In the lower right inset, we show the relative improvement of the total variation distance $\delta(\xi_i)$ compared to the unoptimized case $\delta(\tilde{\xi})$ for the different methods. We find for the current example that the only method that provides improvements is the vacuum-overlap correction (solid black) that coincides with the minimum achievable total variation distance. The other methods, including the fidelity optimization (dashed blue), the Wasserstein-metric optimization (dashed orange), and the low-order photon-number moment corrections (mean: dash-dotted purple; variance: dashed-dotted green) do not provide any improvement compared to the unoptimized case.}
    \label{fig:deltaOverRtilde}
\end{figure}

In Fig.~\ref{fig:deltaOverRtilde} we summarize how the total variation distance $\delta\big(\hat{\rho}(\tilde{\xi}),\hat{\rho}'(\xi_i;\eta)\big)$ depends on the chosen mitigation scheme, which prescribes a corrected or optimized squeezing parameter $\xi_i(\tilde{\xi};\eta)$, for an example loss of $1-\eta=0.5$. Vacuum-overlap correction (solid black curve) yields the smallest total variation distance and coincides with the true minimizer. We also plot the unoptimized case (solid red curve) which trivially coincides with the displacement correction for the current example. All other methods increase $\delta$ relative to the unoptimized reference, including fidelity optimization (dashed blue), Wasserstein metric optimization (dashed orange), and the low-order moment corrections (mean: dash-dotted purple; variance: dash-dotted green). Fidelity optimization can be competitive at small $\tilde{\xi}$ but rapidly deteriorates as $\tilde{\xi}$ increases, eventually becoming the worst-performing method. 
The only other method that undershoots the optimal squeezing parameter is displacement correction, $\xi_{\rm F} \leq \zt = \xi_{\mathrm{DC}} < \xi_{\rmmin}$, yielding $\delta(\xi_{\mathrm{DC}}) = \delta(\zt) < \delta(\xi_{\rmF})$ due to the monotonicity properties of the total variation distance proven in Appendix~\ref{sec:VacEqMin}. Apart from vacuum-overlap correction, the remaining methods systematically overshoot the optimal parameter, $\xi_{\rmmin} = \xi_{\mathrm{vac}}<\xi_{\Delta_{\bar{n}}}<\xi_{\Delta_{\mathcal{N}}}<\xi_{\mathrm{WAS}}$. This ordering of the overshooting squeezing parameters is also reflected in a consistent performance hierarchy among the corresponding methods due to the monotonicity of $\delta$ for $\xi > \xi_{\rmmin}$. Mean matching performs best, followed by variance matching, with Wasserstein-metric optimization performing worst in this setting. Since the squeezing parameters obtained from the other phase-space measures also obey a strict ordering (see Eq.~\eqref{eq:IneqPhaseSpace}), they exhibit the same performance hierarchy, with Wasserstein-metric optimization as the best-performing representative among the phase-space distribution methods. Finally, for large $\tilde{\xi}$ all methods except vacuum-overlap correction and fidelity optimization converge to the total variation distance of the uncorrected state.

Remarkably, if we exclude vacuum-overlap correction, which coincides with the true minimizer, then the best remaining choice for single-mode squeezed vacuum is simply to leave the squeezing unchanged, i.e., $\xi = \zt = \xi_{\mathrm{DC}}$. In this case, optimizing the fidelity, any of the phase-space distances, or low-order photon-number moments yields a larger total variation distance than the unoptimized baseline. Only in the regime of large squeezing and sufficiently strong loss, $\eta < 0.5$, does mean photon-number matching start to outperform the unoptimized choice.

Up to this point, we have restricted the discussion to mitigation schemes that modify only the input squeezing parameter in order to reduce the deviation between the target photon-number distribution and those of the loss-affected state. In principle, one could enlarge the ansatz class and allow for more general (possibly mixed) Gaussian approximating states. We discuss this extension in Appendix~\ref{sec:GeneralGaussian Approximation} and show that general lossy single-mode Gaussian ans\"{a}tze do not yield a smaller total variation distance than the best results obtained within the squeezed-vacuum ansatz.

\section{Single-mode displaced squeezed vacuum state}
\label{sec:smDisplacedSqueezedState}
In this chapter, we extend our analysis of the effectiveness of each mitigation method to displaced squeezed vacuum states. 
Specifically, we consider a target displaced squeezed vacuum state defined by a squeezing parameter $\tilde{\xi}$ and displacement parameter $\tilde{\alpha}$,
\begin{equation}
	\hat{\rho}(\tilde{\xi}, \tilde{\alpha}) = \hat{D}(\tilde{\alpha}) \hat{S}(\tilde{\xi}) \op{0} \hat{S}^\dagger(\tilde{\xi}) \hat{D}^\dagger(\tilde{\alpha}),
\end{equation}
and compare its photon-number distribution to the photon-number distribution of a loss-affected displaced squeezed vacuum state
\begin{equation}
	\hat{\rho}'(\xi_i, \alpha_i; \eta) = \mathcal{E}_{\eta}\big(\hat{D}(\alpha_i) \hat{S}(\xi_i) \op{0} \hat{S}^\dagger(\xi_i) \hat{D}^\dagger(\alpha_i)\big),
\end{equation}
where the parameters $\xi_i$ and $\alpha_i$ depend on the chosen mitigation method.

Although displaced squeezed vacuum is characterized by two complex parameters in general, its photon-number statistics only depend on three real-valued degrees of freedom, the moduli of the squeezing and displacement parameters $|\xi|$ and $|\alpha|$ as well as on the relative phase between both.
This relative phase determines the orientation of the displacement relative to the principal axes of the squeezing ellipse and thereby influences whether the photon-number statistics have a more number-squeezed (sub-Poissonian) or a more phase-squeezed (super-Poissonian) character.
By contrast, a simultaneous rotation of both parameters corresponds to a global phase-space rotation and therefore has no physical consequence as it merely reflects the gauge freedom in fixing the phase-space coordinate system.
We therefore, choose the squeezing $\xi$ to be real and positive, while the displacement $\alpha=\abs{\alpha}\E^{\I \varphi}$ is a complex parameter. 
The explicit expression for the photon-number distribution $\mathcal{P}'_{\xi,\alpha;\eta}(m)$ is given in Appendix~\ref{sec:PofLossySC}.

To determine the parameters $\xi_{\rmmin}$ and $\alpha_{\rmmin}$ that minimize $\delta$, we perform a full grid search over $\xi$, $\abs{\alpha}$, and $\varphi$. Usually, the optimal phase parameter satisfies $\varphi_{\rmmin} \approx \tilde{\varphi}$ as one would naively expect. This behavior is also reflected in the various correction and optimization methods that we discuss below. 
Moreover, we observe that the dependence of the total variation distance on $\varphi$ is often rather flat in the vicinity of the minimum, indicating that small deviations from $\tilde{\varphi}$ have only a weak effect on $\delta$.

In contrast to the squeezed-vacuum case, deriving generic analytical statements for the minimizers $\xi_{\rmmin}$ and $|\alpha_{\rmmin}|$ is substantially more challenging. The photon-number probabilities $\mathcal{P}_{\xi,\alpha;\eta}'(m)$ depend nonlinearly on both parameters, and, crucially, the target distribution no longer exhibits a strict parity constraint. This loss of an exact even-odd structure removes the main simplifications that enabled several analytic arguments in the squeezed-vacuum case.

Based on numerical evidence, we find that the minimizers exhibit distinct trends across different parameter regimes. In particular, for loss exceeding roughly $20\%$ (i.e., $\eta < 0.8$), the squeezing parameters that minimize $\delta$ for number-squeezed target states are typically smaller than the corresponding minimizers for phase-squeezed target states. Conversely, the optimal displacement magnitude tends to be larger for sub-Poissonian target distributions than for super-Poissonian targets. In the former regime, we frequently observe $|\alpha_{\rmmin}| > |\tilde{\alpha}|/\sqrt{\eta}$, indicating that the $\delta$-optimal displacement can exceed the naive rescaling expected from a purely attenuating channel.

A simple heuristic for these trends could be based on the width of the photon-number distribution. For phase-squeezed target states the distribution is broader and places more weight on higher photon numbers. Under loss, this high-$m$ weight is more strongly redistributed towards lower $m$, which typically increases the mismatch in the photon-number distributions. As already observed for the squeezed-vacuum case, a broader target distribution usually requires a larger input squeezing to best compensate loss once the loss is sufficiently strong. This suggests that, for $\eta < 0.8$, the minimizing squeezing parameter for phase-squeezed target states is often larger than for number-squeezed target states. At the same time, the squeezed-vacuum analysis also shows that this intuition does not necessarily carry over to the weak-loss regime. We showed that for $\eta>14/15$ there exist parameter regions in which $\xi_{\rmmin} < \zt$. For displaced squeezed vacuum, we observe analogous low-loss exceptions, resulting in $\xi_{\min}$ that are smaller for the super-Poissonian than for the sub-Poissonian case.

Beyond these trends, our numerical search reveals additional particular examples for number-squeezed target states.
For small $\tilde{\alpha}$ and fixed $\eta$, we find intervals of $\tilde{\xi}$ for which the minimizer satisfies $\alpha_{\min}=0$.
Thus, the photon-number distribution $\mathcal{P}'_{\xi,0;\eta}$ of a loss-affected squeezed vacuum state is closer to the target distribution $\mathcal{P}_{\zt,\tilde\alpha}$ of a displaced squeezed vacuum state than that of any lossy displaced squeezed vacuum with nonvanishing displacement in this case. As the loss increases, this behavior occurs already at smaller values of $\zt$.
For example, for $\tilde\alpha=0.2$, the resulting $\alpha_\mathrm{min}$ is zero at $\tilde{\xi} \approx 0.4$ for $\eta = 0.5$.
In fact, for the same target displacement, $\alpha_\mathrm{min} = 0$ already at $\tilde{\xi} \approx 0.22$ when the loss increases to 80\% ($\eta = 0.2$).

An intuitive explanation for this behavior is that for weak squeezing a squeezed vacuum is well approximated by its lowest Fock components, with most of its weight in $\ket{0}$ and $\ket{2}$ and no odd-photon contribution. Loss transfers now population from $\ket{2}$ to $\ket{1}$. A weakly displaced squeezed vacuum state likewise has its main weight in $\ket{0}$, $\ket{1}$, and $\ket{2}$. Thus, for small $|\tilde{\alpha}|$, the distribution of a lossy squeezed vacuum can approximate those of a displaced squeezed vacuum target state. In this regime the best match is often obtained by slightly increasing the input squeezing to compensate for the loss-induced depletion of the $\ket{2}$ component. For larger $|\tilde{\alpha}|$, however, the displacement-induced odd-photon population cannot be mimicked by loss alone, and the optimum shifts to $\alpha_{\rmmin}\neq 0$.

\begin{figure*}[t]   
    \centering
    \includegraphics[width=1\textwidth]{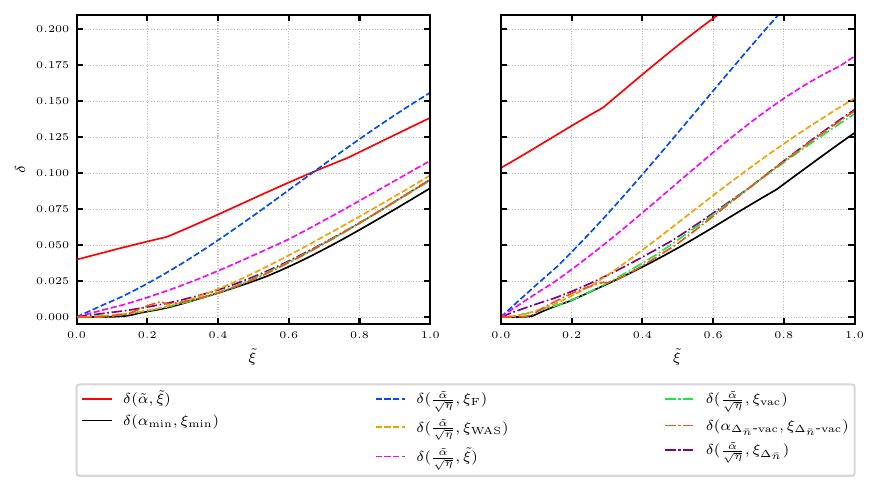}
	\caption{Total variation distance $\delta$ between the photon-number distribution of the target state and different optimized lossy single-mode displaced squeezed vacuum states as a function of the target squeezing $\tilde{\xi}$. Left and right panels correspond to channel transmissivities of $\eta = 0.8$ and $\eta = 0.5$, respectively. The plotted optimization and correction methods are: "no" (solid red), the $\delta$ minimizing (solid black), fidelity optimization (dashed blue), Wasserstein-metric optimization (dashed orange), displacement correction (dashed magenta), vacuum-overlap with displacement correction (dash-dotted green), vacuum-overlap with mean correction (dash-dotted dark orange), and mean with displacement correction (dash-dotted purple).}
	\label{fig:2together}
\end{figure*}

In the following, we consider the mitigation methods introduced in Sec.~\ref{sec:Mitigation} and evaluate whether they provide useful approximations for determining the parameters that minimize the total variation distance between the desired target photon-number distribution and that of a lossy displaced squeezed vacuum state. 
We illustrate this by plotting the total variation distance as a function of the target squeezing $\zt$ for $\tilde{\alpha} = 0.5$ (thus, $\tilde\varphi=0$) for two representative loss scenarios (left panel: $\eta = 0.8$; right panel: $\eta = 0.5$) in Fig.~\ref{fig:2together}. The solid red line depicts $\delta$ for the loss-affected state with the same input parameters as the target state, whereas the solid black line corresponds to squeezing and displacement parameters $\xi_{\rmmin}$ and $\alpha_{\rmmin}$ that minimize $\delta$. We further compare these results to the different methods trying to mitigate the impact of loss. A detailed discussion of the various options is done below.
In the case of $\tilde\xi=0$, all correction and optimization methods properly adjust the photon-number distribution, as this situation corresponds to a coherent state.

\subsection{Fidelity optimization}
As discussed in Sec.~\ref{sec:Fidelity}, the displacement and squeezing optimization to maximize the fidelity between a pure Gaussian target state and an arbitrary (lossy) Gaussian state decouples. 
For the displacement parameter, we obtain $\alpha_\rmF = \tilde\alpha/\sqrt{\eta}$. The remaining optimization over $\xi$ is identical to the squeezed-vacuum problem treated in Eq.~\eqref{eq:zFidelity}, and its solution satisfies $\xi_{\rmF} \leq \tilde{\xi}$.

We again observe that fidelity optimization does not necessarily reduce the total variation distance. For sufficiently large target squeezing $\zt$ at a fixed $\tilde{\alpha}$, we find ${\delta\big(\rho(\tilde{\xi},\tilde\alpha), \rho'(\xi_\mathrm{F},\alpha_\mathrm{F};\eta) \big) \geq \delta\big(\rho(\tilde{\xi},\tilde\alpha), \rho'(\tilde{\xi},\tilde\alpha;\eta) \big)}$.
For sufficiently small squeezing, by contrast, fidelity maximization decreases the dissimilarity between the photon-number distributions of the states. This is consistent with the fact that in this regime the target distribution is close to a Poissonian distribution of a coherent state for which the effect of loss can be compensated. The performance of the fidelity optimization is depicted in Fig.~\ref{fig:2together} as a dashed blue line. Although fidelity optimization can decrease $\delta$ in certain parameter regimes (unlike for squeezed vacuum), it generally performs worse than the other mitigation schemes.

\subsection{Displacement correction}
For the displacement correction, one obtains $\alpha_\mathrm{DC}=\tilde\alpha/\sqrt{\eta}$ and $\xi_\mathrm{DC}=\tilde{\xi}$. 
We again observe that this method achieves a smaller total variation distance (and thus closer photon-number distribution) than fidelity optimization, see the dashed magenta line in Fig.~\ref{fig:2together}. In contrast to the squeezed vacuum case, where displacement correction even outperformed the phase-space optimization methods as well as the low-order photon-number moments corrections over a wide parameter regime, including a nonzero displacement in the target state usually changes this ordering and displacement correction yields larger $\delta$ than the other mitigation methods. Only for sufficiently small displacements of the target state, displacement correction leads to smaller $\delta$ than the other methods. This effect also strongly depends on the relative phase $\tilde\varphi$. For example, for $\zt = 1$, $|\tilde\alpha| = 0.1$, and $\eta = 0.5$, we find that displacement correction performs slightly worse ($\delta \approx 0.186$) than choosing $\alpha_{\mathrm{DC}}$ and additionally fixing the squeezing parameter by matching the mean photon numbers ($\delta \approx 0.183$) for $\tilde\varphi = 0$. This ordering is inverted for $\tilde\varphi = \frac{\pi}{2}$, as ${\delta \approx 0.203}$ for displacement correction and $\delta \approx 0.210$ for additional mean matching. This behavior is consistent with the heuristic picture that displacement correction performs well for broad (super-Poissonian) distributions. This is in line with our squeezed-vacuum results, where the (super-Poissonian) squeezed-vacuum distribution also favored displacement correction over low-order photon-number-moments correction or phase-space distribution optimization for large parameter ranges. For larger $\tilde\alpha$, however, this advantage disappears and displacement correction is outperformed by the other mitigation methods for any $\tilde\varphi$. For instance, additional mean matching yields smaller $\delta$ already for $|\tilde\alpha| = 0.2$ independent of the relative phase.

\subsection{Phase-space distribution optimization}
The task to optimize the displacement and squeezing parameter also decouples for the various phase-space measures.
As in the fidelity and displacement correction case, the optimal displacement is obtained by inverting the loss-induced attenuation, yielding $ \alpha_\mathrm{WAS}=\alpha_\mathrm{BHA}=\alpha_\mathrm{KLD}=\alpha_\mathrm{KLDup}=\alpha_\mathrm{KLDpu}=\tilde\alpha/\sqrt{\eta} $.
All results derived for the optimal squeezing parameters in the squeezed-vacuum case (cf. Sec.~\ref{sec:SMSV-PSO}) apply unchanged in the present case.
In particular, the inequality given in Eq.~\eqref{eq:IneqPhaseSpace} remains valid. 
We exemplarily show the resulting $\delta$ for the minimization of the Wasserstein metric in Fig.~\ref{fig:2together} as a dashed orange line. While the total variation distance obtained by the phase-space optimization methods is typically close to the actual minimum for small squeezing parameters, the photon-number-statistics based correction methods perform better for larger target squeezing parameters.

\subsection{Low-order photon-number moments correction}\label{sec:SCmeanVariance}
Extending the target states from squeezed vacuum to displaced squeezed vacuum states introduces additional degrees of freedom in both the photon-number mean and variance, originating from the displacement.
This leads to ambiguities if we only impose a single photon-number moment correction.
For example, one can match the photon-number means of the target and lossy states and solve for the corresponding squeezing parameter, which leads to
\begin{equation}
	\xi_{\Delta_{\bar{n}}}=\mathrm{arsinh}\sqrt{\frac{\sinh^2(\tilde \xi)+\abs{\tilde\alpha}^2}{\eta}-\abs{\alpha}^2}.
\label{eq:SMDSV-MeanMatching}
\end{equation}
Here, the displacement $\alpha$ of the loss-affected state, is in principle a free parameter as long as $\xi_{\Delta_{\bar{n}}} \geq 0$, i.e. $\frac{\sinh^2(\tilde \xi)+\abs{\tilde\alpha}^2}{\eta}-\abs{\alpha}^2\geq0$.
Equivalently we could have solved the mean condition for the displacement, rendering the squeezing parameter of the loss-affected state a free parameter.
In Sec.~\ref{sec:SCasApprox}, we attempted to fix this freedom by additionally matching the variances, and showed that no physically valid solution exists. 
Following this approach, we use the variance condition to fix $|\alpha|$. This results in a root finding problem of a quartic polynomial in $|\alpha|^2$.
Although a closed-form solution exists for quartic polynomials, it is impractical to verify whether the resulting roots, that still depend on $\varphi$, satisfy the physicality condition $\xi_{\Delta_{\bar{n}}} \geq 0$.
Instead of proceeding analytically, we therefore analyze the condition numerically to determine whether physically valid solutions exist for a given target state.
We find that for small displacement and comparatively large squeezing, physically valid solutions do not exist, which is consistent with the observations made in Sec.~\ref{sec:SCasApprox}.
However, for sufficiently large displacement $\tilde\alpha$ relative to the squeezing $\zt$, valid solutions do exist. 
Nonetheless, simultaneously matching the photon-number mean and variance often suffers from numerical instabilities, since small changes in the input parameters can cause abrupt jumps in the resulting corrected values. In practice, we find more reliable mitigation strategies based on photon-number moments correction that outperform this approach and yield smaller $\delta$ which we discuss in the following.

Motivated by the fact that photon-number mean correction provided the smallest $\delta$ among the photon-number moments corrections for squeezed vacuum, we use the condition of mean matching given by Eq.~\eqref{eq:SMDSV-MeanMatching}. We further use displacement correction, $\alpha_{\Delta_{\bar{n}}}=\tilde\alpha/\sqrt{\eta}$, to fix the remaining ambiguity of the initial state. This choice also implicitly enforces $\varphi_{\Delta_{\bar{n}}}=\tilde{\varphi}$, which is consistent with the full $\delta$-minimization analysis, where the optimum typically occurs for aligned phases. Furthermore note that displacement correction implies that the squeezing parameter gets adjusted as in the squeezed vacuum case, $\xi_{\Delta_{\bar{n}}}=\arsinh\big[ \sinh(\tilde \xi)/\sqrt{\eta}\big]$, cf. Eq.~\eqref{eq:SMSV-MeanMatching}. Overall, this scheme remains close to the minimal achievable $\delta$ for small target squeezing and outperforms the phase-space optimization methods in the high-squeezing regime (see the purple dash-dotted curve in Fig.~\ref{fig:2together} for a representative example). A further option combining photon-number mean correction with vacuum overlap is discussed in the next subsection.

\subsection{Vacuum-overlap correction}
We now turn to vacuum-overlap correction for displaced squeezed states by imposing $\mathcal{P}_{\tilde{\xi}, \tilde{\alpha}}(0)=\mathcal{P}'_{\xi,\alpha;\eta}(0)$, with
\begin{align}
	\mathcal{P}_{\tilde{\xi}, \tilde\alpha}(0)=\frac{\E^{-\abs{\tilde{\alpha}}^{2} \big[1+\tanh(\tilde{\xi})\cos (\frac{\tilde\varphi}{2}) \big]}}{\cosh \tilde{\xi} }
\end{align}
and
\begin{align}
 \mathcal{P}'_{\xi,\alpha;\eta}(0) = \frac{ \exp\!\Big[\!-\! \frac{\eta \abs{\alpha}^2 \big[2-\eta + \eta \cosh(2\xi) + \eta \cos (\frac{\varphi}{2}) \sinh(2\xi)\big]}{2 [ 1+\eta(2-\eta) \sinh^2\xi ]}\!\Big] }{\sqrt{1+\eta(2-\eta) \sinh^2\xi }}.
\end{align}
In contrast to the squeezed vacuum case and analogous to the photon-number moment correction for displaced squeezed vacuum, the vacuum-matching condition does not fix the initial parameters ($\xi$, $\abs{\alpha}$, $\varphi$) of the lossy state uniquely. 
Before we resolve this ambiguity by combining the scalar vacuum-overlap relation with additional constraints that can be extracted from a Gaussian state in a computationally efficient manner and benchmarking these resulting schemes, we first discuss generic structural features of the photon-number statistics of the $\delta$-minimizing lossy displaced squeezed vacuum state and how they relate to vacuum-overlap correction.

\begin{figure}[t]   
    \centering
    \includegraphics[width=1\columnwidth]{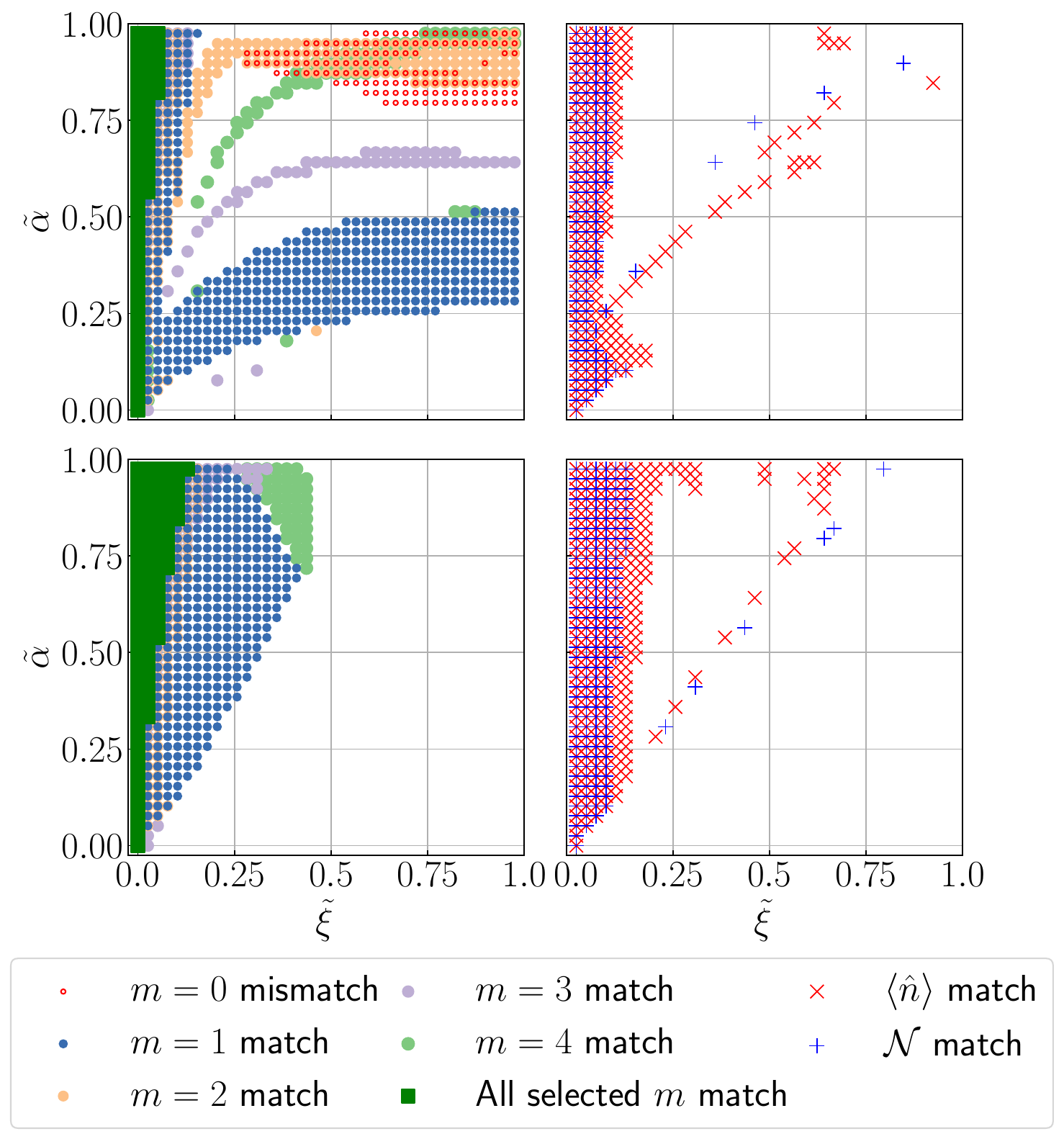}
	\caption{
		Comparison of the photon-number statistics of displaced squeezed vacuum target states with the photon-number statistics of loss-affected states that minimize the total variation distance $\delta$ for $\eta = 0.5$. For each point in the $(\tilde{\xi},\tilde{\alpha})$ grid we compute the lossy displaced squeezed state parameters $(\xi_\mathrm{min},\alpha_\mathrm{min})$ that minimize $\delta\big(\hat{\rho}(\tilde{\xi},\tilde{\alpha}),\hat{\rho}'(\xi,\alpha;\eta)\big)$ and compare the photon-number statistics of both states. The left panels show the agreement on the level of individual photon-number probabilities $\mathcal{P}(m)$. Colored dots indicate grid points where the individual photon-number probabilities match within a relative tolerance of $0.005$. Open red circles highlight locations where the vacuum component $m=0$ deviates beyond this tolerance. The filled green squares indicate points where all selected $m\in\{0,1,2,3,4\}$ satisfy the matching criterion simultaneously. The right panels isolate aggregate statistics, marking regions where the mean photon number $\langle \hat n\rangle$ (red $\times$) and the photon-number variance $\mathcal{N}$ (blue $+$) agree within the same tolerance. The upper plots show a number squeezed case, $\tilde\varphi=0$, while the lower plots show a phase squeezed example with $\tilde\varphi=\pi/2$.
	}
	\label{fig:dotplot}
\end{figure}

Numerical simulations, shown in Fig.~\ref{fig:dotplot}, reveal recurring structural features in the parameter values that minimize the total variation distance.
In a broad region of parameter space, the minimizing state reproduces the target state's vacuum probability to high accuracy. Note that we marked only those points where the vacuum probabilities do not match with a red open circle for better readability, see the upper right region of the upper left plot.
Moreover, regions exist where, in addition to the vacuum component, one or more higher photon-number probabilities coincide between the target and the optimal lossy distributions, which likely becomes possible due to the additional degrees of freedom introduced by displacement. 
These coincidences form bands in parameter space for number-squeezed distributions, whose ordering appears to depend on photon-number parity. 
Fixing $\zt$, the bands of odd photon numbers appear in increasing order of $m$ as $\tilde\alpha$ increases, while the bands of even photon numbers appear in decreasing order of $m$ as $\tilde\alpha$ increases.
We also find bands in the parameter space where the mean photon number or the photon-number variance match between the two distributions.
This is not surprising, as these are aggregate statistics that depend on all photon-number probabilities. 
By contrast, the coincidence sets appear as extended patches (with a pronounced low-$\tilde{\xi}$ ridge) for phase-squeezed target states.

In the squeezed vacuum case, we observed a comparatively narrow region where instead of the vacuum component, other individual photon-number probabilities matched (see Fig.~\ref{fig:regions} and Appendix~\ref{sec:EdgeCase}).
Displacement appears to enhance this non-vacuum-matching behavior for number-squeezed states, leading to a shift of this region to smaller values of $\zt$, see the upper right region in the upper left plot of Fig.~\ref{fig:dotplot}. 
Heuristically, this may be related to the fact that a nonzero displacement transfers probability weight away from the vacuum component and the distribution becomes narrower for number-squeezed states, so that vacuum matching becomes less dominant and other photon-number components can control the optimality condition. 
By contrast, for phase-squeezed target states we find no such deviation within the simulated parameter ranges. This seems to be consistent from the perspective that phase-squeezed targets exhibit a broader photon-number distribution, more akin to the nondisplaced squeezed-vacuum case, so that the non-vacuum-matching region is shifted to larger $\zt$ and therefore lies outside the parameter region explored here.

\begin{figure}[t]   
    \centering
    \includegraphics[width=1\columnwidth]{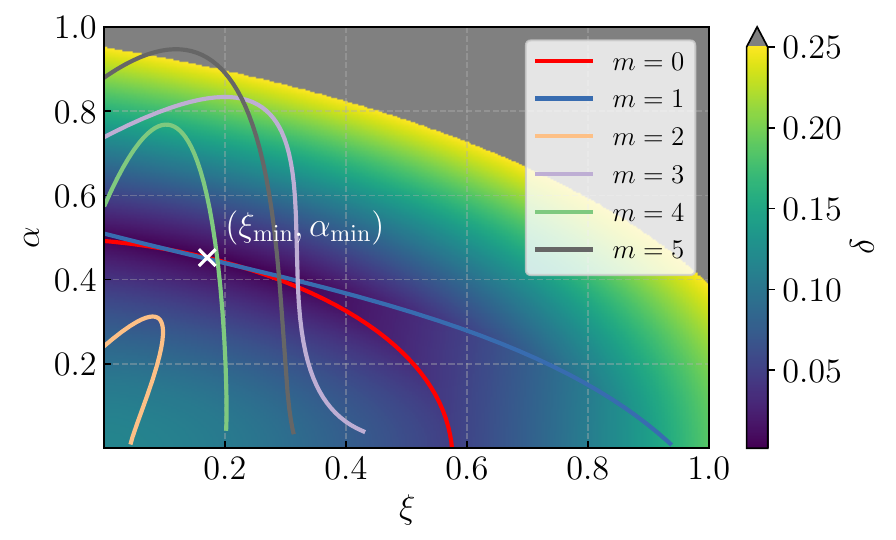}
	\caption{
		This figure illustrates how $\delta$ changes in the $(\xi,\alpha)$ parameter space for a fixed loss of $50\%$. The target displaced squeezed vacuum state has parameters $\tilde{\xi}=0.15$ and $\tilde{\alpha}=0.31$. The color map shows the total variation distance $\delta\big(\hat{\rho}(\tilde{\xi},\tilde{\alpha}),\hat{\rho}'(\xi,\alpha;\eta)\big)$ as a function of the lossy state parameters $(\xi,\alpha)$ for $\varphi = 0$. Superimposed are colored belts indicating locations where individual photon-number probabilities $\mathcal{P}(m)$ of the target and lossy states match.
		Each colored band corresponds to a specific photon number $m$.
		The global minimum of $\delta$ is often located near an intersection of some of these belts, suggesting that simultaneous matching of multiple photon-number probabilities contributes to minimizing the total variation distance.
		In this example the minimum is found at the location marked with a white cross near the intersections of the $m=0$ (red), $m=1$ (blue), and $m=4$ (light green) belts.
	}
	\label{fig:bandplot}
\end{figure}

A closer inspection of $\delta$ in the vicinity of its numerically identified minima reveals valley-like structures in the $(\xi,\alpha)$-parameter space. 
Empirically, these valleys coincide with locations where one specific photon-number probability $\mathcal{P}_{\zt,\tilde{\alpha}}(m)$ of the target state nearly matches the corresponding probability of the lossy state, $\mathcal{P}'_{\xi,\alpha;\eta}(m)$. 
This valley-type behavior is particularly pronounced for the vacuum match ($m=0$), still visible for $m=1$, and barely visible by eye for $m>1$, see Fig.~\ref{fig:bandplot} where we highlighted the various matchings of individual photon-number probabilities of the target and lossy states.
The global minimum of $\delta$ is then often found close to intersections of two (or more of) such valleys, i.e., parameter points where several low-photon-number components are matched simultaneously. 
This behavior provides a heuristic explanation why the minimization of $\delta$ can lead to matching of multiple photon-number probabilities.

Figure~\ref{fig:dotplot} reveals that vacuum overlap correction is also a useful description to mitigate the impact of loss over a wide parameter range for displaced squeezed vacuum states. However, there exists a continuous set of solutions satisfying this scalar relation, see the red curve in Fig.~\ref{fig:bandplot}. To find the optimal initial parameters, we have to perform a minimization over this set of solutions which is, of course, possible for the current example, but becomes a computational intractable task for multimode systems.
In order to have a practical prescription for choosing a unique pair of parameters from the continuous family of solutions defined by the vacuum-matching condition, we consider two options that rely on computationally accessible Gaussian properties. 
We choose $\varphi=\tilde{\varphi}$ and use the relations for equal vacuum overlap, $\mathcal{P}_{\tilde{\xi}, \tilde{\alpha}}(0)=\mathcal{P}'_{\xi,\alpha;\eta}(0)$, as well as equal mean photon number, $\sinh^2 \zt + \abs{\tilde\alpha}^2 = \eta (\sinh^2 \xi + \abs{\alpha}^2)$, to determine $\xi$ and $\abs{\alpha}$. Alternatively, we may choose displacement correction to fix $\alpha = \tilde\alpha/\sqrt{\eta}$ and determine the remaining squeezing parameter via matching vacuum probabilities.
Both prescriptions single out a well-defined pair of parameters and yield comparable reductions of the total variation distance. 
The numerical comparison is shown in Fig.~\ref{fig:2together} as a green dash-dotted curve (vacuum-overlap combined with displacement correction) and as a dark orange dash-dotted curve (vacuum-overlap combined with mean photon-number correction).
Furthermore, these two prescriptions based on the vacuum overlap outperform the other mitigation methods for all tested parameter ranges (albeit photon-number mean correction combined with displacement correction can be surprisingly close for large squeezing parameters).
Even in regions where $\delta$-minimizing parameters do not yield a matching in the vacuum probabilities, implementing vacuum overlap correction still results in only a marginal increase in $\delta$ compared to its minimal value. Thus, even when the $\delta$-minimizing parameters do not exactly satisfy vacuum matching, enforcing $\mathcal{P}_{\tilde{\xi},\tilde{\alpha}}(0)=\mathcal{P}'_{\xi,\alpha;\eta}(0)$ typically yields photon-number distributions that remain close to the $\delta$-optimal ones.

\section{Error mitigation for multimode systems}
\label{sec:MultiMode}
The complexity of the analysis increases substantially with the number of modes for both the lossless and the lossy setting.
In Fig.~\ref{fig:DoktorovSO2MitLoss}, a two-mode system is shown with loss channels placed at physically relevant locations, illustrating where loss acts in a realistic implementation. In this specific example there are four loss segments affecting the output photon-number distribution. In general, the loss segments do not contribute equally. Instead, the impact of a given loss probability $1-\eta^{(\ell)}_j$ (mode $j\in\{1,2\}$ at layer $\ell\in\{1,2\}$) is large when the mean photon number at the corresponding location is large, since the absolute number of photons removed by loss scales with the local photon population. 
The four loss segments  can be grouped into two different layers. A pre-interferometer layer that captures mode-resolved state-preparation imperfections and coupling loss into the interferometer, as well as a post-interferometer layer that models mode-resolved detection inefficiencies. Propagation loss within the interferometric network can be absorbed into the pre- or post-interferometer layer for a two-mode system, but has nontrivial impact for $M>2$.

To properly account for interferometric loss for $M>2$, one may regard the device as a sequence of lossless mixing layers interlaced with mode-dependent loss channels in a generic implementation.
In the phase-space picture this can be implemented by diagonal transmissivity matrices acting at positions $\ell$ in the interferometer,
\begin{equation}
	\ten{T}^{(\ell)}=\diag\Big(\sqrt{\eta^{(\ell)}_1},\sqrt{\eta^{(\ell)}_1},\ldots,\sqrt{\eta^{(\ell)}_M},\sqrt{\eta^{(\ell)}_M}\Big),
\end{equation}
transforming the first and second moments as
\begin{equation}
	\bar{\vb{r}} \mapsto \ten{T}^{(\ell)}\bar{\vb{r}},\quad	\ten{\sigma} \mapsto \ten{T}^{(\ell)}\ten{\sigma}(\ten{T}^{(\ell)})\tran + \frac{1}{2}\big(\ten{\mathbbm{1}}_{2M}-\ten{T}^{(\ell)}(\ten{T}^{(\ell)})\tran\big).
\end{equation}
If several loss segments act sequentially at positions $\ell=1,\dots,n$, they are interlaced with lossless interferometer segments described by symplectic matrices $\ten{S}^{(\ell)}$ characterizing the unitary matrix $\ten{U}$. For a single step, we obtain
\begin{align}\label{eq:finalSigma}
\bar{\vb{r}}^{(\ell)} &= \ten{T}^{(\ell)}\ten{S}^{(\ell)}\,\bar{\vb{r}}^{(\ell-1)},\notag\\
\ten{\sigma}^{(\ell)} &= \ten{T}^{(\ell)}\ten{S}^{(\ell)}\,\ten{\sigma}^{(\ell-1)}\,(\ten{S}^{(\ell)})\tran(\ten{T}^{(\ell)})\tran\\
&\quad+ \frac{1}{2}\big(\ten{\mathbbm{1}}_{2M}-\ten{T}^{(\ell)}(\ten{T}^{(\ell)})\tran\big)\notag,
\end{align}
starting from $(\bar{\vb{r}}^{(0)},\ten{\sigma}^{(0)})=(\bar{\vb{r}},\ten{\sigma})$. The final moments are $(\bar{\vb{r}}',\ten{\sigma}')=(\bar{\vb{r}}^{(n)},\ten{\sigma}^{(n)})$. 
Equivalently, at the level of field operators the overall transformation on the signal modes can be described by a complex contraction $\ten{T}$ (with $\ten{T}^{\dagger}\ten{T} \leq \mathbbm{1}$) having $2M^2$ real parameters characterizing the lossy map $\ten{T}$. 
If one fixes the target lossless unitary $\ten{U}$, the remaining nonredundant degrees of freedom associated with loss alone reduce to $M^2$.

\subsection{Vacuum-overlap correction for multimode systems}
\label{sec:TwoModeVacuum}
Motivated by the effectiveness of vacuum-overlap correction in the single-mode setting, we will focus on investigating whether an analogous strategy can be employed to mitigate the impact of loss in multimode Gaussian systems. 
To this end, we consider the simplest nontrivial multimode system consisting of two modes.

A general $M$-mode nondisplaced pure Gaussian state is obtained by applying a Gaussian operation from the symplectic group $\mathrm{Sp}(2M,\mathbb{R})$ acting on the vacuum.
Since the vacuum is invariant under passive linear optics ($\mathrm{U}(M)$), distinct states are described by the coset space $\mathrm{Sp}(2M,\mathbb{R})/\mathrm{U}(M)$, which has $\big(2M^2+M\big)-M^2=M^2+M$ real degrees of freedom.
These correspond to $M$ single-mode squeezers together with an $M$-mode passive interferometer, where the squeezer phases can be shifted into the interferometer. 
Finally, photon-number statistics are insensitive to output-local phases which can therefore be gauged away, reducing the relevant degrees of freedom to $M^2$. 
Thus, the photon-number statistics of a general two-mode system without displacement can be desribed by four nonredundant physical degrees of freedom which can be parameterized by the real squeezing parameters $\xi_1$ and $\xi_2$ and a passive interferometer given by
\begin{equation}
	\ten{U}(\theta, \gamma) = 
	\begin{pmatrix}
		\cos\theta & -\E^{\I\gamma}\sin\theta \\
		\E^{-\I\gamma}\sin\theta & \cos\theta
	\end{pmatrix}.
\end{equation}

In contrast to the single-mode case, there is no reason to expect that the $\delta$-minimizing lossy state is obtained by only optimizing the input squeezing strengths.
In practice, this is also what we generally observe in our numerical simulations.
%Allowing additional degrees of freedom beyond optimizing the squeezing strengths (in particular the interferometer mixing angle and relative phase shift) can further reduce $\delta$, but the improvement is often small. 
%Allowing optimization over additional degrees of freedom beyond the squeezing strengths, i.e., the interferometer parameters, can further reduce \delta, but the improvement is often small.
Allowing optimization over the interferometer parameters as well can further reduce $\delta$, but the improvement is often small. 
At least for the two-mode setting considered here, we numerically observe that optimizing $\delta$ over the full parameter set $(\xi_1,\xi_2,\theta,\gamma)$ yields a total variation distance that is at most ${\approx}10\%$ smaller than the best value obtained when optimizing over $(\xi_1,\xi_2)$ alone while keeping the interferometer parameters fixed at their target values, $\theta = \tilde{\theta}$ and $\gamma = \tilde{\gamma}$.

The fully optimized states exhibit several notable features.
First, the vacuum probability is matched with high accuracy for most target states, reinforcing its role as the dominant quantity governing the reduction of $\delta$. 
Only for large target squeezing and low loss, we find examples with nonmatching vacuum probabilities, being consistent with the single-mode analysis. 
Moreover, for many different target states, we observe that additional low-photon-number probabilities are also closely matched, also mirroring behavior previously found in the single-mode displaced squeezed vacuum case, see Fig.~\ref{fig:dotplot}.
However, marginal vacuum probabilities of the individual modes are not reproduced by the optimized states, making this approach less suitable for identifying unique input states for vacuum-overlap correction. 
An analogous observation applies to mode-resolved mean-photon-number matching, which is likewise not reproduced in general.
This indicates that the correction preferentially aligns global photon-number statistics while allowing freedom in the local structure of the state. 
Finally, we also observe $\delta_{\mathrm{even}} \approx \delta_{\mathrm{odd}}$ at the minimum, as in the single-mode squeezed-vacuum case.

Given that the global vacuum probability has a dominant impact on $\delta$, vacuum-overlap matching is a natural candidate for mitigating the impact of loss in multimode settings as well. However, vacuum matching only defines a manifold of admissible parameter sets, and it is therefore a priori unclear how to select a near-$\delta$-optimal point on this manifold using only efficiently accessible constraints. 
Concretely, we proceed as follows.
For a fixed target state specified by $(\tilde{\xi}_1,\tilde{\xi}_2,\tilde{\theta},\tilde{\gamma})$ and a fixed loss model, we enforce the vacuum-overlap constraint
\begin{equation}
\mathcal{P}_{\tilde{\xi}_1,\tilde{\xi}_2,\tilde{\theta},\tilde{\gamma}}(\vb{0}) =
\mathcal{P}'_{\xi_1,\xi_2,\theta,\gamma;\vb*{\eta}}(\vb{0}),
\label{eq:VacuumOverlap2Mode}
\end{equation}
using Eq.~\eqref{eq:VacuumOverlapNonDisplaced}. 
For the lossless target state, the vacuum probability is invariant under passive linear optics, so it does not depend on the interferometer parameters $(\tilde\theta,\tilde\gamma)$.
By contrast, in the presence of loss the vacuum probability becomes sensitive to $(\theta,\gamma)$ because the interferometer redistributes photons among the modes before subsequent loss channels act on the state.
The above constraint, Eq.~\eqref{eq:VacuumOverlap2Mode}, defines a three-dimensional manifold within the four-dimensional parameter space $(\xi_1,\xi_2,\theta,\gamma)$. 
As a first test, we randomly sample points on this manifold by varying the squeezing ratio and the interferometer parameters, and compute $\delta$ for each candidate state.
We then compare the resulting values to the total variation distance of the uncorrected lossy target state and to the minimal achievable $\delta$.
Furthermore, we analyze more systematic approaches such as fixing the ratio of the squeezing parameters to match that of the target state, $\xi_1/\xi_2 = \tilde{\xi}_1/\tilde{\xi}_2$, i.e., scaling all squeezing parameters by a constant factor to match global vacuum probabilitiy, or impose a loss-weighted squeezing ratio, while fixing the interferometer parameters to their target values. Moreover, we compare to prescriptions obtained from other methods, such as fidelity maximization, Wasserstein-metric optimization, or mean photon-number correction.

We summarize the numerical values for an illustrative two-mode example with target parameters $(\tilde{\xi}_1,\tilde{\xi}_2,\tilde{\theta},\tilde{\gamma})=(0.4,0.5,0.8,0.44)$ and compare the corresponding target photon-number distribution with the photon-number distribution obtained from an unbalanced loss model with mode- and layer-dependent transmissivities ($\eta^{(1)}_1=0.7$, $\eta^{(1)}_2=0.6$, $\eta^{(2)}_1=0.5$, $\eta^{(2)}_2=0.8$). 
This example is representative in the sense that we observe the same qualitative trends for many other choices of target parameters and loss configurations.
A direct numerical minimization of the total variation distance over all degrees of freedom yields a best achievable value of $\delta_{\rmmin} \approx 0.107$. 
By contrast, if we apply none of the proposed methods to mitigate the impact of loss, we obtain $\delta_\mathrm{uncor} \approx 0.140$ for the lossy target state.
Note that this baseline coincides with displacement correction in this nondisplaced setting. 
Restricting the optimization to sampling from the vacuum-overlap-constrained manifold spans a range of $\delta\in[0.107,0.139]$. %and therefore improves upon the uncorrected parameters.
%Moreover, t
The best vacuum-matched candidates nearly attain the numerically optimized value and even in worst case instances it results in smaller $\delta$ than the uncorrected state. 
Furthermore, the resulting $\delta$ values are not uniformly distributed across this interval but are skewed towards its lower end, so that a randomly chosen vacuum-matched point is more likely to yield a noticeable improvement over the uncorrected baseline. In particular, the sample mean $\langle\delta\rangle \approx 0.117$ lies closer to $\delta_{\rmmin}$ than to $\delta_{\mathrm{uncor}}$. Moreover, the spread of the samples is small, with a standard deviation $\sigma_\delta \approx 0.002$, indicating that vacuum matching typically produces candidates that are close to the optimal value.

Next, we study the performance of vacuum-overlap prescriptions that impose additional constraints on the input squeezing parameters and set the interferometer parameters to their target values to resolve the ambiguities in the multimode case. First, we impose $\xi_1/\xi_2 = \tilde{\xi}_1/\tilde{\xi}_2$ in addition to $\mathcal{P}_{\tilde{\xi}_1,\tilde{\xi}_2,\tilde{\theta},\tilde{\gamma}}(\vb{0}) = \mathcal{P}'_{\xi_1,\xi_2,\tilde\theta,\tilde\gamma;\vb*{\eta}}(\vb{0})$, yielding $\delta\approx0.120$. 
Second, we additionally impose a loss-weighted squeezing-ratio constraint formulated in terms of the squeezing intensities, $\sinh^2\xi_i$. More precisely, we set $\frac{\sinh^2\xi_1}{\sinh^2\xi_2} = \frac{\sinh^2\tilde{\xi}_1}{\sinh^2\tilde{\xi}_2} \frac{\eta_{\mathrm{eff},2}}{\eta_{\mathrm{eff},1}}$, where $\eta_{\mathrm{eff},i}$ is the effective transmissivity of input mode $i$, defined as the total output intensity obtained from unit input in mode $i$.
This results in $\delta\approx0.120$ being close to the previous mitigation scheme for the current example. However, in general, this loss-weighted prescription is expected to better capture the intuition that larger loss in a given mode requires a larger squeezing correction in that mode. Both methods yield $\delta$ values close to $\delta_{\rmmin,\xi} \approx 0.116$ obtained by optimizing only the input squeezings while keeping the interferometer parameters fixed at their target values. Moreover, they offer a straightforward generalization to arbitrary mode numbers.

Finally, we benchmark vacuum-overlap-based prescriptions against the remaining optimization and correction schemes.
As for single-mode squeezed vacuum, the fidelity optimization performs generally worse than the lossy target state. For our current example, the fidelity-optimized state yields $\delta_\mathrm{F} \approx 0.165$. Interestingly, phase-space optimization in terms of the Wasserstein metric performs even worse than fidelity optimization for this concrete example, $\delta_\mathrm{WAS} \approx 0.199$.
However, we stress that the ordering between fidelity- and Wasserstein-based optimizations is not universal and can be reversed depending on the target parameters and loss model. %Furthermore, we investigate low-order photon-number moments correction. 
Similar to vacuum-overlap correction, the photon-number mean correction also has ambiguities in the multimode setting, which we aim to resolve by first fixing the interferometer parameters to their target values and additionally imposing one of three possible conditions. Either matching mode-wise mean photon numbers, combining photon-number mean correction with vacuum-overlap correction, or adjust the squeezings in such a way that the mode-resolved mean photon numbers increase by a common factor, such that the total mean photon number matches the target one.
In many cases the first two approaches have no consistent solution and the third approach yields total variation distances that do not improve upon the uncorrected state, except in the regime of large squeezing parameters and strong loss.

In summary, these observations suggest that key structural features underpinning vacuum-overlap correction in the single-mode analysis carry over to multimode systems.
In this sense, vacuum-overlap correction provides a practically useful and efficiently computable heuristic for mitigating loss-induced distortions of photon-number distributions in the multimode regime.
Our numerical analysis indicates that this is not a peculiarity of the specific two-mode settings discussed here. 
We observe the same qualitative trend for three- and four-mode examples, namely that enforcing vacuum-overlap matching typically yields a noticeable reduction of $\delta$ compared to uncorrected lossy states. 
By contrast, leaving the input parameters unchanged often outperforms both fidelity-based and phase-space-based optimization in terms of the photon-number distribution. This behavior is consistent with our single-mode results, where fidelity and phase-space optimization often increase the distance for squeezed vacuum states when compared to the uncorrected lossy state.

However, the multimode setting introduces additional complexity and degrees of freedom, so that fixing the vacuum probability alone does not uniquely specify the optimizing state. Nevertheless, simple physical heuristics, such as maintaining the squeezing ratios of the target state or weighting them by potential asymmetries in the loss channels, provide robust strategies to select a specific state from the vacuum-matched manifold that consistently improves upon the uncorrected lossy state.

\subsection{Practical application: computing vibronic spectra}

\begin{figure*}[t]   
    \centering
    \includegraphics[width=1\textwidth]{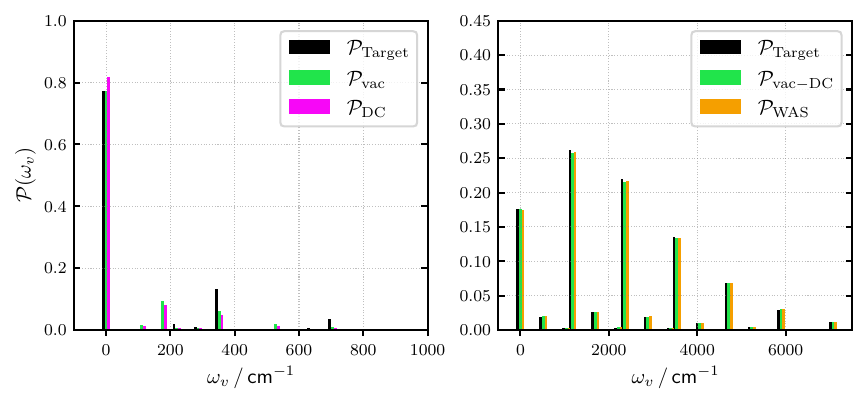}
    \caption{Vibronic spectra for tropolone (left) and SO$_2$ (right) obtained from two-mode Gaussian boson sampling (GBS) simulations with uniform loss ($1-\eta=0.3$ for all 4 positions in the system (cf. Fig.~\ref{fig:DoktorovSO2MitLoss})). Black bars show the target (lossless) distribution $\mathcal{P}_\mathrm{Target}(\omega_v)$. Green bars show vacuum-overlap correction with squeezing ratios fixed to the target ratios and for SO$_2$ we additionally apply displacement correction to fix the input displacements. Magenta bars (tropolone) show the lossy target distribution, being equivalent to displacement correction for this nondisplaced example. Orange bars (SO$_2$) show Wasserstein metric optimization. The magenta and orange references are included because they provide the best performance among the methods considered aside from vacuum-overlap correction for the respective examples. The horizontal axis shows the vibronic transition energy $\omega_v$ (cm$^{-1}$) and the vertical axis the corresponding probability $\mathcal{P}(\omega_v)$.} 
    \label{fig:SO2Tropolone}
\end{figure*}

To demonstrate the practical relevance of our mitigation schemes, we now apply them to the computation of vibronic spectra using Gaussian boson sampling~\cite{huhBosonSamplingMolecular2015}.
Concretely, we compute the FCFs of the $370$~nm transition of tropolone and the transition $\text{SO}_{2}^-\rightarrow \text{SO}_{2}$ of sulfur dioxide, where we restrict our analysis to two-mode subsystems that approximately decouple from the remaining vibronic modes.
In addition, we consider the $1\,^1A' \to 1\,^2A'$ transition for a symmetry block $a'$ of formic acid, which involves seven vibronic modes~\cite{huhBosonSamplingMolecular2015}.
We compute those spectra by simulating the outcomes of the associated GBS systems.
Notably, the sulfur-dioxide and formic-acid transitions include a shift of normal coordinates and therefore require displaced Gaussian inputs, whereas the selected two-mode tropolone subsystem exhibits no shift in the normal coordinates and can be treated without displacement~\cite{huhBosonSamplingMolecular2015,clementsApproximatingVibronicSpectroscopy2018,huhVibronicBosonSampling2017}.
Thus, these examples cover both types of targets, including cases with only squeezing as well as cases with dominant displacement. 
In the following, we investigate how the mitigation methods perform for these specific GBS applications under realistic loss conditions relevant to experimental implementations. 

Before investigating the usefulness of vacuum-overlap correction for the more involved formic-acid example, we first compare different mitigation schemes for both two-mode examples based on the results of the previous subsection~\ref{sec:TwoModeVacuum}. The results are shown in Fig.~\ref{fig:SO2Tropolone}, where we plot the vibronic spectra of the target state and of the mitigated states obtained via vacuum-overlap correction and via the best-performing alternative method.

For tropolone ($\tilde{\xi}_1=0.19$, $\tilde{\xi}_2=0.72$, $\tilde{\theta}=0.27$, $\tilde{\gamma}=0$) under uniform loss $\eta=0.7$ at all four positions (see Fig.~\ref{fig:DoktorovSO2MitLoss}), we obtain a minimal total variation distance of $\delta_{\rmmin,\xi} \approx 0.137$, when optimizing only over the input squeezings $(\xi_1,\xi_2)$ and keeping the interferometer parameters fixed at their target values, for simplicity.
The uncorrected state, which coincides with displacement correction, yields $\delta_\mathrm{DC} = \delta_\mathrm{uncor}\approx0.158$.
Both variants of vacuum-overlap correction, which are constrained to match the target squeezing ratio or the loss-weighted squeezing ratio, yield $\delta_\mathrm{vac}\approx0.139$ due to the symmetric loss profile.
Hence both methods yield an improvement over the uncorrected state, close to the best achievable value under the constraint of fixed interferometer parameters. 
By contrast, fidelity optimization and phase-space optimization increase the total variation distance relative to the uncorrected state, being consistent with the single-mode results. Fidelity optimization results in $\delta_\mathrm{F}\approx0.199$ and phase-space optimization based on the Wasserstein metric yields the same $\delta_\mathrm{WAS}\approx0.199$. Thus, neither of these optimizations provide a useful mitigation strategy in this setting.

\begin{figure*}[t]   
    \centering
    \includegraphics[width=1\textwidth]{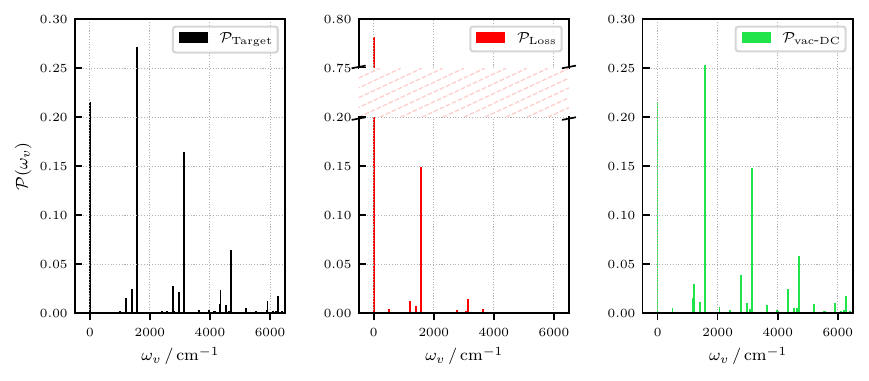}
    \caption{Vibronic spectra for formic acid obtained from GBS simulations. We implement a loss model that is based on realistic assumptions for near-term photonic devices, see the main text for details. In the left panel, we show the target spectrum (black bars), in the middle panel we show the spectrum obtained from the lossy GBS simulation without any mitigation (red bars), and in the right panel we show the spectrum obtained from the lossy GBS simulation with vacuum overlap correction (green bars). The vertical axis of the middle panel includes a break to accommodate the large vacuum peak, $\mathcal{P}_\mathrm{Loss}(0)\approx$ 0.782. The horizontal axis shows the vibronic transition energy $\omega_v$ (cm$^{-1}$) and the vertical axis the corresponding probability $\mathcal{P}(\omega_v)$.
	} 
    \label{fig:FormicAcid}
\end{figure*}

For the transitions of sulfur dioxide ($\tilde{\xi}_1=-0.11$, $\tilde{\xi}_2=-0.05$, $\tilde{\alpha}_1=-0.93$, $\tilde{\alpha}_2=1.01$, $\tilde{\theta}=0.59$, $\tilde{\gamma}=0$), all mitigation strategies significantly improve the FCFs for the same loss model.
Varying only the input squeezing and displacement parameters, the minimal total variation distance is $\delta_{\rmmin,\xi,\alpha}\approx 0.0039$.
The uncorrected state yields $\delta_\mathrm{uncor}\approx 0.305$, while displacement correction already achieves a significant improvement  by an order of magnitude with $\delta_\mathrm{DC}\approx0.027$.
Applying vacuum-overlap correction in addition to displacement correction further reduces the total variation distance to $\delta_{\mathrm{vac}}\approx 0.0076$, both when fixing the squeezing ratios and when using loss-weighted squeezing ratios.
Among the other mitigation approaches, phase-space optimization based on the Wasserstein metric performs best, reaching $\delta_\mathrm{WAS}\approx0.0058$, while the fidelity optimization yields $\delta_\mathrm{F}\approx0.037$. 
Overall, all methods perform well in this instance because the squeezing parameters are small and the target state is largely displacement-driven, so loss acts predominantly as an attenuation that can be compensated efficiently. This also provides a plausible explanation for why the Wasserstein-metric-based phase-space optimization performs particularly well here, as the dominant mismatch is essentially geometric in phase space.

Both examples demonstrate that the vacuum-overlap correction, in particular with the additional constraint of also matching the target squeezing ratios or loss weighted squeezing ratios, can significantly reduce the total variation distance to the target spectrum, even under pronounced loss in these two-mode settings (overall system transmissivity $0.49$).

To demonstrate the applicability of our methods beyond the two-mode setting, we also simulate the vibronic spectrum of formic acid, shown in Fig.~\ref{fig:FormicAcid}, which involves seven vibrational modes, with squeezing and displacement parameters
\begin{center}
	\begin{tabular}{c | c | c}
		Mode $i$ & $\tilde{\xi}_i$ & $\tilde{\alpha}_i$ \\
		\hline
		1 & -0.0972 &  0.6176 \\
		2 & -0.0701 &  0.4362 \\
		3 & -0.0208 & -0.4334 \\
		4 &  0.0597 & -0.5482 \\
		5 &  0.0749 & -0.1207 \\
		6 &  0.1120 &  0.6419 \\
		7 &  0.1867 & -0.0611. \\
	\end{tabular}
\end{center}
The unitary transformation characterizing the interferometer can be found in Ref.~\cite{huhBosonSamplingMolecular2015}.
We implement a loss model motivated by typical assumptions for near-term photonic devices, where the pre-interferometer transmissivities $\eta^{(1)}_{\mathrm{pre},i}$ are randomly drawn from $[0.5,0.6]$, interferometer transmissivities $\eta^{(l)}_{\mathrm{in},i}$ for each internal layer $\ell$ are randomly drawn from $[0.8,0.85]$, and post-interferometer transmissivities $\eta^{(n)}_{\mathrm{post},i}$ are randomly drawn from $[0.7,0.8]$,~\cite{stefszkyBenchmarkingGaussianNonGaussian2025,sauerResolvingPhotonNumbers2023,schapelerElectricalTraceAnalysis2024}. 
We use the decomposition method from Ref.~\cite{clementsOptimalDesignUniversal2016} to split the interferometer into layers of beamsplitters and phase shifters, and assume that after every beamsplitter we have the losses $1-\eta^{(l)}_{\mathrm{in},i}$.
Since pure loss is phase covariant, losses associated with phase shifters can be shifted to adjacent beam-splitter locations.
Thus, it suffices to place loss channels after each beam splitter. 
The interferometer decomposition and the randomly chosen loss is asymmetric, resulting in different loss levels for each mode.
The fraction of input photons injected into mode $j$ that are detected at the output ranges from approximately $0.12$ to $0.29$, with an average system transmissivity of $0.17$.

The results of our simulations are shown in Fig.~\ref{fig:FormicAcid}.
The uncorrected loss-affected FCFs (red bars, middle plot) show a significant deviation from the target FCFs (black bars, left plot), with a large total variation distance of $\delta_\mathrm{uncor}\approx0.573$.
Using the vacuum-overlap correction method with the additional constraint that all squeezer ratios remain unchanged, as discussed in Sec.~\ref{sec:TwoModeVacuum}, and additional displacement correction, we obtain the mitigated spectrum shown in the right panel (green bars), which has a significantly improved total variation distance of $\delta_\mathrm{vac}\approx0.090$. 
We would like to highlight that not only the main peaks are well reproduced, but also finer features such as the structures around $\omega_{\nu} = 3000$ cm$^{-1}$, $\omega_{\nu} = 4500$ cm$^{-1}$, and $\omega_{\nu} = 6000$ cm$^{-1}$.

This example indicates that our proposed mitigation strategies, which rely on correcting efficiently computable Gaussian properties, remain effective for larger, physically relevant multimode scenarios.
Even with complex, nonuniform loss profiles, this approach helps to recovering characteristic features of the vibronic spectrum in experimentally realistic settings.

\section{Conclusions}
\label{sec:Conclusions}
In this work, we investigated various strategies to mitigate the impact of loss on the photon-number distribution of pure Gaussian states. 
Photon loss is a central limitation for GBS as it distorts the output photon-number distribution that encode both computational hardness and application-specific observables. We addressed this problem by redefining the Gaussian input parameters such that the loss-affected photon-number distribution approximates as closely as possible the desired lossless target distribution in terms of the total variation distance. Crucially, all mitigation prescriptions that we study are efficiently computable, as they rely only on the Gaussian description of the involved states and do not require explicit evaluation of the full photon-number distribution, which becomes intractable for large systems.

For single-mode squeezed vacuum, we provided a systematic comparison between fidelity optimization, phase-space distribution optimizations, low-order photon-number-moment corrections, and vacuum-overlap correction. We established explicit performance hierarchies and showed that, among all efficient prescriptions considered, vacuum-overlap correction is optimal over a broad parameter range. In particular, matching the vacuum probability yields an analytic estimator $\xi_{\mathrm{vac}}$ that coincides with the true minimizer $\xi_{\rmmin}$ in a large region of the $(\eta,\xi_{\mathrm{vac}})$-plane. More precisely, this is the case for $\xi_{\mathrm{vac}} < 2.290047\ldots$ (independent of $\eta$) and also for $\xi_{\mathrm{vac}} \geq 2.290047\ldots$ provided $\eta<14/15$. Outside this region the minimizer shifts to $\xi_{\min}<\xi_{\mathrm{vac}}$. Nevertheless, the resulting increase in $\delta$ when using $\xi_{\mathrm{vac}}$ instead of $\xi_{\rmmin}$ is small, so vacuum-overlap correction remains a robust near-optimal prescription even in this narrow parameter region. Besides providing the minimizing parameter, the analysis also allowed us to predict (and lower bound) the minimal achievable error due to loss using only Gaussian information, without computing the full photon-number distribution.

A notable qualitative finding is that fidelity maximization is generally not a good proxy for minimizing $\delta$ for squeezed-vacuum targets. Over wide parameter ranges it performs worse than leaving the input parameters of the Gaussian target state unchanged. This highlights an important conceptual point. State overlap in Hilbert space given by the fidelity can be dominated by features that are not reflected in photon-counting statistics. By contrast, even the seemingly naive baseline of ``do nothing'' can outperform fidelity- and phase-space-based prescriptions for squeezed-vacuum distributions in terms of $\delta$. Albeit we proved that photon-number mean or variance correction perform systematically better than phase-space distribution optimization, also low-order photon-number moments are typically outperformed by the uncorrected baseline, except for large loss and large squeezing configurations.

Extending the analysis to single-mode displaced squeezed vacuum, the absence of a strict parity structure makes general analytic statements about $\xi_{\rmmin}$ and $|\alpha_{\rmmin}|$ substantially more difficult. Nevertheless, our numerical study reveals that the vacuum-overlap mechanism remains structurally relevant. Across broad parameter ranges, the $\delta$-minimizing lossy state reproduces the target vacuum probability to high accuracy, and the $\delta$ landscape additionally exhibits valley-like structures associated with approximate matching of various photon-number probabilities. Because vacuum matching becomes underdetermined once additional state parameters are present, we proposed practical disambiguation strategies that remain efficiently computable, most notably combining vacuum-overlap correction with displacement correction (fixing $\alpha$ by compensating the loss-induced attenuation) or with mean-photon-number matching.

For multimode squeezed-vacuum systems, we found that the same key structural feature persists as for single-mode systems. Fully optimized states typically match the global vacuum probability, while mode-resolved vacuum probabilities and mode-resolved mean photon numbers are generally not reproduced. This indicates that optimizing $\delta$ primarily aligns global photon-number statistics while leaving some freedom in the local structure of the state. Consequently, global vacuum-overlap correction remains a powerful and efficiently computable heuristic in the multimode regime, but it introduces intrinsic ambiguities because the global vacuum constraint defines a manifold of admissible parameters. We showed that a particularly simple and scalable way to resolve this underdetermination for nondisplaced multimode squeezed vacuum is to rescale all squeezing parameters by a common factor until the global vacuum probability matches, thereby preserving the target squeezing ratios, and also discussed loss-weighted variants. For displaced multimode instances, combining vacuum matching with displacement correction provides an equally natural and efficient way to fix the displacement parameters.

We demonstrated the practical relevance of these strategies for concrete GBS applications by computing vibronic spectra (Franck-Condon factors) in settings with both purely squeezed and displaced-squeezed Gaussian targets. 
For states with a comparatively large displacement component many methods performed well, which is consistent with the fact that such spectra can often be accurately approximated using classical coherent-state models~\cite{eickmannFullPowerGaussian2025}. For such essentially classical states, loss merely acts as an attenuation that can be efficiently compensated by rescaling the displacement. 
Furthermore, we demonstrated that combining vacuum-overlap correction (with fixed squeezing ratios) and displacement correction reduced the total variation distance from $\delta \approx 0.573$ (uncorrected) to $\delta\approx0.090$ and preserved not only dominant peaks but also finer spectral features in a seven-mode formic-acid example with a nonuniform loss model. This indicates that the proposed prescriptions remain effective in higher-dimensional, experimentally motivated scenarios.

Beyond photon loss, near-term GBS experiments face additional imperfections such as phase instabilities of the input states, mode mismatch, and interferometer calibration errors. In Appendix~\ref{sec:PhaseDrifts}, we explicitly study input phase noise by randomizing the squeezing phase in one mode for the calculation of FCFs of tropolone and find that vacuum-overlap-based correction continues to reduce $\delta$ compared to the unmitigated baseline even for large phase noise. In this example, loss remains the dominant contribution to deviations in the photon-number distribution, while phase noise mainly becomes visible through an increased spread of $\delta$ once the loss-induced distortion is reduced. This picture is consistent with recent results showing that relative squeezing-phase noise does not, by itself, lead to an efficient classical simulation of GBS in the regimes studied~\cite{paryzkovaInputPhaseNoise2025}. Moreover, stochastic interferometer errors based on unitary averaging over multiple noisy realizations have been investigated for boson-sampling experiments in the literature, yielding rigorous bounds on the distance to the target distribution~\cite{singhCoherentlyMitigatingBoson2025}.

Finally, our multimode results further motivate a systematic search for efficiently computable constraints that select near-$\delta$-optimal points on the vacuum-matched manifold, for instance by combining global vacuum overlap with other correction or optimization schemes and by integrating these prescriptions into device-level calibration and stabilization loops for near-term photonic platforms. Overall, our results provide a concrete foundation for optimizing Gaussian state preparation under loss and thereby improving the reliability of GBS-based quantum simulations in realistic, lossy environments.

\section*{Acknowledgement}
We thank Philip Heinzel and Emil Donkersloot for valuable discussions and are grateful to Fabian Steinlechner for discussions and comments on the manuscript. We acknowledge funding from the German Federal Ministry of Research, Technology and Space (BMFTR) within the PhoQuant project (Grant No. 13N16103). HE is a member of the Max Planck School of Photonics, supported by the Dieter Schwarz Foundation, the BMFTR, and the Max Planck Society.

\appendix

\section{Relations of the phase-space optimizing squeezing parameters}
\label{sec:RelationsPSO}
In this appendix, we sketch the proof of the relation $\xi_\mathrm{WAS} < \xi_\mathrm{KLDup} < \xi_\mathrm{BHA} < \xi_\mathrm{KLDsym} < \xi_\mathrm{KLDpu}$ stated in the main text, Sec.~\ref{sec:SMSV-PSO}. Calculating the first derivatives of the phase-space measures with respect to $\xi$ and manipulating the resulting equations for which these derivatives vanish by multiplying the different equations $0=\partial_\xi D_i$ ($i \in \{\mathrm{WAS}, \mathrm{BHA}, \mathrm{KLDup}, \mathrm{KLDpu}, \mathrm{KLDsym}\}$) with respective strictly positive functions of $\xi$, we obtain the following functions defining the root finding problems 
\begin{widetext}
\begin{align}
 f_\mathrm{KLDpu}(\xi) &= \sinh{(2\xi-2\zt)} - \frac{(1-\eta)\sinh(2\xi)}{\eta^2 + (1-\eta)^2 + 2\eta(1-\eta)\cosh(2\xi)}, \\
 %%%%%
 f_\mathrm{KLDsym}(\xi) &= \sinh{(2\xi-2\zt)} %\notag \\ 
 - \frac{2\eta(1-\eta)\sinh(2\zt) + (1-\eta)^2 \sinh(2\xi+2\zt)}{\eta^2 + \big[\eta^2 + (1-\eta)^2 + 2\eta(1-\eta)\cosh(2\xi)\big]^2}, \\
 %%%%%
 f_\mathrm{KLDup}(\xi) &= \eta^2 \sinh{(2\xi-2\zt)} - (1-\eta)^2 \sinh{(2\xi+2\zt)}
  -2\eta(1-\eta) \sinh(2\zt) + \eta(1-\eta)^2 \sinh(4z)  \notag \\
 &\quad + (1-\eta)\big[\eta^2 + (1-\eta)^2\big] \sinh(2\xi), \\
 %%%%%
 f_\mathrm{BHA}(\xi) &= \eta \sinh(2\xi-2\zt) - (1-\eta) \sinh(2\zt)
 + \frac{2\eta(1-\eta)^2 \sinh^2(\xi) \sinh(2\xi)}{1+ 2(1-\eta) \sinh^2(\xi)}, \\
 %%%%%
 f_\mathrm{WAS}(\xi) &= 2 \sinh(2\xi) + \frac{\E^{-(2\xi+\zt)}}{\sqrt{1-\eta + \eta \E^{-2\xi}}} - \frac{\E^{(2\xi+\zt)}}{\sqrt{1-\eta + \eta \E^{2\xi}}}.
 \label{eq:fWAS}
\end{align}
\end{widetext}
First, we notice that all functions are continuous and fulfill $f_i(0)<0$ and $\lim_{\xi\to\infty} f_i(\xi) \to \infty$. Thus, they exhibit at least one root according to Bolzano's theorem. Moreover, it is possible to show that each function has actually only one positive root and that this root is larger than $\zt$.

For $f_\mathrm{KLDpu}(\xi)$, a positive root has to be larger than $\zt$ as $f_\mathrm{KLDpu}<0$ for every $\xi\leq \zt$. In case $\xi > \zt$, we can reformulate the root finding problem to $\frac{\sinh{(2\xi-2\zt)}}{\sinh(2\xi)} = \frac{(1-\eta)}{\eta^2 + (1-\eta)^2 + 2\eta(1-\eta)\cosh(2\xi)}$. Note that the left-hand side is a monotonically increasing function for $\xi>\zt$ while the right-hand side is monotonically decreasing. Further, we have that the LHS vanishes while the RHS is positive for $\xi\to \zt+$ and the LHS approaches one while the RHS vanishes for $\xi\to \infty$. Thus, there is only one crossing of the LHS and RHS according to the intermediate value theorem, implying that there is only a single positive root  $f_\mathrm{KLDpu}(\xi_\mathrm{KLDpu}) = 0$ fulfilling $\xi_\mathrm{KLDpu} > \zt$. Precisely the same chain of arguments can be used to prove that $f_\mathrm{KLDsym}$ has only a single positive root $\xi_\mathrm{KLDsym}>\zt$.

To show that the function $f_\mathrm{KLDup}$ has only a single positive root, we can equivalently investigate the zero of the function $\frac{f_\mathrm{KLDup}}{\sinh(2\xi)}$ as $\sinh(2\xi)$ is positive for $\xi>0$. Determining $\frac{\mathrm{d}}{\mathrm{d}\xi} \frac{f_\mathrm{KLDup}}{\sinh(2\xi)} = 4\eta(1-\eta)^2\sinh(2\xi) + 2\frac{\sinh(2\zt)}{\sinh(2\xi)} + 2\eta(1-\eta)\frac{\sinh(2\zt)}{\cosh^2(\xi)} > 0$, $\frac{f_\mathrm{KLDup}}{\sinh(2\xi)}$ is monotonically increasing for $\xi>0$. Further, as this function is negative for sufficiently small $\xi$, in particular for $\xi=\zt$, and positive in the limit $\xi\to\infty$. Thus, the function $\frac{f_\mathrm{KLDup}}{\sinh(2\xi)}$ and, therefore, also $f_\mathrm{KLDup}$ has only one root $\xi_\mathrm{KLDup}>\zt$. 
Similarly, the function $f_\mathrm{BHA}$ is strictly monotonically increasing which can be inferred from the fact that $\frac{\mathrm{d}}{\mathrm{d}\xi} f_\mathrm{BHA}(\xi) > 0$ for $\xi > 0$. Thus, there is only one positive root $f_\mathrm{BHA}(\xi_\mathrm{BHA}) = 0$.

Finally, $\frac{f_\mathrm{WAS}(\xi)}{2\sinh(2\xi)}$ is monotonically increasing which can be proven by determining the derivative with respect to $\E^{2\xi}$ and using cylindrical algebraic decomposition with respect to this variable. As $\frac{f_\mathrm{WAS}(\zt)}{2\sinh(2\zt)} < 0$ and $\lim_{\xi\to\infty}\frac{f_\mathrm{WAS}(\xi)}{2\sinh(2\xi)} = 1$, the function $\frac{f_\mathrm{WAS}(\xi)}{2\sinh(2\xi)}$ has a unique positive zero that is larger than $\zt$. Thus, $f_\mathrm{WAS}(\xi)$ has only one positive root $\xi_\mathrm{WAS} > \zt$.

In the following, we will show the root ordering listed in the main text, see~Eq.~\eqref{eq:IneqPhaseSpace}.

\subparagraph{$\boldsymbol{\xi_\mathrm{KLDsym} < \xi_\mathrm{KLDpu}:}$}\quad \\
First, we define ${D(\xi) = \eta^2 + (1-\eta)^2 + 2\eta(1-\eta)\cosh(2\xi)}$ and consider the root $\xi_\mathrm{KLDpu}$ of $f_\mathrm{KLDpu}$, implying $D(\xi_\mathrm{KLDpu}) = \frac{(1-\eta)\sinh(2\xi_\mathrm{KLDpu})}{\sinh{(2\xi_\mathrm{KLDpu}-2\zt)}}$. Evaluating $f_\mathrm{KLDsym}$ at the root $\xi_\mathrm{KLDpu}$ and regrouping terms, one obtains 
\begin{align}
&f_\mathrm{KLDsym}(\xi_\mathrm{KLDpu}) \notag \\
&\quad = \frac{\big( \eta \sinh(2\xi_\mathrm{KLDpu} - 2\zt) - (1-\eta)\sinh(2\zt) \big)^2}{\sinh(2\xi_\mathrm{KLDpu} - 2\zt) \big( \eta^2 + D(\xi_\mathrm{KLDpu})^2 \big)}
\label{eq:fKLDsymATzKLDpu}
\end{align}
where we used the above relation for $D(\xi_\mathrm{KLDpu})$ to manipulate the numerator in Eq.~\eqref{eq:fKLDsymATzKLDpu}. The denominator of Eq.~\eqref{eq:fKLDsymATzKLDpu} is strictly positive. The numerator is also positive, except for the case $\eta \sinh(2\xi_\mathrm{KLDpu} - 2\zt) = (1-\eta)\sinh(2\zt)$. However, this case is incompatible with an actual zero of the function $f_\mathrm{KLDpu}$. Using $\sinh(x) = \sinh(y) \cosh(x-y) + \cosh(y) \sinh(x-y)$, we have $D(\xi_\mathrm{KLDpu}) = (1-\eta) \big[ \frac{\sinh(2\zt) \cosh(2\xi_\mathrm{KLDpu}-2\zt)}{\sinh(2\xi_\mathrm{KLDpu}-2\zt)} +  \cosh(2\zt) \big] = \eta \cosh(2\xi_\mathrm{KLDpu}-2\zt) + (1-\eta)\cosh(2\zt)$. 
Further, $D(\xi_\mathrm{KLDpu}) = \eta^2 + (1-\eta)^2 + 2(\eta-\eta^2) \cosh(2\zt) \cosh(2\xi_\mathrm{KLDpu}-2\zt) + 2(1-\eta)^2 \sinh^2(2\zt)$ where we used $\cosh(x) = \cosh(y) \cosh(x-y) + \sinh(y) \sinh(x-y)$. Equating both relations for $D(\xi_\mathrm{KLDpu})$ and rearranging terms, we have 
\begin{align}
2(1-\eta)\!-\!1 &= [(1-\eta)\cosh(2\zt)+\eta\cosh(2\xi_\mathrm{KLDpu}-2\zt)] \notag \\
&\quad \times[2(1-\eta)\cosh(2\zt)-1].
\label{eq:contradiction}
\end{align}
But this statement is a contradiction as the first factor in square brackets on the right-hand side is always $>1$ (for $\zt>0$) and the second term is always larger than $2(1-\eta)-1$. 
Thus, we obtain $f_\mathrm{KLDsym}(\xi_\mathrm{KLDpu}) > 0$. As $f_\mathrm{KLDsym}$ has only a single root and we know that $f_\mathrm{KLDsym}(\xi)>0$ for $\xi > \xi_\mathrm{KLDsym}$ and $f_\mathrm{KLDsym}(\xi) < 0$ for $\xi < \xi_\mathrm{KLDsym}$ from the previous analysis, we conclude the following root ordering: $\xi_\mathrm{KLDsym} < \xi_\mathrm{KLDpu}$.

\subparagraph{$\boldsymbol{\xi_\mathrm{BHA} < \xi_\mathrm{KLDsym}:}$}\quad \\
Next, we show that $\xi_\mathrm{BHA} < \xi_\mathrm{KLDsym}$ in the same manner by showing that $f_\mathrm{BHA}(\xi_\mathrm{KLDsym}) > 0$. 
From ${f_\mathrm{KLDsym}(\xi_\mathrm{KLDsym}) = 0}$, we obtain
$\sinh{(2\xi_\mathrm{KLDsym}-2\zt)} = \frac{2\eta(1-\eta)\sinh(2\zt) + (1-\eta)^2 \sinh(2\xi_\mathrm{KLDsym}+2\zt)}{\eta^2 + D(\xi_\mathrm{KLDsym})^2}$. 
A second useful relation can be obtained by using $\sinh(x \pm y) = \sinh(x)\cosh(y) \pm \cosh(x)\sinh(y)$ for the left- and right-hand side of this relation and regrouping terms leading to 
$\sinh(2\xi_\mathrm{KLDsym})\cosh{2\zt} = \frac{[ \eta^2 + (1-\eta)^2 + D(\xi_\mathrm{KLDsym})^2 ]\cosh(2\xi_\mathrm{KLDsym}) + 2(\eta-\eta^2)}{\eta^2 - (1-\eta)^2 + D(\xi_\mathrm{KLDsym})^2} \sinh(2\zt)$. 
Inserting the relation for $\sinh{(2\xi_\mathrm{KLDsym}-2\zt)}$ into $f_\mathrm{BHA}(\xi_\mathrm{KLDsym})$ and using the second relation to eliminate the remaining $\sinh(2\xi_\mathrm{KLDsym})\cosh{2\zt}$ term, leads to 
\begin{widetext}
\begin{align}
 f_\mathrm{BHA}(\xi_\mathrm{KLDsym}) &= \frac{\eta(1-\eta)^2(\cosh(2\xi_\mathrm{KLDsym})-1)}{[\eta^2-(1-\eta)^2 + D(\xi_\mathrm{KLDsym})^2][\eta + (1-\eta)\cosh(\xi_\mathrm{KLDsym})]}  \notag \\
 &\quad \times \Big( [\eta^2-(1-\eta)^2 + D(\xi_\mathrm{KLDsym})^2] \sinh(2\xi_\mathrm{KLDsym}) \notag \\
 &\qquad\quad - 2\sinh(2\zt)D(\xi_\mathrm{KLDsym})[\eta + (1-\eta)\cosh(\xi_\mathrm{KLDsym})] \Big)
\end{align}
\end{widetext}
Note that the factor separated in the first line is strictly positive. Thus, the sign of $f_\mathrm{BHA}(\xi_\mathrm{KLDsym})$ corresponds to the sign of the term in parentheses in the second and third line. The term in the second line can be replaced by $\frac{F(\xi_\mathrm{KLDsym})}{\cosh(2\zt)} \sinh(2\zt)$ with $F(\xi) {= [ \eta^2 + (1-\eta)^2 + D(\xi)^2 ]\cosh(2\xi)} + 2(\eta-\eta^2)$ via our second identity discussed above. Factoring $\sinh(2\zt) > 0$, we have to show $\frac{F(\xi_\mathrm{KLDsym})}{\cosh(2\zt)} > {2 D(\xi_\mathrm{KLDsym})[\eta + (1-\eta)} \cosh(\xi_\mathrm{KLDsym})]$. As both sides of this inequatlity are positive, we can square this expression. Then, we can remove the $\cosh^2(2\zt)$ dependency via $\frac{1}{\cosh^2(2\zt)} = 1-\tanh^2(2\zt) = \frac{F(\xi_\mathrm{KLDsym})^2 - (\eta^2 - (1-\eta)^2 + D(\xi_\mathrm{KLDsym})^2)^2\sinh^2(2\xi_\mathrm{KLDsym})}{F(\xi_\mathrm{KLDsym})^2}$ using again our second identity derived from $f_\mathrm{KLDsym}(\xi_\mathrm{KLDsym})=0$. Thus, we have to show $F^2 - (\eta^2 - (1-\eta)^2 + D^2)^2\sinh^2(2\xi_\mathrm{KLDsym}) > 4D^2[\eta + (1-\eta)\cosh(\xi_\mathrm{KLDsym})]^2$ where the functions $F$ and $D$ are evaluated at $\xi_\mathrm{KLDsym}$. As $F^2 - (\eta^2 - (1-\eta)^2 + D^2)^2\sinh^2(2\xi_\mathrm{KLDsym}) - 4D^2[\eta + (1-\eta)\cosh(\xi_\mathrm{KLDsym})]^2 = \big(D - D^2\big)^2 > 0$ (because $D>1$), we eventually have shown that $f_\mathrm{BHA}(\xi_\mathrm{KLDsym}) > 0$. Further, we know that $f_\mathrm{BHA}$ is a continuous function with a unique positive root and $f_\mathrm{BHA} > 0$ for $\xi>\xi_\mathrm{BHA}$ such that $f_\mathrm{BHA}(\xi_\mathrm{KLDsym}) > 0$ implies $\xi_\mathrm{BHA} < \xi_\mathrm{KLDsym}$.

\subparagraph{$\boldsymbol{\xi_\mathrm{KLDup} < \xi_\mathrm{BHA}:}$}\quad \\
While we could try to continue with the same strategy of the previous two proofs, writing $f_\mathrm{KLDup}(\xi_\mathrm{BHA})$ as a total square, we follow a different strategy by using quantifier elimination via cylindrical algebraic decomposition. A similar strategy can also be used for the previous two proofs. To show that $f_\mathrm{KLDup}(\xi_\mathrm{BHA}) > 0$, we introduce the notation $s_\mathrm{BHA} = \sinh(2\xi_\mathrm{BHA}) > 0$, $\tilde{s} = \sinh(2\zt) >0$, $c_\mathrm{BHA} = \cosh(2\xi_\mathrm{BHA}) >1$, $\tilde{c} = \cosh(2\zt) >1$, allowing us to write 
\begin{align}
  f_\mathrm{KLDup}(\xi_\mathrm{BHA}) &= 
  \eta^2 (s_\mathrm{BHA}\tilde{c}-c_\mathrm{BHA}\tilde{s}) \notag \\
  &\quad - (1-\eta)^2 (s_\mathrm{BHA}\tilde{c}-c_\mathrm{BHA}\tilde{s})  \notag \\
  &\quad-2\eta(1-\eta) \tilde{s} + 2\eta(1-\eta)^2 s_\mathrm{BHA} c_\mathrm{BHA}  \notag \\
 &\quad + (1-\eta)\big[\eta^2 + (1-\eta)^2\big] s_\mathrm{BHA}. 
\end{align}
With the additional constraints $s_\mathrm{BHA}\tilde{c}-c_\mathrm{BHA}\tilde{s} >0$, $c_\mathrm{BHA}^2-s_\mathrm{BHA}^2=1$, $\tilde{c}^2-\tilde{s}^2=1$, and 
\begin{align}
0 &= [\eta+(1-\eta)c_\mathrm{BHA}]f_\mathrm{BHA}(\xi_\mathrm{BHA}) \notag \\
&= [\eta+(1-\eta)c_\mathrm{BHA}] [\eta(s_\mathrm{BHA}\tilde{c}-c_\mathrm{BHA}\tilde{s}) - (1-\eta\tilde{s})] \notag\\
&\quad + \eta(1-\eta)^2(c_\mathrm{BHA}-1)s_\mathrm{BHA},
\end{align}
we have encoded the problem as polynomial constraints in the variables $s_\mathrm{BHA}$, $\tilde{s}$, $c_\mathrm{BHA}$, $\tilde{c}$, and $\eta$, which can be analyzed by cylindrical algebraic decomposition. Using a computer algebra system to perform the cylindrical algebraic decomposition, we obtain $f_\mathrm{KLDup}(\xi_\mathrm{BHA}) > 0$. Once more, as we know that $f_\mathrm{KLDup} > 0$ for $\xi > \xi_\mathrm{KLDup}$, we have shown that $\xi_\mathrm{KLDup} < \xi_\mathrm{BHA}$.

\subparagraph{$\boldsymbol{\xi_\mathrm{WAS} < \xi_\mathrm{KLDup}:}$}\quad \\
Finally, we show the remaining inequality of the root ordering, $\xi_\mathrm{WAS} < \xi_\mathrm{KLDup}$, that can also be addressed via quantifier elimination. A convenient strategy is to reformulate the two functions of the root finding problems in terms of exponentials of the squeezing parameters instead of hyperbolic functions. The only additional complexity compared to the previous case comes from the fact that $f_\mathrm{WAS}(\xi)$ contains square-roots of terms evolving exponentials. We will introduce new positive variables encoding these structures and then eliminate all quantifiers. More precisely, we use the following notation, ${e = \E^{\xi_\mathrm{KLDup}}>1}$, $\tilde{e} = \E^{\zt} > 1$, $g_{\pm} = \sqrt{1-\eta + \eta e^{\pm 2} } {>0}$ and have the additional constraints $g_{\pm}^2 = 1-\eta + \eta e^{\pm 2}$, $e > \tilde{e}$, and $2e^4\tilde{e}^2 f_\mathrm{KLDup}(\xi_\mathrm{KLDup}) = 0$. For the latter relation, we have multiplied $f_\mathrm{KLDup}(\xi_\mathrm{KLDup})$ by $2e^4\tilde{e}^2$ to formulate the constraint in terms of a polynomial with respect to the variable $e$ and parameters $\tilde{e}$ and $0<\eta<1$. Writing now 
\begin{align}
 f_\mathrm{WAS}(\xi_\mathrm{KLDup}) = \frac{(e^4-1)\tilde{e}g_+g_- + g_+ - e^4\tilde{e}^2g_-}{e^2\tilde{e} g_+ g_-},
 \end{align}
 the sign of $f_\mathrm{WAS}(\xi_\mathrm{KLDup})$ corresponds to the sign of the numerator. Using cylindrical algebraic decomposition, one can show $(e^4-1)\tilde{e}g_+g_- + g_+ - e^4\tilde{e}^2g_- > 0$ under the above listed constraints. Thus, we obtain $f_\mathrm{WAS}(\xi_\mathrm{KLDup}) > 0$. As in the previous cases, we already have shown that $f_\mathrm{WAS}$ is only positive for $\xi > \xi_\mathrm{WAS}$ implying $\xi_\mathrm{WAS} < \xi_\mathrm{KLDup}$.

Therefore, we have proven the chain of inequalities for the parameters optimizing the different phase-space measures in the main text, see Eq.~\eqref{eq:IneqPhaseSpace}.

\section{Relation between $\xi_{\mathrm{WAS}}$ and $\xi_{\Delta_{\mathcal{N}}}$}
\label{sec:RelationWAS-Var}
In this appendix, we prove the inequality $\xi_{\Delta_{\mathcal{N}}} < \xi_{\mathrm{WAS}}$ stated in the main text.
As shown in App.~\ref{sec:RelationsPSO}, the function $f_{\mathrm{WAS}}(\xi)$ given by Eq.~\eqref{eq:fWAS} has a unique positive root $\xi_{\mathrm{WAS}}$, and $f_{\mathrm{WAS}}(\xi) < 0$ holds for $\xi < \xi_{\mathrm{WAS}}$.
Thus, it suffices to show that $f_{\mathrm{WAS}}(\xi_{\Delta_{\mathcal{N}}})<0$. 
Recall that $\xi_{\Delta_{\mathcal{N}}}>0$ is defined via the variance-matching condition (cf. Eq.~\eqref{eq:xiDeltaVar} in the main text)
\begin{align*}
 \sinh^2(\xi_{\Delta_{\mathcal{N}}})
 = \frac{1+\eta}{4\eta}
 \Bigg[
 \sqrt{1+ \frac{16\sinh^2(\zt)\cosh^2(\zt)}{(1+\eta)^2} } -1
 \Bigg],
\end{align*}
where $\zt>0$ and $0<\eta<1$. 
Regrouping terms, we can give an explicit expression for $\zt$,
\begin{align}
 \sinh^2(2\zt)
 &= 2\eta \sinh^2 \xi_{\Delta_{\mathcal{N}}} \big(1-\eta + 2\eta \sinh^2 \xi_{\Delta_{\mathcal{N}}}\big) \notag \\
 &= \eta \big[\cosh(2\xi_{\Delta_{\mathcal{N}}})-1\big] \big[1+\eta\cosh(2\xi_{\Delta_{\mathcal{N}}})\big], \notag \\
 &\equiv \eta (c_{\Delta_\mathcal{N}} - 1)(1 + \eta c_{\Delta_\mathcal{N}})
 \label{eq:sinhZT}
\end{align}
where we have used $\sinh^2(x) = [\cosh(2x) - 1]/2$ in the second line and defined $c_{\Delta_\mathcal{N}} = \cosh(2\xi_{\Delta_{\mathcal{N}}})$ in the third line. Further, we define $s_{\Delta_\mathcal{N}} = \sinh(2\xi_{\Delta_{\mathcal{N}}})$ for notational convenience.

The basic idea is to use Eq.~\eqref{eq:sinhZT} to bound the terms $\E^{\pm \zt} \gtrless \sqrt{1-\eta + \eta \E^{\pm 2\xi_{\Delta_{\mathcal{N}}}}} $ in $f_{\mathrm{WAS}}(\xi_{\Delta_{\mathcal{N}}})$. If these inequalities hold, we have 
\begin{align}
 f_{\mathrm{WAS}}(\xi_{\Delta_{\mathcal{N}}})
 &= 2\sinh(2\xi_{\Delta_{\mathcal{N}}})
 + \E^{-2\xi_{\Delta_{\mathcal{N}}}} \frac{\E^{-\zt}}{\sqrt{1-\eta+\eta\E^{-2\xi_{\Delta_{\mathcal{N}}}}}} \notag \\
 &\quad 
 - \E^{-2\xi_{\Delta_{\mathcal{N}}}} \frac{\E^{\zt}}{\sqrt{1-\eta+\eta\E^{2\xi_{\Delta_{\mathcal{N}}}}}} \notag\\
 &< \bigl(\E^{2\xi_{\Delta_{\mathcal{N}}}}-\E^{-2\xi_{\Delta_{\mathcal{N}}}}\bigr)
 + \E^{-2\xi_{\Delta_{\mathcal{N}}}} - \E^{2\xi_{\Delta_{\mathcal{N}}}}
 = 0,
\label{eq:fWASbound}
\end{align}
which implies $\xi_{\Delta_{\mathcal{N}}}<\xi_{\mathrm{WAS}}$ due to the uniqueness and sign structure of the root of $f_{\mathrm{WAS}}$.

To prove the stated inequalities, we use Eq.~\eqref{eq:sinhZT} to first show $ \cosh(2\zt) < 1-\eta + \eta c_{\Delta_\mathcal{N}}$, as 
\begin{align*}
 &(1-\eta + \eta c_{\Delta_\mathcal{N}} )^2 - \cosh^{2}(2\zt) \\
 &\quad = (1-\eta + \eta c_{\Delta_\mathcal{N}} )^2 - \big(1 + \sinh^{2}(2\zt) \big) \\
 &\quad = \eta (1-\eta) (c_{\Delta_\mathcal{N}} - 1) >0. 
\end{align*}
Similarly, $\sinh(2\zt) > \eta s_{\Delta_\mathcal{N}}$, as
\begin{align*}
 &\sinh^{2}(2\zt) - \eta^2 s_{\Delta_\mathcal{N}}^2  \\
 &\quad = \eta (c_{\Delta_\mathcal{N}} - 1)(1 + \eta c_{\Delta_\mathcal{N}}) - \eta^2 (c_{\Delta_\mathcal{N}}^2 -1) \\
 &\quad = \eta (c_{\Delta_\mathcal{N}} - 1) (1 - \eta) >0.
\end{align*}
Therefore, 
\begin{align}
 \E^{-2\zt} &= \cosh(2\zt) - \sinh(2\zt)  \notag \\
 &< 1-\eta + \eta c_{\Delta_\mathcal{N}} - \eta s_{\Delta_\mathcal{N}} 
 = 1-\eta + \eta \E^{-2\xi_{\Delta_{\mathcal{N}}}},
\label{eq:Bound1AppB}
\end{align}
allowing us to upper bound the second term on the right hand side of Eq.~\eqref{eq:fWASbound}. To lower bound the third term in Eq.~\eqref{eq:fWASbound}, we use that we have already shown that $\sinh^{2}(2\zt) - \eta^2 s_{\Delta_\mathcal{N}}^2 = (1-\eta + \eta c_{\Delta_\mathcal{N}} )^2 - \cosh^{2}(2\zt) > 0$. As $\cosh(2\zt) > \sinh^{2}(2\zt)$ and $1-\eta + \eta c_{\Delta_\mathcal{N}} > \eta s_{\Delta_\mathcal{N}}$, we obtain
$\sinh(2\zt) - \eta s_{\Delta_\mathcal{N}} 
= \frac{(1-\eta + \eta c_{\Delta_\mathcal{N}} )^2 - \cosh^{2}(2\zt)}{\sinh(2\zt) + \eta s_{\Delta_\mathcal{N}} } 
< \frac{(1-\eta + \eta c_{\Delta_\mathcal{N}} )^2 - \cosh^{2}(2\zt)}{\cosh(2\zt) + 1-\eta + \eta c_{\Delta_\mathcal{N}} } 
= 1-\eta + \eta c_{\Delta_\mathcal{N}} - \cosh(2\zt)$. 
Thus, 
\begin{align}
 \E^{2\zt} &= \sinh(2\zt) + \cosh(2\zt) \notag \\
 &> 1-\eta + \eta c_{\Delta_\mathcal{N}} + \eta s_{\Delta_\mathcal{N}} = 1-\eta + \eta\E^{2\zt}.
\label{eq:Bound2AppB}
\end{align}
Using the inequalities~\eqref{eq:Bound1AppB} and \eqref{eq:Bound2AppB}, we arrive at the third line of Eq.~\eqref{eq:fWASbound}. Therefore, $ f_{\mathrm{WAS}}(\xi_{\Delta_{\mathcal{N}}}) < 0$ and, thus, $\xi_{\Delta_{\mathcal{N}}} < \xi_{\mathrm{WAS}}$.

\section{Proof $\xi_\mathrm{vac}=\xi_\mathrm{min}$ for single-mode squeezed vacuum}
\label{sec:VacEqMin}
In this appendix, we prove that $\xi_\mathrm{vac}$ minimizes the total variation distance $\delta$ for a single-mode squeezed vacuum target state for all $\xi_\mathrm{vac} < 2.290047\ldots$ (independently of $\eta$), and also for $\xi_\mathrm{vac} \ge 2.290047\ldots$ provided that $\eta < 14/15$. 
We first separate $\delta$ in contributions for which $\mathcal{P}_{\zt}(m)\geq \mathcal{P}_{\xi;\eta}'(m)$ and in contributions for which $ \mathcal{P}_{\zt}(m)<  \mathcal{P}_{\xi;\eta}'(m)$.
We define $I_{+}=\{m:\mathcal{P}_{\zt}(m)-\mathcal{P}_{\xi;\eta}'(m)\geq 0\}$ and $I_{-}=\{m:\mathcal{P}_{\zt}(m)-\mathcal{P}_{\xi;\eta}'(m)<0\}$. Then
\begin{align}
	0&=\sum_{m=0}^\infty \mathcal{P}_{\zt}(m) - \sum_{m=0}^\infty\mathcal{P}_{\xi;\eta}'(m)  \notag\\
	&=\underbrace{\sum_{m\in I_{+}}(\mathcal{P}_{\zt}(m)-\mathcal{P}_{\xi;\eta}'(m))}_{\equiv S_{+}}- \underbrace {\sum_{m\in I_{-}}(\mathcal{P}_{\xi;\eta}'(m)-\mathcal{P}_{\zt}(m))}_{\equiv S_{-}}%\\
%	&=S_{+}-S_{-}\notag,
\end{align}
which implies $S_{+}=S_{-}$. For $\delta$, we then find
\begin{equation}
	\delta=\frac{1}{2}\sum_{m=0}^\infty|\mathcal{P}_{\zt}(m)-\mathcal{P}_{\xi;\eta}'(m)|=\frac{1}{2}(S_{+}+S_{-})=S_{-}.
\end{equation}

As a next step, we examine the value of $S_{-}$ for a fixed $\zt>0$ at $\xi=\xi_{\mathrm{vac}}$, which corresponds to $\mathcal{P}_{\zt}(0)=\mathcal{P}_{\xi;\eta}'(0)$.
In the following, we assume that $S_{-}$ consists only of odd photon-number contributions, i.e., $I_{-}=\{1,3,5,\ldots\}$, at this particular point, which we explicitly prove in the next section \ref{sec:proofIminus}.
Thus,
\begin{align}
\delta = S_{-}=\sum_{m=0}\mathcal{P}_{\xi_\mathrm{vac};\eta}'(2m+1) = 2\delta_{\mathrm{odd}},
\end{align}
at $\xi = \xi_{\mathrm{vac}}$.
Furthermore, we are able to bound $\delta$ for $\xi \neq \xi_{\mathrm{vac}}$.
For $\xi < \xi_{\mathrm{vac}}$, the vacuum component of the lossy distribution is larger than the target distribution, $\mathcal{P}_{\zt}(0)<\mathcal{P}'_{\xi;\eta}(0)$. 
This implies that $m=0$ contributes to the set $I_{-}$. 
All odd indices also belong to $I_{-}$ in general, since $\mathcal{P}_{\zt}(2n+1) = 0$ while the lossy state contains odd photon-number population for any $\xi$.
Therefore, we can lower bound $\delta\geq2\delta_{\mathrm{vac}}+2\delta_{\mathrm{odd}}$ for $\xi<\xi_\mathrm{vac}$.
For $\xi\geq \xi_\mathrm{vac}$, we have $\delta\geq2\delta_{\mathrm{odd}}$ as discussed in the main text.

In the proof that $I_{-}$ only contains odd photon-number probabilities at $\xi= \xi_\mathrm{vac}$, we show the strict inequality $\mathcal{P}_{\zt}(2m) > \mathcal{P}_{\xi_{\mathrm{vac}};\eta}'(2m)$ for $m>0$, see Sec.~\ref{sec:proofIminus}.
By continuity of the photon-number probabilities, there exists an $\epsilon>0$ such that the ordering of the even photon-number probabilities remains unchanged and one still has $\mathcal{P}_{\zt}(2m) > \mathcal{P}_{\xi+\epsilon;\eta}'(2m)$ for all $m>0$.
So for values slightly larger than $\xi_\mathrm{vac}$ we still find $\delta = 2\delta_{\mathrm{odd}}$.
A completely analogous argument applies to values of $\xi$ slightly smaller than $\xi_{\mathrm{vac}}$.
Consequently for these values,
\begin{equation}
	I_{-} = \{0\} \cup \{1,3,5,\ldots\},
\end{equation}
and thus $\delta$ acquires an additional contribution from the vacuum term
\begin{equation}
	\delta = 2\delta_{\mathrm{vac}} + 2\delta_{\mathrm{odd}}.
\end{equation}
Summarizing, $\delta$ transitions from $2\delta_{\mathrm{vac}} + 2\delta_{\mathrm{odd}}$ at $\xi = \xi_{\mathrm{vac}} - \epsilon$, to $2\delta_{\mathrm{odd}}$ at $\xi = \xi_{\mathrm{vac}}$, and remains $2\delta_{\mathrm{odd}}$ for $\xi = \xi_{\mathrm{vac}} + \epsilon$.
Hence, it remains to show that $2\delta_{\mathrm{odd}}$ is an increasing function for $\xi > \xi_{\mathrm{vac}}$, and that $2\delta_{\mathrm{vac}} + 2\delta_{\mathrm{odd}}$ is decreasing for $\xi < \xi_{\mathrm{vac}}$. 
Once these monotonicities are established, it follows that $\delta$ indeed has a minimum at $\xi = \xi_{\mathrm{vac}}$.

It is straightforward to show that $\delta_{\mathrm{odd}} =  \frac{1}{2} \Big(1 - \frac{1}{\sqrt{1 + 4\eta(1-\eta)\sinh^2 \xi}} \Big)$ is strictly increasing.
However, for $2\delta_{\mathrm{vac}} + 2\delta_{\mathrm{odd}}$, see Eq.~\eqref{eq:lowerbounddeltaSMSV}, the monotonicity analysis is more involved. 
One finds that its derivative with respect to $\xi$ is strictly negative for all $\xi < 2.290047\dots$, ensuring that the function is decreasing in this regime.
For larger values of $\xi$, the sign of the derivative depends on $\eta$ and can be determined by a polynomial inequality. It stays strictly negative provided $\eta < 14/15$.
For larger transmissivities, the derivative can become positive leading to $\xi_{\rmmin} \lesssim \xi_{\mathrm{vac}}$.
A detailed analysis is provided in Appendix~\ref{sec:EdgeCase}.

\subsection{Proof that $I_{-}=\{1,3,5,\ldots\}$}
\label{sec:proofIminus}
We consider $\Delta\mathcal{P}(m)=\mathcal{P}_{\zt}(m)-\mathcal{P}_{\xi;\eta}'(m)$  and show that $\Delta\mathcal{P}(2m) > 0$ at $\xi = \xi_\mathrm{vac}$ for all $m>0$ when the vacuum probabilities are matched.
We have
\begin{equation}
	\mathcal{P}_{\zt}(2m)=\frac{(2m)!}{2^{2m}(m!)^2} \frac{\tanh^{2m}(\zt)}{\cosh(\zt)} \equiv C_{m} \frac{\tanh^{2m}(\zt)}{\cosh(\zt)}
\end{equation}
for the target photon-number probabilities, and
\begin{equation}
	\mathcal{P}_{\xi;\eta}'(2m)=\sum_{j=m}^\infty \mathcal{P}_{\xi}(2j)\binom{2j}{2m}\eta^{2m}(1-\eta)^{2j-2m}
\end{equation}
for the lossy photon-number probabilities.
For matching vacuum probabilities, i.e., $\mathcal{P}_{\zt}(0)=\mathcal{P}_{\xi_\mathrm{vac};\eta}'(0)$, we get
\begin{equation}
	\frac{1}{\cosh(\zt)}=\frac{1}{\cosh(\xi_\mathrm{vac})}\sum_{j=0}^\infty C_{j}\tanh^{2j}(\xi_\mathrm{vac})(1-\eta)^{2j}.
\end{equation}
For $\mathcal{P}_{\zt}(2m)-\mathcal{P}_{\xi_\mathrm{vac};\eta}'(2m)$, we then have
\begin{align}
	\Delta\mathcal{P} & = \frac{C_{m}\tanh ^{2m}\zt}{\cosh \xi_\mathrm{vac}} \sum_{j=0}^\infty C_{j} (1-\eta)^{2j} \tanh^{2j} \xi_\mathrm{vac} \notag \\
	&\quad - \frac{\eta^{2m}}{\cosh \xi_\mathrm{vac}}\sum_{j=m}^\infty C_{j} \binom{2j}{2m} (1-\eta)^{2j-2m} \tanh ^{2j}\xi_\mathrm{vac}.
\end{align}

Next, we express $\tanh^{2}(\zt)$ in terms of $\tanh^{2}(\xi_\mathrm{vac})$ by using the explicit solution for $\xi_\mathrm{vac} = \arsinh\Big(\frac{\sinh\zt}{\sqrt{\eta(2-\eta)}}\Big)$, yielding
\begin{equation}
	\tanh^{2}(\zt)= \frac{(2\eta-\eta^{2})\tanh^{2}(\xi_\mathrm{vac})}{1-\tanh ^{2}(\xi_\mathrm{vac})(1-\eta)^{2}}.
\end{equation}
Defining $z=(1-\eta)^{2}\tanh ^{2}(\xi_\mathrm{vac})$, we combine terms to
\begin{align}
	\Delta\mathcal{P}&=\frac{\binom{2 m}{m}}{2^{2m}\sqrt{1-z} \cosh \xi_\mathrm{vac}}\Bigg[\!\left(\frac{(\eta -2) \eta  \tanh^2 \xi_\mathrm{vac}}{z-1}\right)^m\notag\\
	&\quad-(\eta  \tanh\xi_\mathrm{vac})^{2 m} \left(z-1\right)^{-2 m}\! \sum_{k=0}^m \binom{m}{k}^{2} \frac{k!}{\left( \frac{1}{2} \right)^{(k)}}z^{k}\Bigg] \notag \\
	&=\frac{\binom{2 m}{m} \eta^m\tanh ^{2m} \xi_\mathrm{vac}}{2^{2m}(1-z)^{2m}\sqrt{1-z}\cosh\xi_\mathrm{vac} }\notag\\
	&\quad\times\underbrace{\Bigg(\!(2-\eta)^{m}(1-z)^{m}-\eta^{m}\! \sum_{k=0}^m \binom{m}{k}^{2} \frac{k!}{\left( \frac{1}{2} \right)^{(k)}}z^{k}\!\!\Bigg)}_{B(z)}\!.
\end{align}
The finite sum in $B(z)$ could equivalently be written as a hypergeometric function,
\begin{equation}
\sum_{k=0}^m \binom{m}{k}^{2} \frac{k!}{\left( \frac{1}{2} \right)^{(k)}}z^{k}
={}_2F_{1}\!\left(-m,-m;\frac12;z\right),
\end{equation}
highlighting the same hypergeometric structure that appears in the photon-number probabilities in Eq.~\eqref{eq:PNSsqth}.
As $z<1$ for $\eta \in(0,1)$ and $\xi_\mathrm{vac} < \infty$, the factor separated in the first line on the right-hand side is positive.
Thus, it suffice to show $B(z) > 0$.

The argument $z$ ranges from $z=0$ for $\xi_\mathrm{vac} = 0$ to $z=z_{\mathrm{max}}=(1-\eta)^{2}$ for $\xi_\mathrm{vac}\to\infty$.
We now examine $B(z)$ at these two boundary values and study its monotonicity.
For $z=0$ we get
\begin{equation}
	B(0)=(2-\eta)^{m}-\eta^{m}>0.
\end{equation}
The derivative
\begin{equation}
	B'(z)=-m(2-\eta)^{m}(1-z)^{m-1}-\eta^{m}\sum_{k=1}^m k \binom{m}{k}^{2} \frac{k!}{\left( \frac{1}{2} \right)^{(k)}}z^{k-1}
\end{equation}
is strictely negative on $[0,z_{\mathrm{max}}]$.
Hence, $B(z)$ is monotonically decreasing and it now sufficies to show that $B(z_{\mathrm{max}}) > 0$. 
By factoring $\eta^m$, we obtain
\begin{equation}
	B(z_{\mathrm{max}})=\eta^{m}\left(\!(2-\eta)^{2m}-\sum_{k=0}^m \binom{m}{k}^{2} \frac{k!}{\left( \frac{1}{2} \right)^{(k)}}(1-\eta)^{2k}\right),
\end{equation}
yielding a sufficient expression for $\Delta\mathcal{P}(2m) > 0$,
\begin{equation}
	\sum_{k=0}^m \binom{m}{k}^{2} \frac{k!}{\left( \frac{1}{2} \right)^{(k)}}(1-\eta)^{2k}<(2-\eta)^{2m}.
\end{equation}
To show this, we reexpress the sum on the left-hand side as
\begin{align}
	\sum_{k=0}^m \binom{m}{k}^{2} \frac{k!}{\left( \frac{1}{2} \right)^{(k)}}(1-\eta)^{2k}&=\sum_{k=0}^m \binom{m}{k}^{2} \frac{4^{k}(k!)^{2}}{(2k)!}(1-\eta)^{2k}\notag \\
	&=\sum_{k=0}^m \binom{m}{k}^{2} \frac{4^{k}}{\binom{2k}{k}}(1-\eta)^{2k}.
\label{eq:sumComparison1}
\end{align}
Next, we use the binomial series to express $(2-\eta)^{2m}$ and split the sum into even and odd contributions
\begin{align}
	(2-\eta)^{2m}&=\sum_{k=0}^{2m}\binom{2m}{k}(1-\eta)^{k}\notag\\
	&=\sum_{k=0}^{m}\binom{2m}{2k}(1-\eta)^{2k} \notag\\
	&\quad+\sum_{k=0}^{m-1}\binom{2m}{2k+1}(1-\eta)^{2k+1}\notag\\
	&\equiv S_{\mathrm{even}}+S_{\mathrm{odd}}.
\end{align}
We can now lower bound this by multiplying another factor $(1-\eta)$ to the odd terms, yielding
\begin{equation}
	S_{\mathrm{even}}+S_{\mathrm{odd}}>S_{\mathrm{even}}+(1-\eta)S_{\mathrm{odd}}.
\end{equation}
This can be reexpressed using properties of the binomial coefficient 
\begin{align}
	&S_{\mathrm{even}}+(1-\eta)S_{\mathrm{odd}} \notag \\
	& =\sum_{k=0}^{m}\binom{2m}{2k}(1-\eta)^{2k}
	+\sum_{k=0}^{m-1}\binom{2m}{2k+1}(1-\eta)^{2k+2} \notag\\
	&=1+\sum_{k=1}^{m}\Bigg[\binom{2m}{2k}
	+\binom{2m}{2k-1}\Bigg](1-\eta)^{2k} \notag\\
	&=1+\sum_{k=1}^{m}\binom{2m+1}{2k}(1-\eta)^{2k} \notag\\
	&=\sum_{k=0}^{m}\binom{2m+1}{2k}(1-\eta)^{2k}.
\label{eq:sumComparison2}
\end{align}
As we lower bounded $(2-\eta)^{2m}$, we can now compare coefficients of the two sums given in Eqs.~\eqref{eq:sumComparison1} and \eqref{eq:sumComparison2},
\begin{align}
	&\quad &\binom{2m+1}{2k}&>\binom{m}{k}^{2} \frac{4^{k}}{\binom{2k}{k}}\notag\\
	&\Leftrightarrow &\binom{2m+1}{2k}\binom{2k}{k}&>\left(\binom{m}{k}2^{k}\right)^{2} \notag\\
	&\Leftrightarrow &\frac{\binom{2m+1}{2k}\binom{2k}{k}}{\left(\binom{m}{k}2^{k}\right)^{2}}&>1 
\end{align}
Lastly we express the binomial coefficients using factorials
\begin{align}
	\frac{\binom{2m+1}{2k}\binom{2k}{k}}{\left(\binom{m}{k}2^{k}\right)^{2}}&=\frac{(2m+1)!\,(2k)!\,(k!)^{2}\,((m-k)!)^{2}}{(m!)^{2}\,(2k)!\,(2m+1-2k)!\,k!\,k!\,4^{k}} \notag\\
	&=\frac{\frac{(2m+1)!}{(2m+1-2k)!}}{\left[\frac{m!}{(m-k)!}2\right]^{2}}=\frac{ \prod_{i=0}^{2k-1}(2m+1-i) }{\left( \prod_{i=0}^{k-1}2(m-i) \right)^{2}} \notag\\
	&=\frac{ \prod_{i=0}^{k-1}(2m+1-2i)(2m-i) }{\prod_{i=0}^{k-1}4(m-i)(m-i)} \notag\\
	&=\prod_{i=0}^{k-1} \frac{\left( m-i+\frac{1}{2} \right)}{m-i} \notag\\
	&=\prod_{i=0}^{k-1}\left( 1+ \frac{1}{2(m-i)} \right)>1 \quad\forall\, k>0
\end{align}
This is true as $m>i$, since $i$ runs from $0$ to $k-1$ and $k$ runs from $0$ to $m$.
This implies that at the point where $\xi=\xi_\mathrm{vac}$, all even photon-number probabilities of the loss affected state are strictly smaller than the corresponding target state, i.e., $I_{-}=\{1,3,5,\ldots\}$.

\subsection{Monotonicity and edge case}\label{sec:EdgeCase}
Here, we analyze the monotonicity of $2\delta_{\mathrm{vac}}+2\delta_{\mathrm{odd}}$ and show that it is strictly decreasing in $\xi$ for all $\eta$ whenever $\xi<2.290047\ldots$ holds.
For larger values of $\xi$, we derive conditions on $\eta$ under which this monotonicity persists.
We further discuss the edge case, where these conditions fail, and show that the minimizer then satisfies $\xi_{\min}<\xi_{\mathrm{vac}}$.

Starting by computing the derivative of $2\delta_{\mathrm{vac}} + 2\delta_{\mathrm{odd}}$ with respect to $\xi$ and demanding that it is negative, we arrive at a polynomial inequality
\begin{align}
	0&< 4-3 \eta + 12 (2-\eta)(1-\eta) \sinh^2\xi \notag\\
	&\quad + 36 \eta  (2-\eta)^2(1-\eta)^2 (\sinh^2\xi)^2 \notag\\
	&\quad +4 \eta^2 (14 -15\eta) (2-\eta)^2(1-\eta)^2 (\sinh^2\xi)^3 \notag\\
	& \equiv P(\eta, \sinh^2 (\xi)).
\label{eq:monotonicity_condition}
\end{align}
where we ignored an overall positive factor $\eta$, which does not affect the inequality and is therefore irrelevant for the argument.
This defines a region where $2\delta_{\mathrm{vac}}+2\delta_{\mathrm{odd}}$ is strictly decreasing in the $(\eta, \xi)$-plane, bounded by ${P(\eta, \sinh^2 \xi)=0}$.

Interpreting $P$ as a polynomial in $\sinh^2(\xi)$ with coefficients depending on $\eta$, we see that all coefficients are positive for $\eta < 14/15$, ensuring that $P(\eta, \sinh^2 (\xi)) > 0$ for all $\xi$ and $\eta<14/15$.
Thus for $\eta < 14/15$, the function $2\delta_{\mathrm{vac}} + 2\delta_{\mathrm{odd}}$ is strictly decreasing for all $\xi$.
For $\eta > 14/15$, the leading-order coefficient becomes negative and the polynomial $P$ exhibits one positive root in $\sinh^2(\xi)$ according to Descartes' rule of signs. Thus, for $\sinh^2(\xi)$ larger than this root, the function $2\delta_{\mathrm{vac}} + 2\delta_{\mathrm{odd}}$ is increasing.
Fixing $\eta$, one can numerically solve for the root of $P$ depending on $\eta \in (\frac{14}{15},1)$. 
We find that the minimum root is given by $\xi=2.290047\ldots$.
Subsequently, for $\xi < 2.290047\ldots$, the function $2\delta_{\mathrm{vac}} + 2\delta_{\mathrm{odd}}$ is strictly decreasing for all $\eta$.

We now examine in more detail the parameter regime in which $\xi_{\rmmin} \neq \xi_{\mathrm{vac}}$ and develop intuition for the behavior of the minimizer.
To do this, we first fix $\eta$ for an arbitrary value with $\eta>14/15$.
There exists a $\bar{\xi}$ that minimizes $2\delta_{\mathrm{vac}} + 2\delta_{\mathrm{odd}}$, i.e., $2\delta_{\mathrm{vac}} + 2\delta_{\mathrm{odd}}$ is decreasing for $\xi < \bar{\xi}$ and increasing for $\xi > \bar{\xi}$. Note that the minimum of $2\delta_{\mathrm{vac}} + 2\delta_{\mathrm{odd}}$ does not necessarily coincide with the minimum of the full total variation distance $\delta$. If $\bar{\xi}$ is slightly smaller than $\xi_{\mathrm{vac}}$, then $2\delta_{\mathrm{vac}} + 2\delta_{\mathrm{odd}}$ starts increasing already before $\xi_{\mathrm{vac}}$. In this case the global minimum of $\delta$ is attained at a value $\xi_{\min}<\xi_{\mathrm{vac}}$, typically only slightly below $\xi_{\mathrm{vac}}$.

Next, we consider how this picture changes as $\tilde{\xi}$, and thus $\xi_{\mathrm{vac}}$, is increased.
In this case, the separation between $\bar{\xi}$ and $\xi_{\mathrm{vac}}$ becomes larger as $\bar{\xi}$ does not depend on $\zt$, and it may occur that the index set $I_{-}$ no longer has the form $I_- = \{0\}\cup\{1,3,5,\ldots\}$, but instead also contains $2$.
This introduces a new candidate for the minimizer $\xi_{\min}$, the value $\xi = \xi_2$ at which the two-photon-number probabilities coincide,
\begin{equation}
	\mathcal{P}_{\xi;\eta}'(2) = \mathcal{P}_{\zt}(2).
\end{equation}

For $\xi > \xi_2$, the quantity $2\delta_{\mathrm{vac}} + 2\delta_{\mathrm{odd}}$ is increasing.
Therefore, to determine whether $\xi_2$ indeed corresponds to the global minimum, one must analyze whether the extended expression
\begin{equation}
	2\delta_{\mathrm{vac}} + 2\delta_{\mathrm{odd}} + \bigl(\mathcal{P}_{\xi;\eta}'(2) - \mathcal{P}_{\zt}(2)\bigr)
\end{equation}
is strictly decreasing for all $\xi \leq \xi_2$.
This again leads to a condition on $\eta$ analogous to the one obtained previously.

From a more general perspective, increasing $\tilde{\xi}$ enlarges the index set $I_{-}$ in a stepwise manner by successively adding even photon-number indices.
Suppose that, for a given value of $\tilde{\xi}$, the set $I_{-}$ contains the even indices $2,4,\dots2m$.
In this case, the relevant quantity governing the total variation distance is
\begin{equation}
	2\delta_{\mathrm{vac}} + 2\delta_{\mathrm{odd}} + \sum_{k=1}^{m} \bigl(\mathcal{P}_{\xi;\eta}'(2k) - \mathcal{P}_{\zt}(2k)\bigr).
\end{equation}
For this partial sum one can again derive a monotonicity condition in terms of $\eta$ (similar to Eq.~\ref{eq:monotonicity_condition}).
If this condition is satisfied, the minimum of the total variation distance is attained at the value $\xi_{\min} = \xi_{2m}$.
If, on the other hand, the monotonicity condition is violated, but the next even index $2(m+1)$ is not yet contained in $I_{-}$, an intermediate regime arises.
In this regime, the minimum is determined by the minimum of the above partial sum for $\xi < \xi_{2m}$.

To quantify the deviation from the $I_{-} = \{0,1,3,5,\dots \}$ scenario, we evaluate numerically the parameter regimes in which even photon-number contributions enter $I_{-}$.
We fix, $\eta= 0.97$.
Then the smallest $\zt$, such that $I_{-}$ also includes $2$ is around $\zt\approx2.38$ and we have $\xi_\mathrm{min}=\xi_2\approx2.317$.
Note that the minimum $\xi_\mathrm{min}$ is smaller than $\zt$, which is counterintuitive, as loss generally reduces the number of photons and thus one would expect the optimal squeezing to be larger than $\zt$ to compensate for this effect.
Nevertheless, we still find that $\xi_\mathrm{vac}$ is a good estimator as the total variation distance at $\xi_\mathrm{vac}$ is $\delta(\xi_\mathrm{vac})\approx0.2600$, while the total variation distance at the actual minimum $\xi_\mathrm{min}$ is $\delta(\xi_\mathrm{min})\approx0.2598$.

\section{General single-mode Gaussian states as approximating states}
\label{sec:GeneralGaussian Approximation}

\subsection{Displaced squeezed vacuum as approximating state for squeezed vacuum}\label{sec:SCasApprox}
In Sec.~\ref{sec:smSqueezedState} of the main text, we considered changing the input squeezing parameter to reduce the difference in photon-number distribution between the target state (single-mode squeezed vacuum) and a loss-affected single-mode squeezed vacuum state.
We now extend the optimization domain to include general single-mode pure Gaussian states affected by loss, i.e., lossy displaced squeezed vacuum states $\mathcal{E}_{\eta}(\hat{D}(\alpha)\hat{S}(\xi)\op{0}\hat{S}^\dag(\xi)\hat{D}^\dag(\alpha))$, such that the minimization of $\delta\big(\hat{\rho}(\tilde{\xi}),\hat{\rho}'(\xi,\alpha;\eta)\big)$ is performed jointly over the squeezing parameter $\xi$ and the displacement $\alpha$.

As discussed in the main text at the beginning of Sec.~\ref{sec:smDisplacedSqueezedState}, the photon-number statistics of displaced squeezed vacuum does only depend on three real-valued degrees of freedom, the moduli of the squeezing and displacement parameters as well as on the relative phase between both. Therefore, we again choose the squeezing $\xi$ to be real and positive in this section, while the displacement $\alpha=\abs{\alpha}\E^{\I \varphi}$ is a complex parameter.

Due to the additional displacement, any loss-affected displaced squeezed vacuum state generally has a nonzero quadrature mean $\bar{r}'$.
However, the phase-space measures as well as the fidelity optimization and the displacement correction, force an equal quadrature mean $\bar{r} = \bar{r}'$ (see Eqs.~\eqref{eqn:Fidelity}-\eqref{eqn:BHA}) which uniquely determines the displacement $\alpha$ independent of the squeezing parameters. Since $\bar{r} = 0$ for the target state, we obtain $\alpha = 0$.
Consequently, extending the optimization domain to include loss-affected displaced squeezed vacuum states does not lead to any improvement for these measures compared to the previously discussed nondisplaced case.
Similarly, $\delta_\mathrm{PS} $ would increase if the displacement of the loss-affected state deviates from that of the target, as any misalignment in phase space reduces the overlap of their corresponding Wigner functions.
Hence, also for the phase-space optimization based on the total variation distance between the Wigner functions, the inclusion of displacement does not lead to any improvement over the displacement-free case.

For the photon-number-moment correction schemes and the vacuum-overlap correction, we have an intrinsic ambiguity because the relevant quantities now depend on both the input squeezing $\xi$ and the displacement $\alpha$. Enforcing a single correction condition, e.g., matching mean, variance, or vacuum probability, does not uniquely determine the Gaussian input state.
To resolve this ambiguity, one could impose further correction conditions, e.g., matching mean photon number and photon-number variance simultaneously. 
However, it turns out that it is not possible to match the mean photon number and the photon-number variance of a single-mode squeezed vacuum and a lossy single-mode displaced squeezed vacuum at the same time for any admissible physical input parameters.
This is not immediately obvious as the additional degrees of freedom introduced by the displacement would allow for a more flexible shaping of the photon-number distribution, which could in principle enable simultaneous matching of both mean and variance.
Instead we impose either the mean or the variance correction condition, which defines a two dimensional surface in the three dimensional parameter space spanned by  $\xi$ and $\alpha$ and then minimize the resulting $\delta\big(\hat{\rho}(\tilde{\xi}),\hat{\rho}'(\xi,\alpha;\eta)\big)$ over this remaining surface.
This minimization results in $\alpha = 0$ for all correction schemes.

All different mitigation methods suggest that $\alpha = 0$.
However, this is not guaranteed for the minimum of $\delta$ a priori. 
Analytically, one can show that when extending from a single-mode squeezed vacuum to a single-mode displaced squeezed vacuum state, the previously identified global minimum persists as (at least) a local minimum in the enlarged parameter space $(\xi,\alpha)$.
Using the fact that $\mathcal{P}_{\zt}(2m) > \mathcal{P}'_{\xi_\mathrm{vac};\eta}(2m)$, a small perturbation in form of a displacement of the state does not lead to a change of the set $I_{-}$, i.e., it still consists only of the odd photon numbers.
However, this perturbation does increase the value of $\delta_\mathrm{odd}$ and thus leads to a larger $\delta$. 
Moreover, numerical tests indicate that the global minimum remains at $\alpha = 0$.

\subsection{Mixed Gaussian state as approximating state for single-mode squeezed vacuum}\label{sec:SthasApprox}
Next, we consider a second extension of the state class by considering squeezed thermal states as an ansatz for $\hat{\rho}'$.
Any single-mode (mixed) Gaussian state $\hat{\rho}'$ with zero mean quadrature can be expressed as
\begin{equation}
\hat{\rho}'(\xi, \mu; \eta) = \hat{S}(\xi) \, \hat{\rho}_{\mathrm{th}}(\mu) \, \hat{S}^\dagger(\xi),
\end{equation}
where $\hat{\rho}_{\mathrm{th}}(\mu)$ denotes a thermal state with mean photon number $\mu$. 
As in the previous case, we analyze whether extending the parameter space in this way leads to a better approximation of the target distribution in terms of the total variation distance $\delta\big(\hat{\rho}(\tilde{\xi}), \hat{\rho}'(\xi, \mu; \eta)\big)$.

For the phase-space measures and the fidelity, we do not expect any improvement, since adding thermal noise to the approximating state can only degrade these similarity quantifiers for a squeezed-vacuum target. Indeed, evaluating the phase-space measures and the fidelity as functions of the squeezing $\xi$ and the thermal occupation $\mu$ via the covariance matrix and setting the derivatives with respect to $\xi$ and $\mu$ to zero, we find that all physically admissible extrema occur at $\mu=0$. The extrema therefore reduce to the previously analyzed case of nonthermal squeezed states. Any additional stationary points are outside the physical parameter domain, either corresponding to $\mu < 0$ or to $\eta \notin (0,1)$, and are thus unphysical.

For the photon-number mean, the photon-number variance, and the vacuum overlap, we face a similar ambiguity as in the case of displaced squeezed vacuum.
As in the displaced case, it is not possible to match the photon-number mean and the photon-number variance simultaneously for physically admissible parameters.
Imposing a chosen correction prescription results this time in a one-parameter family of solutions of $\xi$ and $\mu$ and we minimize $\delta$ along this curve. 
The optimal solution is consistently found at $\mu = 0$.

When a small thermal perturbation is added to the state that minimizes $\delta$ under the constraint that squeezed thermal states are used as approximating states, the set $I_{-}$ again contains only odd indices. Consequently, the minimum is attained at $\xi_{\mathrm{vac}}$ with $\mu = 0$.
We conjecture that extending the class of input states to include squeezed thermal states does likely not yield a better match with the photon-number distribution of the target state.
Numerical minimization of $\delta$ supports this conjecture as it consistently returns $\mu = 0$ (under the constraint $\mu\geq0$).
This confirms that within the considered parameter space, thermal contributions do not provide any benefit.

Combining the findings from Sec.~\ref{sec:SCasApprox} and Sec.~\ref{sec:SthasApprox}, it is evident that neither the inclusion of displacement nor thermal contributions lead to any improvement in phase-space-based measures, the fidelity, or photon-statistics based measures.
We performed a direct minimization of $\delta\big(\hat{\rho}(\tilde{\xi}),\hat{\rho}'(\xi,\alpha, \mu;\eta)\big)$, treating $\xi$, $\alpha$, and $\mu$ as free variables. The minimum is still found at $\alpha = \mu = 0$ and $\xi = \xi_{\mathrm{vac}}$, within numerical accuracy. 
All analytical and numerical results consistently indicate that, within the class of Gaussian states considered, extending beyond the loss-affected squeezed vacuum does not improve the approximation of the target state.

\subsection{Mixed Gaussian states as approximating state for displaced squeezed vacuum}
In Sec.~\ref{sec:smDisplacedSqueezedState}, we have only considered pure states for mitigating the impact of loss on the photon-number distribution of displaced squeezed states. 
In the following, we extend this discussion to mixed Gaussian states, i.e., displaced squeezed thermal states
\begin{equation}
	\hat{\rho}(\xi, \alpha, \mu)=\hat{D}(\alpha)\hat{S}(\xi)\hat{\rho}_\mathrm{th}(\mu)\hat{S}\dg(\xi)\hat{D}\dg(\alpha).
\end{equation}
For the fidelity and phase-space optimizations, the task of finding optimal quadrature means and covariances of the Gaussian state decouples. It is straightforward to impose the condition $\bar{r} = \bar{r}'$ implying again $\alpha = \tilde\alpha/\sqrt{\eta}$. This reduces the optimization over $\xi$ and $\mu$ to the corresponding squeezed-vacuum problem, for which we found that these measures admit no physically admissible solution.

As shown in Sec.~\ref{sec:SCmeanVariance}, it is possible to simultaneously match the photon-number mean and variance of the target and loss-affected states. 
This implies that physically valid solutions exist also within the extended class of input states.
Numerical results show that these new solutions do not yield any improvement in the total variation distance $\delta\big(\hat{\rho}(\tilde{\xi}, \tilde{\alpha}), \hat{\rho}'(\xi, \alpha, \mu; \eta)\big)$ compared to the case without extending the input state class to include thermal contributions. 
Direct numerical minimization of the total variation distance confirms that including thermal contributions in the input state does not improve the approximation of the lossy target compared to pure displaced squeezed states.

\section{Photon-number probabilities for lossy displaced squeezed vacuum states}\label{sec:PofLossySC}
To compute the photon-number distribution of a lossy displaced squeezed vacuum state with squeezing parameter ${\xi=\abs{\xi}\E^{\I \varphi_\xi}}$ and displacement $\alpha=\abs{\alpha}\E^{\I \varphi_\alpha}$, we exploit an equivalent Gaussian-state representation of the loss channel. In particular, the lossy state can be written as a displaced squeezed thermal state, $ \mathcal{E}_\eta\big( \hat{D}(\alpha)\hat{S}(\xi) \op{0} \hat{S}^\dagger(\xi)\hat{D}^\dagger(\alpha)\big) = \hat{D}(\sqrt{\eta}\alpha) \hat{S}(\zeta)\hat{\rho}_{\mathrm{th}}(\mu)\hat{S}^\dagger(\zeta) \hat{D}^\dagger(\sqrt{\eta}\alpha) $, generalizing the correspondence between a lossy squeezed vacuum and a squeezed thermal state.

The effect of the loss is absorbed into effective parameters as in Eqs.~\eqref{eq:ThermalMean} and~\eqref{eq:zeta}, which are derived by matching the first and second quadrature moments of the lossy state to those of the squeezed displaced thermal state. In particular, the phases of the displacement and squeezing parameters remain unchanged.
The resulting Gaussian state can be written as an exponential of a quadratic form in the bosonic ladder operators.
Expressing this operator in normal-ordered form~\cite{banDecompositionFormulasSu11993} and evaluating the diagonal Fock-basis elements $\langle m | \hat{\rho} | m \rangle$, we obtain the photon-number probabilities.
The resulting expression is given by
\begin{widetext}

\begin{equation}
\begin{aligned}
	\mathcal{P}'_{z,\alpha;\eta}(m)&=\mathrm{exp}\left(\frac{\abs{\breve\alpha}^2 \left(-2 \mu+\sinh (2 \abs{\zeta} ) \cos (2 \gamma  -\varphi_\xi )-2 \cosh ^2(\abs{\zeta} )\right)}{2 \left(\mu \left(\mu+\cosh (2 \abs{\zeta} )+1\right)+\cosh ^2(\abs{\zeta} )\right)}\right) \\
	&\quad\times \sum _{k=0}^m\sum _{k'=0}^m\sum _{\ell=0}^{\left\lfloor \frac{m-k}{2}\right\rfloor }\sum _{\ell'=0}^{\left\lfloor \frac{m-k'}{2}\right\rfloor }\left( \frac{(-1)^{\ell+\ell'}\;2^{k+k'-2\ell-2\ell'} m!\;\delta _{k+2 \ell,k'+2 \ell'}}{\sqrt{(m-k-2 \ell)!\;(m-k'-2 \ell')!}\;k!\; k'! \;\ell!\;\ell'!} \right.\\
	&\quad\times\left(\E^{-i \varphi_\xi } \sinh (2\abs{\zeta} ) \left(2 \mu+1\right)\right)^\ell
	\left(\abs{\breve\alpha}  \E^{-i \gamma } \cosh (\abs{\zeta} ) \left(\mu+1\right)+\abs{\breve\alpha}  \E^{i (\gamma-\varphi_\xi) } \sinh (\abs{\zeta} ) \mu\right)^k\\
	&\quad\times\left(\E^{i \varphi_\xi } \sinh (2\abs{\zeta} )  \left(2 \mu+1\right)\right)^{\ell'}
	\left( \abs{\breve\alpha}  \E^{ i \gamma } \cosh (\abs{\zeta} ) \left(\mu+1\right)+\abs{\breve\alpha}  \E^{i (\varphi_\xi-\gamma) } \sinh (\abs{\zeta} ) \mu\right)^{k'}\\
	&\quad\times\left(\frac{\sinh ^2(\abs{\zeta} )}{\mu+1}+\frac{\cosh ^2(\abs{\zeta} )}{\mu}+1\right)^{k-m+2\ell}\left(\mu \left(\mu+\cosh (2 \abs{\zeta} )+1\right)+\cosh ^2(\abs{\zeta} )\right)^{-(k+k'+\ell+\ell'+1/2)},
\end{aligned}
\end{equation}
\end{widetext}
where 
\begin{align*}
	\mu &= \frac{1}{2} \sqrt{ \eta^2 + (1-\eta)^2 + 2\eta(1-\eta)\cosh(2\abs{\xi})} - \frac{1}{2}, \\
	\zeta &= \frac{1}{2} \arsinh\bigg(\frac{\eta \sinh(2\abs{\xi})}{2\mu + 1} \bigg)\E^{\I \varphi_\xi}\equiv\abs{\zeta}\E^{\I \varphi_\xi}, \\
	\abs{\breve\alpha}&=\sqrt{\eta}\abs{\alpha} \sqrt{\sinh (2 \abs{\zeta} ) \cos (2 \varphi_\alpha -\varphi_\xi )+\cosh (2 \abs{\zeta} )}, \\
	\gamma&=\arg \Big(\sqrt{\eta}\abs{\alpha} \E^{-i (\varphi_\alpha -\varphi_\xi )} \sinh ( \abs{\zeta} ) \\
	&\qquad\quad\; +\sqrt{\eta}\abs{\alpha} \E^{i \varphi_\alpha } \cosh ( \abs{\zeta} )\Big).
\end{align*}

\section{Phase instabilities}
\label{sec:PhaseDrifts}

\begin{figure}[t]   
    \centering
    \includegraphics[width=1\columnwidth]{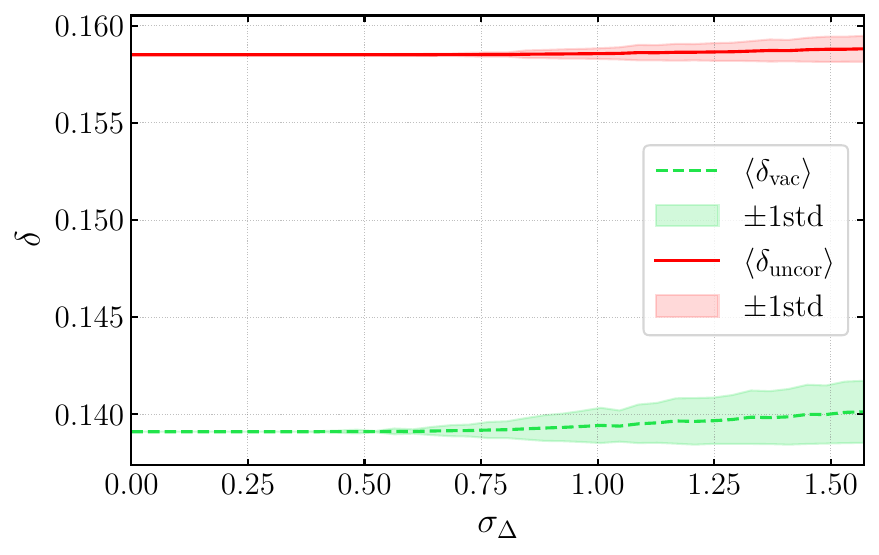}
	\caption{Total variation distance $\delta$ between the target photon-number distribution and the one obtained by the mitigation method based on the vacuum-overlap correction with fixed squeezing ratio, as a function of the standard deviation $\sigma_\Delta$ of the phase noise. The loss model is the same as in Fig.~\ref{fig:SO2Tropolone}, i.e., symmetric loss $1-\eta=0.3$ at all four positions. The results show that even with increasing phase noise, the mitigation method still outperforms not correcting for the effects induced by loss.
	}
	\label{fig:SO2PhaseNoise}
\end{figure}

In experimental settings, photon loss is not the only source of imperfection. Another limitation in GBS experiments are input phase instabilities, i.e., fluctuations or drifts of the relative phases between the squeezing parameters. Such instabilities can arise from temperature variations or mechanical vibrations that change optical path lengths. Phase instabilities randomize the orientation of the squeezing in phase space across modes. When averaged over, this suppresses phase-sensitive interference effects and leads to deviations in the output photon-number statistics. Without access to the actual phase fluctuations during an experiment, they cannot be corrected for directly on a shot-by-shot basis.

Nevertheless, the mitigation prescriptions considered here can still be beneficial in the presence of unknown phase noise, since loss remains the dominant source of imperfection. To illustrate this, we revisit the tropolone example of Fig.~\ref{fig:SO2Tropolone} and extend the symmetric loss model ($1-\eta=0.3$ at all four loss segments) by additional phase noise. 
We model phase noise by applying a random phase shift $\Delta\varphi_\xi$ to the squeezing phase of the first mode, with $\Delta\varphi_\xi$ drawn from a normal distribution of zero mean and standard deviation $\sigma_\Delta$. For each $\sigma_\Delta$, we generate $1000$ samples of $\Delta\varphi_\xi$ and compute for each sample the output photon-number distribution and its total variation distance $\delta$ to the noise-free target. We perform this procedure both for the unmitigated case (no parameter adjustment to account for loss) and for vacuum-overlap correction with an additional squeezing-ratio constraint, which compensates for loss via a redefinition of the input parameters while leaving the phase instabilities uncorrected.

We plot the sample mean and standard deviation of $\delta$ as a function of $\sigma_\Delta$ in Fig.~\ref{fig:SO2PhaseNoise}.
The case $\sigma_\Delta=0$ reproduces the left panel of Fig.~\ref{fig:SO2Tropolone}. The results for $\sigma_\Delta>0$ demonstrate that vacuum-overlap correction continues to yield a clear reduction in $\delta$ relative to the unmitigated baseline. Photon loss remains the dominant contribution to deviations in the photon-number distribution in this example. Even for comparatively large phase fluctuations $\sigma_\Delta$, the uncorrected baseline $\langle\delta_{\mathrm{uncor}}\rangle$ changes only moderately and shows a small spread, whereas vacuum-overlap correction consistently reduces the mean $\delta$ across the full range of $\sigma_\Delta$. At the same time, once the loss-induced distortion is reduced, the residual sensitivity to phase noise becomes more visible as an increased standard deviation of $\delta$, but the resulting fluctuations remain moderate compared to the loss-driven error level. 
This is consistent with recent work on input phase noise in GBS, which models relative squeezing-phase fluctuations as a dephasing mechanism and finds that such phase noise does not by itself render the problem easy to simulate classically, suggesting that loss remains the dominant practical limitation in many regimes~\cite{paryzkovaInputPhaseNoise2025}. For small phase errors of a few degree, which can be achieved with active phase stabilization, we observe no noticeable deviation from the $\sigma_\Delta = 0$ case.

\bibliography{draft}

\end{document}